\documentclass[
final
]{dmtcs-episciences}

\usepackage[utf8]{inputenc}
\usepackage{subfigure}

\usepackage{amsmath,amssymb,stmaryrd,dsfont,paralist,amsthm}
\usepackage{graphicx}
\usepackage{todonotes}
\usepackage{multicol}
\usepackage{tikz}

\usetikzlibrary{calc}

\newtheorem{theorem}{Theorem}[section]
\newtheorem{lemma}[theorem]{Lemma}

\newtheorem{corollary}[theorem]{Corollary}
\newtheorem{observation}[theorem]{Observation}
\newtheorem{example}[theorem]{Example}
\newtheorem{definition}[theorem]{Definition}
 
\allowdisplaybreaks[3]

\newcommand{\mmon}[0]{M-monoid}

\DeclareMathOperator{\pop}{pop}
\DeclareMathOperator{\push}{push}

\newcommand{\myrm}[1]{\mathrm{#1}}

\newcommand{\Nat}{\mathbb N}

\newcommand{\rhs}{{\myrm{rhs}}}
\newcommand{\maxrk}{{\myrm{maxrk}}}

\newcommand{\rank}{{\myrm{rk}}}
\newcommand{\pos}{{\myrm{pos}}}

\newcommand{\all}{\mathrm{all}}

\newcommand{\height}{{\myrm{height}}}

\newcommand{\lBert}{[\![}% better call it: llbracket
\newcommand{\rBert}{]\!]}% better call it: rrbracket
\newcommand{\seml}{[\![}
\newcommand{\semr}{]\!]}

\newcommand{\beh}[1]{|\!| #1 |\!|}
\newcommand{\sem}[1]{[\![ #1 ]\!]}

\newcommand{\G}{\mathcal G}
\renewcommand{\H}{\mathcal H}

\newcommand{\rk}{\mathrm{rk}}
\newcommand{\pr}{\mathrm{pr}}

\newcommand{\id}{\mathrm{id}}
\newcommand{\true}{\mathrm{true}}

\newcommand{\Reg}{\mathrm{Reg}}
\newcommand{\Regnc}{\mathrm{Reg}_{\mathrm{nc}}}

\newcommand{\TRIV}{\mathrm{TRIV}}
\newcommand{\COUNT}{\mathrm{COUNT}}

\newcommand{\Def}{\mathrm{Def}}
\newcommand{\MExp}{\mathrm{MExp}}

\newcommand{\MSO}{\mathrm{MSO}}

\newcommand{\edge}{\mathrm{edge}}

\newcommand{\M}{\mathrm{M}}

\newcommand{\ourpath}{\mathrm{path}}

\newcommand{\B}{\mathcal B}

\newcommand{\lhs}{\mathrm{lhs}}

\newcommand{\mul}{\mathrm{mul}}

\newcommand{\DS}{\langle \Delta, \Sigma\rangle}

\def\ui#1{^{(#1)}}

\DeclareMathOperator{\PD-op}{\mathrm{P}}
\DeclareMathOperator{\bottom}{\mathrm{bottom}}
\DeclareMathOperator{\ttop}{top}
\DeclareMathOperator{\test}{test}
\DeclareMathOperator{\inc}{inc}
\DeclareMathOperator{\dec}{dec}
\DeclareMathOperator{\zero}{zero}

\newcommand{\Ops}{\myrm{Ops}}

\newcommand{\wt}{\myrm{wt}}

\newcommand{\supp}{\myrm{supp}}
\newcommand{\ycut}{\lfloor\bar{y}\rfloor_{\mathrm{dg}(\bar{a})}}
\newcommand{\yycut}{\lfloor\bar{y}_1\rfloor_{\mathrm{dg}(\bar{a})}}
\newcommand{\yicut}{\lfloor\bar{y}_i\rfloor_{\mathrm{dg}(\bar{a})}}
\newcommand{\ykcut}{\lfloor\bar{y}_k\rfloor_{\mathrm{dg}(\bar{a})}}
\newcommand{\tikzhighlight}[2][]{%
	\tikz[baseline=0, anchor=base, inner sep=0, text height=, text depth=]%
			{\node[#1]{#2};}%
}

\newcommand{\mathhighlight}[2][]{%
	\mathchoice
		{\tikzhighlight[#1]{\(\displaystyle{#2}\)}}
		{\tikzhighlight[#1]{\(\textstyle{#2}\)}}
		{\tikzhighlight[#1]{\(\scriptstyle{#2}\)}}
		{\tikzhighlight[#1]{\(\scriptscriptstyle{#2}\)}}
}

\newcommand{\mhl}[2][]{%
	\mathhighlight[fill=black!20, rounded corners=0.1em, inner sep=0.2em,text height=0.7em,text depth=0.15em,#1]{#2}%
}

\newcommand{\p}{\tikz{
\node(p)[draw,inner sep=0.2em,text height=0.5em,text depth=0.25em,text width=1.3em]{$\gamma_0$};
\draw ($(p.north west) + (0,0.3em)$) -- ($(p.south west) - (0,0.3em)$);
}}
\newcommand{\pp}{\tikz{
\node(p)[draw,inner sep=0.2em,text height=0.5em,text depth=0.25em,text width=1.3em]{$\gamma_0$};
\node(p1)[draw,inner sep=0.2em,text height=0.5em,text depth=0.25em,text width=1.3em,right=-0.1em of p]{$\gamma$};
\draw ($(p.north west) + (0,0.3em)$) -- ($(p.south west) - (0,0.3em)$);
}}
\newcommand{\ppp}{\tikz{
\node(p)[draw,inner sep=0.2em,text height=0.5em,text depth=0.25em,text width=1.3em]{$\gamma_0$};
\node(p1)[draw,inner sep=0.2em,text height=0.5em,text depth=0.25em,text width=1.3em,right=-0.1em of p]{$\gamma$};
\node(p2)[draw,inner sep=0.2em,text height=0.5em,text depth=0.25em,text width=1.3em,right=-0.1em of p1]{$\gamma$};
\draw ($(p.north west) + (0,0.3em)$) -- ($(p.south west) - (0,0.3em)$);
}}

\author{Zolt\'an F\"ul\"op\affiliationmark{1}\thanks{Supported by the NKFI grant no~K~108448}
  \and Luisa Herrmann\affiliationmark{2}\thanks{Supported by DFG Graduiertenkolleg 1763 (QuantLA)}
  \and Heiko Vogler\affiliationmark{2}}
\title[Weighted Regular Tree Grammars with Storage]{Weighted Regular Tree Grammars\\ with Storage}
\affiliation{
  Department of Foundations of Computer Science, University of Szeged, Hungary\\
  Faculty of Computer Science, Technische Universität Dresden, Germany}
\received{2017-5-19}
\revised{2018-6-11}
\accepted{2018-6-22}
\begin{document}
\publicationdetails{20}{2018}{1}{26}{3664}
\maketitle
\begin{abstract}
 We introduce weighted regular tree grammars with storage as combination of (a) regular tree grammars with storage and (b) weighted tree automata over multioperator monoids. Each weighted regular tree grammar with storage generates a weighted tree language, which is a mapping from the set of trees to the multioperator monoid. We prove that, for multioperator monoids canonically associated to particular strong bimonoids, the support of the generated  weighted tree languages can be generated by (unweighted) regular tree grammars with storage.  We characterize the class of all generated weighted tree languages by the composition of three basic concepts. Moreover, we prove results on the elimination of chain rules and of finite storage types, and we characterize weighted regular tree grammars with storage by a new weighted MSO-logic.  
\end{abstract}

\section{Introduction}

In automata theory, weighted string automata with storage are a very recent topic of research \cite{hervog15,hervog16,vogdroher16}. This model generalizes finite-state string automata in two directions. On the one hand, it is based on the concept of ``sequential program + machine", which Scott \cite{sco67} proposed in order to harmonize the wealth of upcoming automaton models. Each of them has a finite control (sequential program) and allows to control computations by means of some storage (machine), like pushdown, stack, nested-stack, or counter. The storage contains configurations which can be tested by predicates and transformed by instructions.
 On the other hand, finite-state automata have been considered in a weighted setting in order to investigate quantitative aspects of formal languages \cite{eil74,salsoi78,kuisal86,berreu88,sak09,drokuivog09}. Weighted automata have been defined over several weight structures such as semirings \cite{eil74}, strong bimonoids \cite{drostuvog10,cirdroignvog10,drovog12}, and valuation monoids \cite{dromei10,dromei11,dromei12}. In \cite{drovog13,drovog14} weighted pushdown automata were investigated where the weights are taken from a unital valuation monoid. The automata studied in \cite{hervog15,vogdroher16} are weighted string automata with arbitrary storage in which the weights are taken from unital valuation monoids.  In \cite{hervog16} weighted symbolic automata with data storage were introduced, which cover weighted string automata with storage and, e.g.,  weighted visibly pushdown automata and weighted timed automata.

Finite-state tree automata and regular tree grammars generalize finite-state string automata and regular grammars, respectively (cf.  \cite{eng75-15,gecste84,gecste97} for surveys\footnote{We use the numbering in the arXiv-version of \cite{gecste84} published in  2015.}).
 Also for tree automata and tree grammars there is a rich tradition of adding storages, like pushdown tree automata \cite{gue83} and tree pushdown automata \cite{schgal85}.  In \cite{eng86} the general concept  of regular tree $S$ grammar was introduced, where $S$ is an arbitrary storage type, e.g., iterated pushdown \cite{eng86,engvog86,engvog88}.
Moreover, tree automata and tree grammars have also  been considered in a weighted setting, in particular, for commutative semirings \cite{aleboz87}, for continuous semirings \cite{esikui03}, for complete distributive lattices \cite{esiliu07}, for fields \cite{berreu82}, for tree valuation monoids \cite{drogoe11}, and multioperator monoids \cite{kui99e,mal04,stuvogfue09}. For a survey on weighted tree automata we refer to \cite{fulvog09new}. 

In this paper we investigate the combination of both generalizations of regular tree grammars, and  we introduce \textit{weighted regular tree grammars with storage}. We consider an arbitrary storage type $S$ with configurations, predicates, and instructions. The generated trees are built up over some ranked alphabet $\Sigma$. The weights are taken from a complete multioperator monoid $K$, which is a commutative  monoid $(K,+,0)$ with a set of operations on $K$. We call such a device an $(S,\Sigma,K)$-regular tree grammar, for short: $(S,\Sigma,K)$-rtg. Each rule of an $(S,\Sigma,K)$-rtg $\G$ has one of the following two forms:
\begin{flalign*}
&A(p) \rightarrow \sigma(A_1(f_1),\ldots,A_k(f_k)) \tag{1}\\
&A(p) \rightarrow B(f) \tag{2}
\end{flalign*}
where  $A$, $A_1$, $\ldots$, $A_k$, and $B$ are nonterminals, $\sigma$ is a terminal in $\Sigma$, $p$ is a predicate of $S$, and $f$, $f_1$, $\ldots$, $f_k$ are instructions of $S$.  We call rules of type (2) chain rules. Each rule is equipped with an operation on $K$, of which the arity is the rank of the symbol $\sigma$ (for rules of type (1)) or  one (otherwise). 

\begin{figure}
\footnotesize\centering
\begin{tikzpicture}[sibling distance=4em, level distance=4.5em,
  every node/.style = {align=center}]]

  \pgfdeclarelayer{bg}    % declare background layer
  \pgfsetlayers{bg,main}  % set the order of the layers (main is the standard layer)

\begin{scope}[sibling distance=12em]
 \node {$A(p)\to\sigma(B(f_1),C(f_2))$}
    child { node[] {$B(p_1)\to\gamma(D(f_4))$}
      child { node {$D(p_2)\to\alpha$} } }
    child { node {$C(p_3)\to E(f_3)$}
      child { node {$E(p_4)\to\beta$} } };
 \end{scope}

\begin{scope}[sibling distance=12em, xshift=7.5em,yshift=0.8em,black!50]
\begin{pgfonlayer}{bg}
 \node[draw=black!50] {$c_0$}
   child[edge from parent/.style={draw,densely dashed,black!50,->}] { node[draw=black!50,xshift=-2em] {$c_1$} 
      child { node[draw=black!50,solid] {$c_2$} edge from parent node[right] {$f_4$}} edge from parent node[right, xshift=0.6em] {$f_1$}}
    child[edge from parent/.style={draw,densely dashed,black!50,->}] { node[draw=black!50,,xshift=-2.7em] {$c_3$}
      child { node[draw=black!50,solid] {$c_4$} edge from parent node[right] {$f_3$}} edge from parent node[right, xshift=0.3em] {$f_2$}};
\end{pgfonlayer}
 \end{scope}

\end{tikzpicture}
\caption{\label{fig:derivation-tree-intro} An example of a derivation tree, where $c_1=f_1(c_0)$, $c_2=f_4(f_1(c_0))$, $c_3=f_2(c_0)$, and $c_4=f_3(f_2(c_0))$.}
\end{figure}
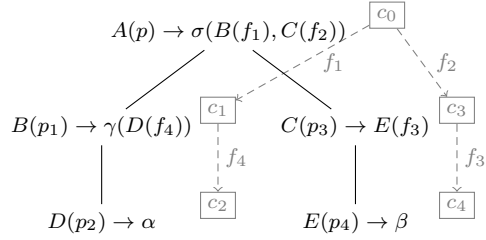

The semantics of $\G$ is based on the concept of \emph{derivation tree}.
A derivation tree $d$ is a  parse tree for derivations of $\G$ in the sense of \cite[Sect.~3.1]{gogthawagwri77} (where we view $\G$ as a context-free grammar with extra symbols for parentheses and comma; cf., e.g., \cite[Def.~3.18]{eng75-15}).  Figure \ref{fig:derivation-tree-intro} without the grey shaded part shows an example of a derivation tree.  Each position of $d$ is labeled by a rule of $\G$, and the nonterminals occurring in the right-hand side match with the nonterminals in the left-hand sides of the children.  In addition, there is a requirement which refers to the storage type. To each position of $d$ a storage configuration is associated (cf. the grey part of Figure \ref{fig:derivation-tree-intro}): the root is associated with the initial configuration $c_0$ of $S$, and the other configurations are computed successively  by applying the corresponding instructions $f_i$. Moreover, at each position,  the predicate of the local rule has to be satisfied. From the viewpoint of attribute grammars \cite{KN68}, one can consider $\G$ as an attribute grammar with one inherited attribute (cf. \cite[Sect.~1.2]{eng86} for a discussion of this viewpoint). Each derivation tree represents a derivation of a terminal tree where this latter is obtained by reading off the terminal symbols from the rules of type (1) and disregarding rules of type (2). For instance,
the derivation tree in Figure  \ref{fig:derivation-tree-intro} represents a derivation of the terminal tree $\sigma(\gamma(\alpha),\beta)$.  By composing the operations which are associated to the rules we obtain an element in $K$ and we call it the weight of that derivation tree. Finally, by summing up, in the monoid $(K,+,0)$ the weights of all derivation trees of a terminal tree, we obtain the weight of that terminal tree. We call the mapping, which takes a terminal tree to its weight, the \emph{weighted tree language generated by $\G$}.

Two special cases of weighted regular tree grammars with storage are obtained by (i)~choosing $S$ to be the trivial storage type $\TRIV$ and (ii) choosing $K$ to be the Boolean multioperator monoid $\mathbb{B}$. The trivial storage type contains exactly one configuration and, hence, essentially no information can be stored. Thus, for every multioperator monoid $M_K$ associated to a semiring $K$, $(\TRIV,\Sigma,M_K)$-rtgs are extensions of $K$-weighted regular tree grammars of \cite{aleboz87} (because in \cite{aleboz87} there are no chain rules).
The Boolean multioperator monoid  $\mathbb{B}$ is an extension of the 
monoid $(\{0,1\},\vee,0)$ of truth values with disjunction as binary 
operation. The extension amounts to equip this monoid with an 
$n$-ary conjunction for each $n \in \mathbb{N}$.
If $K=\mathbb{B}$, then such grammars are essentially (unweighted) regular tree grammars with storage~\cite{eng86}. 

The reader might wonder why each rule of an  $(S,\Sigma,K)$-rtg has either zero or one terminal, and not an arbitrary number of terminals (as it is usual for regular tree grammars). On the one hand, there is no technical problem to define such generalized  $(S,\Sigma,K)$-rtgs. On the other hand, the proofs of most of our results are based on the restricted form of rules (as in case of regular tree grammars
\cite{bra69,gecste84,gecste97}). To achieve a normal form lemma which takes a generalized  $(S,\Sigma,K)$-rtg and transforms it into an equivalent  $(S,\Sigma,K)$-rtg requires rather technical conditions (in fact restrictions) on the M-monoid (the operation of the original rule has to be decomposed into approriate operations). Thus we refrain from dealing with generalized  $(S,\Sigma,K)$-rtgs. Indeed, our $(S,\Sigma,K)$-rtgs might also be called weighted tree automata with $\varepsilon$-moves (and storage).

In this paper we start to develop a theory of the class of weighted tree languages generated by $(S,\Sigma,K)$-rtg.  In Section \ref{sec:wrtg}, after introducing our new grammar model, we  show that $(S,\Sigma,\mathbb{B})$-rtgs have the same power as $(\TRIV,\Sigma,\mathbb{B})$-rtgs assuming that $S$ is a finite storage type containing the always true predicate and the identity instruction. 

In Section~\ref{support} we prove  that the supports of $(S,\Sigma,K)$-regular tree languages are $(S,\Sigma,\mathbb{B})$-regular provided that $K$  is associated to  a complete, zero-sum free, and commutative strong bimonoid. For the proof  we employ the approach and the technique of \cite{K}.
 
In Section \ref{sec:decomp} we deal with two decompositions of the weighted tree language generated by an $(S,\Sigma,K)$-rtg. First, we represent every $(S,\Sigma,K)$-regular tree language as the composition of some particular mapping $\B_\Delta$ and some $(\TRIV,\Theta,K)$-regular tree language. The mapping $\B_\Delta$ is based on the concept of storage behaviour of $S$: roughly speaking, it enriches each terminal tree with a possible behaviour of $S$. This result was inspired by the decomposition of CFT($S$)-transducers into an approximation and a macro tree transducer in \cite[Thm.~3.26]{engvog86}. Our second result is based on \cite[Lm.~3 and Lm.~4]{drovog13} and shows that the weights of an $(S,\Sigma,K)$-rtg can be encoded into an alphabetic mapping, which is applied to an unambiguous, chain-free $(S,\Theta,\mathbb{B})$-rtg. As a consequence of this  and the fact  that finite storage can be eliminated from $(S,\Sigma,\mathbb{B})$-rtgs, we can prove that one can drop finite storage types from $(S,\Sigma,K)$-rtgs (for arbitrary $K$) without loosing power. Finally,  we combine the two decomposition results and obtain a characterization of the class of $(S,\Sigma,K)$-regular tree languages in terms of three simple and basic components: the mapping $\B_\Delta$, an unambiguous and chain-free $(\TRIV,\Theta,\mathbb{B})$-rtg, and an alphabetic mapping. We illustrate the decomposition results by showing an example in detail.

In Section  \ref{elimchain} we show that, even for the Boolean multioperator monoid, chain rules increase the power of (weighted) regular tree grammars with storage. Moreover, we study under which restrictions  on the multioperator monoid $K$ chain rules can be eliminated from $(S,\Sigma,K)$-rtgs.

 In Section \ref{sec:BET} we prove a characterization of $(S,\Sigma,K)$-regular tree languages in terms of weighted monadic second order logic (MSO). For this, we introduce a weighted MSO-logic with storage behaviour based on M-expressions of \cite{fulstuvog12,fulvog15} and on the MSO-logic introduced in \cite{vogdroher16}. However, our new logic generalizes the weighted MSO-logic with storage behaviour of \cite{vogdroher16} by considering trees as models and by allowing chain rules. 

Apart from Section \ref{sec:BET} we have tried to write the paper in such a way that it is self-contained. In Section \ref{sec:BET} we have given detailed references to the literature where the reader can find the relevant definitions. Readers who are familiar with algebraic structures (like semirings and multioperator monoids) and concepts concerning trees (like tree transformations, weighted tree languages, and term rewriting systems) can skip Sections \ref{sec:M-monoids} and \ref{sec:trees-prel} on first reading and consult them later if necessary.

%%%%%%%%%%%%%%%%%%%%%%%%%%%%%%%%%%%%

\section{Preliminaries}

\subsection{Notations}

We denote the set $\{0,1,2,\ldots\}$ of natural numbers by $\Nat$ and the set $\{1,2,\ldots\}$ by $\Nat_+$. 
For every $n\in \Nat$  we define $[n]=\{1,\ldots,n\}$.

Let $A_1$, $A_2$ be two sets and let $a=(a_1,a_2)\in A_1\times A_2$. Then, for each $i\in[2]$, the \emph{$i$-th projection of $a$}, denoted by $\pr_i(a)$, is defined by $\pr_i(a)=a_i$.

Let $A$ be a set. Then $A^*$ denotes the set of all finite sequences over $A$ including the empty sequence denoted by $\varepsilon$. We denote the  set of all subsets of $A$ by $\mathcal P(A)$. Moreover, we denote by $\id_A$ the identity mapping over $A$. If $A$ contains exactly one element, then sometimes  $A$ is identified with that element. For every $B$, $C\subseteq A^*$ we let $BC=\{bc\mid b\in B,c\in C\}$.

Let $u$, $w\in A^*$. Wey say that \emph{$u$ is a prefix of $w$}, denoted by $u\preceq w$, if there is a $v\in A^*$ with $uv=w$. Moreover, for every $a\in A$ we denote by $|w|_a$ the number of occurrences of $a$ in $w$.

%Let $n \ge 1$ and $A_1,\ldots,A_n$ subsets of $A$. The family $(A_i \mid i \in [n])$ is a {\em partitioning of $A$} if $A_i \cap A_j = \emptyset$ for each $i,j \in [n]$ with $i\not= j$, and $A= \bigcup_{1\le i \le n}A_i$. We use this concept also for finite index sets different from $[n]$.\todo{We do not use the concept 'partitioning'.} 

A set is \emph{countable} if its cardinality coincides with that of a subset of the natural numbers.

\subsection{Algebraic structures}
\label{sec:M-monoids}

We recall  the concept of strong bimonoids and semirings from \cite{drostuvog10,cirdroignvog10} and \cite{hebwei98,gol99,eil74}, respectively,
 and that of multioperator monoids from \cite{kui99e}.

A monoid $(K,+,0)$ is  \emph{commutative} if $a+b=b+a$, \emph{zero-sum free} if $a+b=0$ implies $a=b=0$, and \emph{idempotent} if $a+a=a$ for every $a,b\in K$. Moreover, $K$ is {\em complete} if it has a
sum operation $\sum_I\colon K^I \rightarrow K$ for each countable index  set $I$ which coincides with + when $I$ is finite (for the axioms cf.~\cite[p. 124]{eil74}).

A \emph{strong bimonoid} \cite{drostuvog10,cirdroignvog10} is an algebra $(K,+,\cdot,0,1)$, where $(K,+,0)$ is a commutative monoid, $(K,\cdot,1)$ is a monoid, and $0\cdot a=a\cdot 0=0$ for each $a\in K$. We assume that $0\not=1$.  We call $K$ \emph{commutative} if $(K,\cdot,1)$ is commutative. 
The strong bimonoid $K$ is called a \emph{semiring} if $a\cdot (b+c)=a\cdot b +a\cdot c$ and
$(b+ c)\cdot a=b\cdot a +c\cdot a$ for every $a,b,c\in K$. We refer to \cite[Ex.~1]{drostuvog10} for a number of examples of strong bimonoids, e.g., each bounded lattice is a strong bimonoid.

 A semiring $(K,+,\cdot,0,1)$ is \emph{complete} if the monoid $(K,+,0)$ is complete and the generalized distributivity law holds for infinite sums (cf. \cite[p. 124]{eil74}).
The semiring $(\{0,1\},\vee,\wedge,0,1)$ is called the \emph{Boolean semiring}, where 
0 and 1 stand for `false' and `true'; and $\vee$ and $\wedge$ are the usual disjunction and conjunction of truth values, respectively.
The Boolean semiring and the semiring $(\Nat_\infty,+,\cdot,0,1)$ with $\Nat_\infty=\Nat\cup\{\infty\}$ and the natural extension of + to $\Nat_\infty$ are complete (cf. \cite[Example 1]{esilei02},\cite[Chap. VI, Sec. 2]{eil74}).

Let $k\in \Nat$ and $K$ be a set. We denote the set of all $k$-ary operations (all operations) on $K$ by $\Ops^{(k)}(K)$ (resp. $\Ops(K)$). For every $k\in\mathbb N$, the $k$-ary constant zero function $0_k\in \Ops^{(k)}(K)$ is defined by $0_k(a_1,\ldots, a_k)=0$ for every $a_1,\ldots, a_k\in K$.  For a set $\Omega\subseteq \Ops(K)$ we define $\Omega^{(k)}=\Omega\cap \Ops^{(k)}(K)$.

A \emph{multioperator monoid} (for short: \emph{\mmon{}}) \cite{kui99e,mal04,stuvogfue09} is a tuple
$(K,+,0,\Omega)$ such that $(K,+,0)$ is a commutative monoid and
$\{0_k \mid  k\in\mathbb N\} \subseteq\Omega\subseteq \Ops(K)$.
 Moreover, we require that for every $k\in\Nat$,  each operation $\omega\in \Omega^{(k)}$ is absorptive, which means that for every  $a_1,\ldots, a_k\in K$, and $i\in[k]$,
the equality $a_i=0$ implies $\omega(a_1,\ldots, a_k) = 0$. We note that in \cite{stuvogfue09,fulstuvog12} such an M-monoid was called absorptive.   An \mmon{}  $(K,+,0,\Omega)$ is called \textit{complete} if $(K,+,0)$ has this property. For instance, the structure $(\mathbb{N}\cup\{\infty\},+,0,\Omega)$ is a complete M-monoid, where + is extended to sum over countable index sets in the obvious way,  $\Omega=\{0_k \mid  k\in\mathbb N\} \cup \{\min_k \mid  k\in\mathbb N\}$, and $\min_k$ is the $k$-ary minimum function (both extended to $\mathbb{N}\cup\{\infty\}$). 

The M-monoid $(\{0,1\},\vee,0,\Omega)$ is called the \emph{Boolean M-monoid}, denoted by $\mathbb{B}$, where for every $k\in \Nat$, we have
$\Omega\ui k=\{0_k,\all_{k,1}\}$ and for every  $a_1,\ldots,a_k\in \{0,1\}$, we have
$\all_{k,1}(a_1,\ldots,a_k)=1$ if and only if $a_1=\ldots=a_k=1$. Then $\mathbb{B}$, equipped with the the operations $\big(\bigvee_I\colon \{0,1\}^I \rightarrow \{0,1\}\mid I \text{ is a countable index set}\big)$, is a complete M-monoid, where for every $I$ and $f\in \{0,1\}^I$,  we have $\bigvee_I(f)=1$ if there is an $i\in I$ with $f(i)=1$, and $0$ otherwise.

To every strong bimonoid $(K,+,\cdot,0,1)$ we associate the M-monoid $(M_K,+,0,\Omega)$, where $M_K=K$ and
$\Omega\ui k= \{\mul_{k,a} \mid a \in K\}$
and $\mul_{k,a}(a_1,\ldots,a_k)=a_1\cdot \ldots\cdot a_k\cdot a$ for every $a_1,\ldots,a_k \in K$. (Note that $\mul_{k,0} = 0_k$ for every $k \ge 0$.)
The corresponding construction for semirings can be found in~\cite[Def. 8.5]{fulmalvog09} and \cite[page 261]{fulstuvog12}.
Note that the Boolean M-monoid is equal to the 
M-monoid associated to the Boolean semiring.

\begin{quote}
    \textit{For the remainder of the paper, if $K$ is left unspecified, then it stands for  an arbitrary complete M-monoid $(K,+,0,\Omega)$.}
\end{quote}

\subsection{Trees, tree transformations, weighted tree languages, and term rewriting systems}
\label{sec:trees-prel}

By an alphabet we mean a finite, non-empty set.
A \emph{ranked alphabet} is an alphabet $\Sigma$ together
with a mapping $\rk_\Sigma\colon\Sigma\rightarrow \Nat$; the natural number $\rk_\Sigma(\sigma)$ is called the \emph{rank of   $\sigma$}.  For every $k\in\Nat$ we let $\Sigma^{(k)} = \rk_\Sigma^{-1}(k)$. 
To avoid obvious cases, we assume that $\Sigma^{(0)}\ne 0$.
If $\sigma\in \Sigma^{(k)}$, i.e., the rank of $\sigma$ is $k$, then we write briefly $\sigma\ui k$.
We denote the maximal rank which occurs in $\Sigma$, i.e., the number $\max\{k \mid \Sigma^{(k)} \not= \emptyset\}$, by $\maxrk(\Sigma)$.

Let $\Sigma$ be a ranked alphabet  and $H$ be a set. The \emph{set of trees over $\Sigma$ indexed by $H$}, denoted by $T_\Sigma(H)$, is the smallest set $T$ such that (i) $H \subseteq T$ and (ii) for every $k\in\Nat$, $\sigma\in\Sigma^{(k)}$, and $\xi_1,\ldots,\xi_k\in T$ also $\sigma(\xi_1,\ldots, \xi_k)\in T$. We abbreviate $T_\Sigma(\emptyset)$ by $T_\Sigma$. Also,  for $\sigma \in \Sigma^{(1)}$ and $\xi \in T_\Sigma$ we abbreviate $\sigma(\ldots \sigma(\xi)\ldots)$ with $n$ occurrences of $\sigma$ by $\sigma^n(\xi)$. For each $\sigma \in \Sigma^{(0)}$ we abbreviate $\sigma()$  by $\sigma$. We note that each $\xi\in T_\Sigma$ has the form $\sigma(\xi_1,\ldots,\xi_k)$ for some $k\in \mathbb{N}$, $\sigma \in \Sigma\ui k$, and $\xi_1,\ldots,\xi_k\in T_\Sigma$.

We define the mappings $\pos:T_\Sigma\rightarrow \mathcal P(\Nat^*)$, $\ourpath: T_\Sigma \to {\cal P}(\Sigma^*)$, and 
$\height:T_\Sigma \to \mathbb{N}$ by induction on the structure of their argument. For this, let $\xi= \sigma(\xi_1,\ldots,\xi_k)$ in 
$T_\Sigma$. If
$k= 0$, then we let $\pos(\xi) =\{\varepsilon\}$, $\ourpath(\xi)= \{\sigma\}$, and $\height(\xi)=1$.
If  $k \ge 1$, then we let $\pos(\xi)=\{\varepsilon\}\cup\{iw\mid i\in[k], w\in \pos(\xi_i)\}$,
$\ourpath(\xi)=\bigcup_{i\in[k]} \{\sigma w \mid w \in \ourpath(\xi_i)\}$, and $\height(\xi)=1 + \max\{\height(\xi_i)\mid i\in[k]\}$.
We call $\pos(\xi)$ the \emph{set of positions in $\xi$} and denote the cardinality of $\pos(\xi)$ by $\mathrm{size}(\xi)$. 
Moreover,  we can define the \emph{lexicographic order} $\le_{\mathrm{lex}}$ on $\pos(\xi)$ as usual. 

Let $\xi\in T_\Sigma$.  For every $w\in\pos(\xi)$, we define the \emph{label $\xi(w)\in \Sigma$ of $\xi$ at position $w$} and the \emph{subtree $\xi|_w\in T_\Sigma$ of $\xi$ at position $w$} in the usual way (cf. e.g. \cite{fulstuvog12}).  Moreover, we abbreviate $\rk_\Sigma(\xi(w))$ by $\rk_\xi(w)$. For every $\Theta\subseteq \Sigma$, we let $\pos_\Theta(\xi)=\{w\in \pos(\xi)\mid \xi(w)\in \Theta\}$.

Let $\Sigma$ and $\Delta$ be two ranked alphabets. 
A \emph{tree transformation from $\Sigma$ to $\Delta$} is a mapping $\tau: T_\Sigma \rightarrow \mathcal{P}(T_\Delta)$. Tree relabelings are particular tree transformations. For their definition, let
$\tau\colon \Sigma \rightarrow \mathcal{P}(\Delta)$ be a mapping such that $\tau(\sigma) \subseteq
\Delta^{(k)}$ for every $k \ge 0$ and $\sigma\in\Sigma^{(k)}$. This mapping is extended to a
mapping $\tau'\colon T_\Sigma \rightarrow \mathcal{P}(T_\Delta)$ by defining inductively
\[
\tau'(\sigma(\xi_1,\ldots,\xi_k)) = \{\gamma(\zeta_1,\ldots,\zeta_k) \, | \, \gamma \in
\tau(\sigma), \zeta_1 \in \tau'(\xi_1),\ldots, \zeta_k \in \tau'(\xi_k)\}
\]
for every $k \ge 0$, $\sigma \in \Sigma^{(k)}$, and $\xi_1,\ldots,\xi_k \in T_\Sigma$.
Obviously, $\tau'(\sigma(\xi_1,\ldots,\xi_k))$ is a finite set.
Next we extend $\tau'$ to a mapping $\tau''\colon {\cal P}(T_\Sigma)
\rightarrow {\cal P}(T_\Delta)$ by defining $\tau''(L) = \{\zeta \in
T_\Delta \mid \zeta \in \tau'(\xi), \xi \in L\}$ for
every $L \in {\cal P}(T_\Sigma)$. We call $\tau'$ and $\tau''$ the  \emph{tree relabeling induced by $\tau$}. In the sequel we will drop the primes from $\tau'$ and $\tau''$.

\begin{quote}
	\textit{For the remainder of the paper, if $\Sigma$ is unspecified, then it stands for an arbitrary ranked alphabet.} 
\end{quote}

We call each mapping $s: T_\Sigma \rightarrow K$ a {\em weighted tree language}. The \emph{support of $s$} is defined
by the formula $\supp(s)=\{\xi\in T_\Sigma \mid s(\xi)\ne 0 \}$. A weighted tree language $s$ is called a \emph{monome} if  $\supp(s)$ is the empty set or a singleton. If $\supp(s) \subseteq \{\xi\}$ for some $\xi \in T_\Sigma$, then we also write $s(\xi).\xi$ instead of $s$. In particular, for each $\xi \in T_\Sigma$ the expression $0.\xi$ denotes the monome $s$ with $\supp(s)=\emptyset$. 
 We denote the set of all monomes of type $T_\Sigma \rightarrow K$ by $K[T_\Sigma]$. 

Let $(s_i \mid i \in I)$ be a family of weighted tree languages of type $T_\Sigma \rightarrow K$ with a countable index set $I$. The \emph{sum of $(s_i \mid i \in I)$}, denoted by $\sum_{i \in I}s_i$, is the weighted tree language $(\sum_{i \in I}s_i): T_\Sigma \rightarrow K$ defined for each $\xi \in T_\Sigma$ by 
\[
(\sum_{i \in I}s_i)(\xi) = \sum_{i \in I} s_i(\xi)\enspace.
\]
Note that  this sum is well defined because $K$ is complete. We write  $s_1 + s_2$ for $\sum_{i \in \{1,2\}}s_i$.

Let $\tau: T_\Sigma \rightarrow \mathcal{P}(T_\Delta)$ be a tree transformation and $s: T_\Delta \rightarrow K$ be a weighted tree language.  We define the \emph{composition of $\tau$ and $s$}, denoted by $(\tau;s)$, to be the weighted tree language $(\tau;s): T_\Sigma \rightarrow K$ defined for each $\xi \in T_\Sigma$ by 
\[
(\tau;s)(\xi) = \sum_{\zeta \in \tau(\xi)} s(\zeta)\enspace. 
\]
Note that $\tau(\xi)$ can be infinite; however, the sum is well defined because $K$ is complete. Whenever  there is no confusion we write $\tau;s$ instead of $(\tau;s)$.

We fix a countable set $X = \{x_1,x_2,\ldots\}$ of \emph{variables} and let $X_k = \{x_1,\ldots,x_k\}$ for each $k \in \mathbb{N}$. We assume that $X$ is disjoint from each ranked alphabet considered in this paper. 

A \emph{term rewriting system} \cite{baanip99} is a tuple $\mathcal R = (\Sigma,R)$ where $\Sigma$ is a ranked alphabet and $R$ is a finite set. Each  element of $R$ is called a \emph{rule} and it has the form $l \rightarrow r$ where $l,r \in T_\Sigma(X_k)$ for some $k \in \mathbb{N}$ such that $l \not\in X$, and each variable which occurs in $r$ also occurs in $l$. The \emph{rewrite relation induced by $\mathcal R$}, denoted by $\Rightarrow_\mathcal R$, is the binary relation $\Rightarrow_{\mathcal R} \subseteq T_\Sigma \times T_\Sigma$ defined for each $\xi_1, \xi_2 \in T_\Sigma$ as follows: $\xi_1 \Rightarrow_\mathcal R \xi_2$ if 
\begin{compactitem}
\item there is a $\theta \in T_\Sigma(X_1)$ in which $x_1$ occurs exactly once,
\item there is a rule $l \rightarrow r$ in $R$ with $l,r \in T_\Sigma(X_k)$ for some $k \in \mathbb{N}$, and
\item there are $\theta_1,\ldots,\theta_k \in T_\Sigma$ 
\end{compactitem}
such that $\xi_1$ is obtained from $\theta$ by replacing $x_1$ by $l'$, and $l'$ is obtained from $l$ by replacing each occurrence of $x_i$ by $\theta_i$ (for each $i \in [k]$); $\xi_2$ is obtained from $\theta$ by replacing $x_1$ by $r'$, and $r'$ is obtained from $r$ in the same way as $l'$ is obtained from $l$.
As usual for binary relations, we denote the reflexive, transitive closure of $\Rightarrow_\mathcal R$ by~$\Rightarrow_\mathcal R^*$. 

%%%%%%%%%%%%%%%%%%%%%%%%%%%%%%%%%%%%%%%%%
   
\subsection{Storage types and behaviours}\label{subsec:storage}
 
We recall the concept of storage type from \cite{eng86} with a slight modification (cf. \cite{hervog15}).
 
A \emph{storage type} is a tuple $S = (C,P,F,c_0)$, where $C$ is a set
(\emph{configurations}), $c_0 \in C$ ({\em initial configuration}), $P$ is a non-empty set of total functions each having the type $p: C \rightarrow \{0,1\}$ (\emph{predicates}), and $F$ is a non-empty set of partial functions $f: C \rightarrow C$ (\emph{instructions}).
A storage type is \emph{finite} if $C$ is a finite set.

 The \emph{identity instruction} is the total function $\id_C$. The \emph{always true predicate}, denoted by $\true_C$, is the predicate such that $\true_C(c)=1$ for each $c \in C$. For the storage type $S=(C,P,F,c_0)$, we denote by $S_{\true,\id}$ the storage type $(C,P\cup \{\true_C\}, F \cup \{\id_C\},c_0)$. Thus, $S$ contains $\true_C$ and  $\id_C$ if and only if $S_{\true,\id} = S$.

We note that our definition of storage type is a special case of the one in \cite{eng86} (and also in \cite{engvog86,engvog88}) in the sense that there the initial configuration $c_0$ is replaced by a set $I$ of inputs,  a set $E$ of encoding symbols, and a meaning function $m$. Each encoding symbol $e$ is interpreted as a partial function $m(e): I \rightarrow C$ and allows to define machines with {\em input and} output. Thus, our storage type $(C,P,F,c_0)$ is the storage type $(C,P',F',I,E,m)$ in the sense of \cite{eng86} with $I = \{i\}$, $E = \{e\}$, $m(e)(i) = c_0$, $P' = \{p' \mid p \in P\}$, and $F'=\{f' \in F \mid f\in F\}$ are sets of names for elements in $P$ and $F$, respectively, and $m(p')=p$ and $m(f')=f$. 

\begin{quote}
	\textit{For the remainder of the paper, if $S$ is unspecified, then it stands for an  arbitrary storage type $S = (C,P,F,c_0)$.}
\end{quote}

\paragraph{Particular storage types}\sloppy Here we recall three particular storage types from \cite{eng86,engvog86}: the trivial storage type, the pushdown storage type, and the counter storage type.

 The {\em trivial storage type}, denoted by $\TRIV$, is the storage type $(\{c\},\{\true_{\{c\}}\},\{\id_{\{c\}}\},c)$ for some arbitrary but fixed symbol $c$.

Let $\Gamma$ be a fixed infinite set ({\em pushdown symbols}) and let $\gamma_0 \in \Gamma$ be a fixed symbol. The {\em pushdown of $S$}
is the storage type $\PD-op(S) = (C',P',F',c_0')$ where
\begin{compactitem}
\item $C' = (\Gamma \times C)^+$ (there is no empty pushdown),
\item $c_0' = (\gamma_0,c_0)$,

\item $P' = \{\bottom\} \cup \{(\ttop = \gamma) \mid \gamma \in \Gamma\} \cup 
\{\test(p) \mid p \in P\}$ such that for every $(\delta,c) \in \Gamma \times C$
and $\alpha \in (\Gamma \times C)^*$ we have  
\[
\begin{array}{rcl}
\bottom\big((\delta,c)\alpha\big) & = & 1  \text{ iff }  \alpha =
\varepsilon\\
(\mathrm{top}=\gamma)\big((\delta,c)\alpha\big) & = &
1  \text{ iff }  \gamma = \delta \\
(\mathrm{test}(p)\big((\delta,c)\alpha\big) & = & p(c) \enspace,
\end{array}
\]

\item $F' = \{\mathrm{pop}\} \cup \{\mathrm{stay}(\gamma) \mid \gamma \in
\Gamma\} \cup \{ \mathrm{push}(\gamma,f) \mid \gamma \in \Gamma, f\in F\} \cup
\{\id_{C'}\}$ such that for every $(\delta,c) \in \Gamma \times C$ and
$\alpha \in (\Gamma \times C)^*$ we have 
\[
\begin{array}{rcl}
\mathrm{pop}\big((\delta,c)\alpha\big) & = &\alpha \ \text{ if } \alpha \not=
\varepsilon\\
\mathrm{push}(\gamma,f)\big((\delta,c)\alpha\big) & = &
(\gamma,f(c))(\delta,c)\alpha \ \text{ if } f(c) \text{ is defined}
\end{array}
\]
and undefined in all other situations. Moreover, 
\[
\begin{array}{rcl}
\mathrm{stay}(\gamma)\big((\delta,c)\alpha\big) & = & (\gamma,c)\alpha \enspace.
\end{array}
\]
 Recall that $\id_{C'}$ is the identity instruction on $C'$.
\end{compactitem}

For each $n \ge 0$ we define the storage type $\PD-op^n(S)$ inductively as follows: $\PD-op^0(S) = S$ and $\PD-op^n(S) = \PD-op(\PD-op^{n-1}(S))$ if  $n \ge 1$.  The \emph{$n$-iterated pushdown storage}, denoted by $\PD-op^n$, is the storage type $\PD-op^n(\TRIV)$.
We note that $\PD-op^n$ contains the always true predicate $\underbrace{\test(\ldots \test}_{n}(\true_{\{c\}}\underbrace{)\ldots)}_{n}$  and the identity instruction. Thus, $(\PD-op^n)_{\true,\id} = \PD-op^n$. Moreover, for the 1-iterated pushdown storage, we write $\Gamma^+$ for $(\Gamma\times \{c\})^+$ and abbreviate instructions by ignoring the part of the trivial storage type, e.g., we write $\push(\gamma)$ instead of $\push(\gamma,\id_{\{c\}})$. In this case we abbreviate the predicate $\test(\true_{\{c\}})$ by $\true$.

The storage type \emph{counter}, denoted by $\COUNT$, is defined by $(\Nat,\{\true_\mathbb{N},\zero\},\{\id_\mathbb{N},\inc,\dec\},0)$ where for each $c\in \Nat$
\begin{compactitem}
	\item $\zero(c)=1$ iff $c=0$, 
	\item $\inc(c)=c+1$, and $\dec(c)=c-1$ if $c\ge 1$ and undefined otherwise.
\end{compactitem}

\paragraph{Behaviour} Let  $P' \subseteq P$ and $F' \subseteq F$ be finite non-empty subsets. Moreover, let $n \in \mathbb{N}$.
We define the ranked alphabet $\Delta=\bigcup_{0\le k\le n} \Delta^{(k)}$ 
 with $\Delta^{(k)} =  P'\times (F')^k$. We call $\Delta$ the \emph{ranked alphabet $n$-corresponding to $P'$ and $F'$}. We write elements $(p,f_1, \ldots, f_k)$ of $\Delta$ 
in the shorter form $(p,f_1\ldots f_k)$.

\begin{figure}
% 	\footnotesize\centering
% \begin{tikzpicture}[sibling distance=12em, level distance=3.5em,
%   every node/.style = {align=center}]]

% \node(t) {$(\ttop=\gamma_0,\push(\gamma))$}
%    child { node {$(\ttop=\gamma,\push(\gamma)\push(\gamma)$}
%     child { node {$(\ttop=\gamma,\pop)$}
%       child { node {$(\ttop=\gamma,\pop)$}
%         child { node {$(\ttop=\gamma_0,\varepsilon)$} } } }
%     child { node {$(\ttop=\gamma,\push(\gamma))$}
%       child { node {$(\ttop=\gamma,\pop)$}
%         child { node {$(\ttop=\gamma,\pop)$}
%         	child { node {$(\ttop=\gamma,\varepsilon)$} } } } } };

% \hspace*{-15mm}
% \begin{scope}[sibling distance=11em, xshift=30em,yshift=0em]

%  \node {\p}
% child[edge from parent/.style={draw,densely dashed,->}] { node {\pp}
%     child { node[solid] {$\ppp$}
%       child { node[solid] {$\pp$}
%         child { node[solid] {$\p$} } } }
%     child { node[solid] {$\ppp$}
%       child { node[solid] {$\pppp$}
%         	child { node[solid] {$\ppp$} 
% 		child { node[solid] {$\pp$} }} }} };

%  \end{scope}
% \end{tikzpicture}
\centering
 \includegraphics{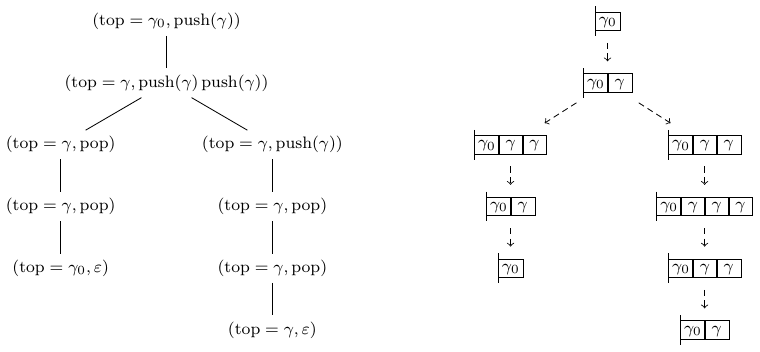}
\caption{On the left-hand side, a $(\Delta,\gamma_0)$-behaviour $b$ for $\Delta$ $2$-corresponding to  $P' = \{\ttop=\gamma_0,\ttop=\gamma\}$ and $F'= \{\push(\gamma),\pop\}$, and on the right-hand side, the family $(c_w \mid w \in \pos(b))$ of configurations determined by $b$ and $\gamma_0$.}\label{pic:behav}
\end{figure}

The concept of behaviour is inspired by and closely related to the concept of approximation \cite[Def.~3.23]{engvog86}. Formally, let $c \in C$, $n \in \mathbb{N}$, and $\Delta$ be the ranked alphabet $n$-corresponding to some finite sets $P'\subseteq P$ and $F'\subseteq F$. Then a tree $b\in T_{\Delta}$ is a {\em $(\Delta,c)$-behaviour}  if there is a family $(c_w \in C \mid w \in \pos(b))$ of configurations such that $c_\varepsilon=c$ and  for every $w\in \pos(b)$: if $b(w)=(p,f_1\ldots f_k)$, then 
\begin{compactitem}
\item  $p(c_w)=1$ and  
\item for every $1\le i\le k$, the configuration $f_i(c_w)$ is defined and $c_{wi}=f_i(c_w)$.
\end{compactitem}
If $b$ is a $(\Delta,c)$-behaviour, then
we call $(c_w \in C \mid w \in \pos(b))$ the \emph{family of configurations determined by $b$ and $c$}.
In Fig.~\ref{pic:behav} we illustrate these concepts for the storage type~$\PD-op^1$ and for $n=2$.

A \emph{$\Delta$-behaviour} is a $(\Delta,c_0)$-behaviour. 
We  denote the set of all $(\Delta,c)$-behaviours by $\B(\Delta,c)$, and the set of all $\Delta$-behaviours by $\B(\Delta)$. 

We will use the concept of corresponding ranked alphabet only in particular scenarios, for which we will now define a more convenient notion. 
Let $\Sigma$ be a ranked alphabet,  $P' \subseteq P$ and $F' \subseteq F$ be finite subsets. Then we call the ranked alphabet $\maxrk(\Sigma)$-corresponding to $P'$ and $F'$ the {\em  ranked alphabet corresponding to $\Sigma$, $P'$, and~$F'$}.

%%%%%%%%%%%%%%%%%%%%%%%%%%%%%%%%%%%%%

\section{Weighted regular tree grammars with storage}
\label{sec:wrtg}

In this section we combine the concept of regular tree grammar with storage \cite{eng86} with the weighting technique using multioperator monoids \cite{kui99e,mal05a,stuvogfue09,fulstuvog12}.

\subsection{Definition of the concept, examples, and special cases}

We recall that $K$ is a complete \mmon{} $(K,+,0,\Omega)$, $S=(C,P,F,c_0)$ is a storage type, and $\Sigma$ is a ranked alphabet.

A {\em regular tree grammar over $\Sigma$ with storage $S$ and weights in $K$} (for short: $(S,\Sigma,K)$-rtg)
is a tuple $\G = (N,Z,R,\wt)$, where 
\begin{compactitem}
\item $N$ is a finite set (\emph{nonterminals}) such that $N \cap \Sigma = \emptyset$,
\item $Z \subseteq N$ (set of \emph{initial nonterminals}), 
\item $R$ is a finite and non-empty set of {\em rules}; each rule has one of the following forms:
\begin{flalign*}
&A(p) \rightarrow \sigma(A_1(f_1),\ldots,A_k(f_k)) \tag{1}\\
&A(p) \rightarrow B(f) \tag{2}
\end{flalign*}
where $k \ge 0$, $A,A_1,\ldots,A_k,B \in N$, $p \in P$, $\sigma \in \Sigma^{(k)}$, and  $f_1,\ldots,f_k,f \in F$, and 
\item $\wt: R \rightarrow \Omega$ is the \emph{weight function} such that each rule of the form (1) is mapped to an element in $\Omega^{(k)}$ and each rule of the form (2) is mapped to an element in~$\Omega^{(1)}$. 
\end{compactitem}

If $r$ is a rule of the form (1), then we denote its parts $A$ and $A_\ell$ (with $1 \le \ell \le k$)  by $\lhs_N(r)$ and $\rhs_{N,\ell}(r)$, respectively; if $r$ is a rule of the form (2), then we denote  $A$ and $B$ by $\lhs_N(r)$ and $\rhs_{N,1}(r)$, respectively. Rules of type (2) are called \emph{chain rules}. If $\G$ does not contain chain rules, then we call it \emph{chain-free}.

In \cite{eng86,engvog86,engvog88} a binary derivation relation was defined for (unweighted) regular tree grammars with storage. Intuitively, the application of a rule follows the principle of context-free rewriting. In each sentential form each occurrence of a nonterminal keeps a configuration $c$ of $S$ and a rule may only be applied if its predicate holds on $c$. Each instruction $f$ occurring in the right-hand side of the rule is replaced by the configuration $f(c)$ if this is defined.

Here we formalize derivations in a different, but equivalent way using derivation trees (cf. Figure \ref{fig:derivation-tree-intro}).
Let $\G = (N,Z,R,\wt)$ be an $(S,\Sigma,K)$-rtg and $P_\G \subseteq P$ and $F_\G \subseteq F$ be the finite sets of predicates and instructions, respectively, which occur in $R$. Moreover, let $\Delta_\G$ be the ranked alphabet corresponding to $\Sigma$, $P_\G$, and $F_\G$. We note that $\Delta_\G$ is non-empty because we required that $R \not= \emptyset$. (In particular, if $R$ contains one rule of the form (1) with $k=0$ and no rule of the form (2), then $P_\G=\{p\}$ and $F_\G=\emptyset$.  Hence $\Delta_\G = \Delta_\G^{(0)}=\{(p,\varepsilon)\}$.)

In the following we will consider $R$ as a ranked alphabet by associating rank $k$ with a rule if its right-hand side contains $k$ nonterminals.
Let $d \in T_R$. We can retrieve from $d$ a tree in $T_\Sigma$ by using the  mapping $\pi \colon T_R \to T_\Sigma$ \label{page:pi}defined by induction on its argument such that for each $d=r(d_1,\ldots,d_k)$ in $T_R$ we have
\[
\pi(r(d_1,\ldots,d_k)) = 
\left\{
\begin{array}{ll}
\sigma(\pi(d_1),\ldots,\pi(d_k)) & \text{ if $r$ is of type (1)}\\
\pi(d_1) & \text{ if $r$ is of type (2)}\enspace.
\end{array}
\right.
\]

Also we can retrieve from $d$ a tree over $\Delta_\G$ by using the tree relabeling $\beta$. This tree relabeling is induced by the mapping $\beta: R \rightarrow \Delta_\G$ defined by $\beta(r) = (p,f_1\ldots f_k)$ if $r$ has the form (1), and by $\beta(r) = (p,f)$ if $r$ has the form (2).

Let $\xi \in T_\Sigma$, $N' \subseteq  N$, and $c \in C$. An \emph{$(N',c)$-derivation tree of $\G$ for $\xi$} is a tree $d \in T_R$ such that 
\begin{compactitem}
\item $\lhs_N(d(\varepsilon))\in N'$,
\item $\pi(d) = \xi$,
\item for each $w\in \pos(d)$ and each $\ell \in [\rank_d(w)]$  we have 
 $\rhs_{N,\ell}(d(w))=\lhs_N(d(w\ell))$, and
\item the tree $\beta(d)$ is in $\B(\Delta_\G,c)$.
\end{compactitem}
We denote the set of all such trees by $D_\G(N',\xi,c)$ and we abbreviate $D_\G(Z,\xi,c_0)$ by $D_\G(\xi)$.
We note that, if $\G$ is chain-free, then $\pos(d) = \pos(\xi)$ for each $d \in D_\G(\xi)$,  and  hence, $D_\G(\xi)$ is finite for each $\xi \in T_\Sigma$. Finally, a {\em derivation tree of $\G$ for a $\xi\in T_\Sigma$} is an $(N',c)$-derivation tree of $\G$ for $\xi$ for some $N'$ and $c$.

Let $d\in D_\G(N',\xi,c)$. We define $\wt'(d) \in K$  by 
$\wt'(d) = \wt''(d,\varepsilon)$ and, in its turn,  for each $w \in \pos(d)$ we define the value $\wt''(d,w) \in K$ inductively on $w$ as follows: 
\[
\wt''(d,w) = \wt(d(w)) \Big(\wt''(d,w\,1), \ldots, \wt''(d,w\,\rk_d(w))\Big) .
\] 
We note that $\wt(d(w))$ is an operation in $\Omega$ of arity $\rk_d(w)$.
For notational convenience we will drop in the sequel the primes from $\wt'$ and $\wt''$ and simply write $\wt$.

Then the {\em weighted tree language generated by $\G$} is the mapping $\seml \G\semr: T_\Sigma \rightarrow K$ defined for each $\xi \in T_\Sigma$ by 
\[
\seml \G\semr(\xi) = \sum_{d \in D_\G(\xi)} \wt(d).
\]
Since $\G$ may contain chain rules, the set $D_\G(\xi)$ can be infinite. However, this sum is well defined because $K$ is complete.

A weighted tree language $s: T_\Sigma \rightarrow K$ is called {\em $(S,\Sigma,K)$-regular}  if there is an $(S,\Sigma,K)$-rtg $\G$ such that $s = \seml \G\semr$,
and we call $s$ \emph{chain-free $(S,\Sigma,K)$-regular} if additionally $\G$ is chain-free.
The class of all $(S,\Sigma,K)$-regular tree languages and of all chain-free  $(S,\Sigma,K)$-regular tree languages
are denoted by $\Reg(S,\Sigma,K)$ and $\Reg_{\text{nc}}(S,\Sigma,K)$, respectively.\footnote{By ``nc'' we would like to express that there are ``\emph{n}o \emph{c}hain rules''.}

\begin{example}\rm\label{ex:run}
	Let $(M_{\Nat_\infty},+,0,\Omega)$ be the complete M-monoid associated to the complete semiring $(\Nat_\infty,+,\cdot,0,1)$ of natural numbers. Moreover, let $\Sigma=\{\alpha\ui 0,\delta\ui 1,\sigma\ui 2\}$.  We consider the weighted tree language $s\colon T_\Sigma\to M_{\Nat_\infty}$  with 
	\[s(\xi)=
	\begin{cases}8^n &\text{ if } \xi=\sigma(\delta^n(\alpha),\delta^n(\alpha)) \text{ for some } n\geq0\\
	0 &\text{ otherwise}
	\end{cases}\]
	for each $\xi\in T_\Sigma$. 

Then $s$ can be generated by a $(\PD-op^1,\Sigma,M_{\Nat_\infty})$-rtg $\G$. Intuitively, $\G$ first generates, using chain rules, a pushdown configuration of length $n$, then it  generates $\sigma$ and makes two copies of this configuration, and finally it turns each copy of each pushdown cell  into a $\delta$. For this, recall that $\Gamma$ is the infinite set of pushdown symbols and that $\gamma_0\in \Gamma$ is the initial pushdown symbol. Let $\gamma\in\Gamma$ with $\gamma\ne\gamma_0$.

We construct the $(\PD-op^1,\Sigma,M_{\Nat_\infty})$-rtg $\G=(N,Z,R,\wt)$, where $N=\{Z,A\}$ and $R$ consists of the rules 
	% \begin{align*}
	% r_1&=Z(\true)\to Z(\push(\gamma)) &r_2&=Z(\ttop=\gamma?)\to Z(\push(\gamma))\\
	% r_3&=Z(\ttop=\gamma?)\to \sigma(A(\pop),A(\pop))
	% &r_4&=A(\ttop=\gamma?)\to \delta(A(\pop))\\
	% r_5&=A(\ttop=\gamma_0?)\to \alpha()
	% \end{align*}
	\begin{align*}
	r_1&:Z(\true)\to Z(\push(\gamma)) & r_2&:Z(\true)\to \sigma(A(\id_{\Gamma^+}),A(\id_{\Gamma^+}))\\
        r_3&:A(\ttop=\gamma)\to \delta(A(\pop)) & 
	r_4&:A(\ttop=\gamma_0)\to \alpha
	\end{align*}
	where $\wt(r_1)=\wt(r_3)=\mul_{1,2}$, $\wt(r_2)=\mul_{2,1}$, and $\wt(r_4)=\mul_{0,1}$.

For each $n \ge 0$ and  $\xi = \sigma(\delta^n(\alpha), \delta^n(\alpha))$, we have $D_{\G}(\xi) = \{d_n\}$ where
$d_n = r_1^{n}r_2(r_3^nr_4, r_3^nr_4)$.
Figure~\ref{pic:deriv} shows $d_2$ for the tree $\xi = \sigma(\delta^2(\alpha),\delta^2(\alpha))$, the $\Delta_\G$-behaviour $b = \beta(d_2)$, and the family of configurations determined by $b$ and $\gamma_0$.

\begin{figure}[t]
	\tiny\centering
\begin{tikzpicture}[sibling distance=4em, level distance=3.5em,
  every node/.style = {align=center}]]

\pgfdeclarelayer{bg}    % declare background layer
  \pgfsetlayers{bg,main}  % set the order of the layers (main is the standard layer)
\begin{scope}[yshift=-5.5em]
  \node {$\mhl{\sigma}$}
    child { node {$\mhl{\delta}$}
      child { node {$\mhl{\delta}$} 
	child { node {$\mhl{\alpha}$}} }}
     child { node {$\mhl{\delta}$}
      child { node {$\mhl{\delta}$} 
	child { node {$\mhl{\alpha}$}}}};
\end{scope}
 \begin{scope}[sibling distance=20em, xshift=0em,yshift=25em, level distance=4em]
 \node {$Z(\true)\to Z(\push(\gamma))$}
 child { node {$Z(\true)\to Z(\push(\gamma))$}
   child { node {$Z(\true)\to \mhl{\sigma}(A(\id_{\Gamma^+}),A(\id_{\Gamma^+}))$}
    child { node {$A(\ttop=\gamma)\to \mhl{\delta}(A(\pop))$}
      child { node {$A(\ttop=\gamma)\to \mhl{\delta}(A(\pop))$}
        child { node {$A(\ttop=\gamma_0)\to \mhl{\alpha}()$} } } }
     child { node {$A(\ttop=\gamma)\to \mhl{\delta}(A(\pop))$}
      child { node {$A(\ttop=\gamma)\to \mhl{\delta}(A(\pop))$}
        child { node {$A(\ttop=\gamma_0)\to \mhl{\alpha}()$} } } } }};

\draw[|->] (0,-4) to node[right]{\normalsize$\pi$} (0,-4.9);
\draw[|->] (3.5,-3.5) to node[above]{\normalsize$\beta$} (4.5,-4);
 \end{scope}

\begin{scope}[sibling distance=15em, xshift=35em,yshift=5em,level distance= 4.5em]
 \node {$(\true,\push(\gamma))$}
child { node {$(\true,\push(\gamma))$}
   child { node {$(\true,\id_{\Gamma^+}\id_{\Gamma^+})$}
    child { node {$(\ttop=\gamma,\pop)$}
      child { node {$(\ttop=\gamma,\pop)$}
        child { node {$(\ttop=\gamma_0,\varepsilon)$} } } }
    child { node {$(\ttop=\gamma,\pop)$}
      child { node {$(\ttop=\gamma,\pop)$}
        	child { node {$(\ttop=\gamma_0,\varepsilon)$} } } } } };
 \end{scope}

\begin{scope}[sibling distance=15.5em, xshift=43em,yshift=6.em,black!50, level distance=4.5em]
\begin{pgfonlayer}{bg}
 \node {\p}
child[edge from parent/.style={draw,densely dashed,black!30,->}] { node {\pp}
   child { node[solid] {\ppp}
    child { node[solid] {$\ppp$}
      child { node[solid] {$\pp$}
        child { node[solid] {$\p$} } } }
    child { node[solid] {$\ppp$}
      child { node[solid] {$\pp$}
        	child { node[solid] {$\p$} } } } } };
\end{pgfonlayer}
 \end{scope}

\end{tikzpicture}
\caption{The derivation tree $d_2 \in D_\G(\xi)$, the tree $\xi = \pi(d_2) = \sigma(\delta^2(\alpha),\delta^2(\alpha))$, and the $\Delta_\G$-behaviour $b=\beta(d_2)$ together with the family $(c_w \mid w \in \pos(b))$ of configurations determined by $b$ and $\gamma_0$. }\label{pic:deriv}
\end{figure}
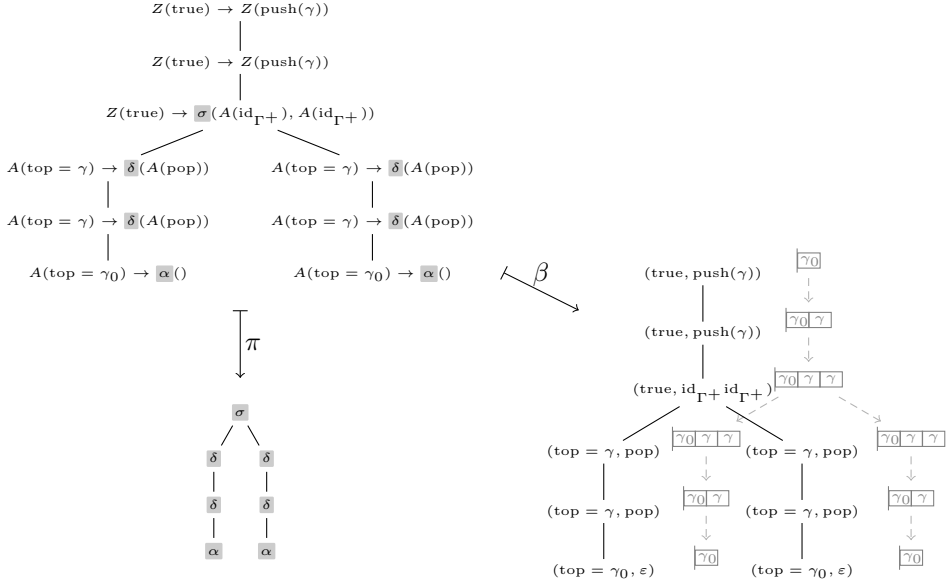

 Then 
\[\wt(d_n) = \underbrace{\mul_{1,2}(\ldots \mul_{1,2}(}_{n}
\mul_{2,1}(u,u) \underbrace{)\ldots)}_{n}
\]
with $u = \underbrace{\mul_{1,2}(\ldots \mul_{1,2}(}_{n}\mul_{0,1}\underbrace{) \ldots )}_{n}$ for each $n \ge 0$. Since $u=2^n$, we have $\wt(d_n) = 8^n$. 
Thus, for each $\xi = \sigma(\delta^n(\alpha), \delta^n(\alpha))$, we have $\seml \G\semr(\xi) = \wt(d_n) = 8^n$. Since for each tree $\xi \not\in \supp(s)$ we have $D_\G(\xi) = \emptyset$ and hence $\seml \G \semr(\xi) = \sum_{d \in \emptyset} \wt(d) = 0$, we finally obtain that $\G$ generates the weighted tree language $s$, i.e.,  $\seml \G \semr = s$.
\end{example}

In the next lemma we will prove that the restriction to exactly one initial nonterminal has no effect on the generating power of $(S,\Sigma,K)$-rtg. This kind of  initial state (or nonterminal) normal form has been proved for semiring-weighted tree automata (cf., e.g., \cite[p.~517]{bor04} or \cite[Thm.3.6]{fulvog09new}) using the following classical idea: take a new initial nonterminal and derive nondeterministically each right-hand side of an original rule that has an initial nonterminal at its left-hand side, and the weight of this new initial rule is the sum of the weights of the original initial rules. Generalizing this idea to the case of $(S,\Sigma,K)$-grammars requires to sum up operations and thus, to assume that the set of operations of  $K$ is closed under summation. We avoid this by employing a slightly more complicated construction.

\pagebreak

\begin{lemma}\label{lm:one-initial} For each $(S,\Sigma,K)$-rtg $\G$ there is an $(S,\Sigma,K)$-rtg $\G'$ such that $\sem{\G} = \sem{\G'}$ and $\G'$ has exactly one initial nonterminal.
\end{lemma}

\begin{proof}
	Let $\G = (N,Z,R,\wt)$ be an $(S,\Sigma,K)$-rtg. We may assume that $Z\neq\emptyset$ since otherwise $\supp(\lBert \G\rBert)=\emptyset$ and thus our statement is obvious. We construct an $(S,\Sigma,K)$-rtg $\G'$ that uses  only one initial nonterminal $Z_0$. Intuitively, $\G'$ encodes the initial nonterminals of $\G$ as second component in its nonterminals. Then any $(Z_0,c_0)$-derivation tree $d'$ of $\G'$ encodes an $(A_0,c_0)$-derivation tree $d$ of $\G$ for some initial nonterminal $A_0$ by keeping $A_0$ in the nodes of $d'$. In this way,  for each tree $\xi$ the sets $D_\G(\xi)$ and $D_{\G'}(\xi)$ are in a weight preserving one-to-one correspondence.

	So let $Z_0$ be a symbol not in $(N \times Z) \cup \Sigma$. We construct the $(S,\Sigma,K)$-rtg $\G'=(N',Z_0,R',\wt')$ such that $ \lBert \G\rBert=\lBert \G'\rBert$ as follows. We let $N'=(N\times Z)\cup\{Z_0\}$. Moreover, for each $A_0\in Z$,
	\begin{compactitem}
		\item if $r=(A_0(p) \rightarrow \sigma(A_1(f_1),\ldots,A_k(f_k)))\in R$,\\
 then $r'=(Z_0(p) \rightarrow \sigma((A_1,A_0)(f_1),\ldots,(A_k,A_0)(f_k)))\in R'$,
		\item if $r=(A_0(p)\to B(f))\in R$, then $r'=(Z_0(p)\to(B,A_0)(f))\in R'$,
		\item if $r=(A(p) \rightarrow \sigma(A_1(f_1),\ldots,A_k(f_k)))\in R$,\\
 then $r'=((A,A_0)(p) \rightarrow \sigma((A_1,A_0)(f_1),\ldots,(A_k,A_0)(f_k)))\in R'$,
		\item if $r=(A(p)\to B(f))\in R$, then $r'=((A,A_0)(p)\to(B,A_0)(f))\in R'$,
	\end{compactitem}
	and in each case we let $\wt'(r')=\wt(r)$.

We can prove  that $\seml \G \semr = \seml \G'\semr$ as follows. For every $\xi \in T_\Sigma$ we define the mapping  $g: D_{\G}(\xi)\to D_{\G'}(\xi)$ (note that $D_{\G'}(\xi) =D_{\G'}(Z_0,\xi,c_0)$).  For this let $\xi=\sigma(\xi_1,\ldots,\xi_k)$ for some $k\geq 0$ and $d \in D_{\G}(A_0,\xi,c_0)$ for some $A_0\in Z$.

 Then either 
\begin{enumerate}
\item[(1)] $d=r(d_1,\ldots,d_k)$ for some rule $r=(A_0(p) \rightarrow \sigma(A_1(f_1),\ldots,A_k(f_k)))$  and $d_i \in D_{\G}(A_i,\xi_i,f_i(c_0))$ for each  $i\in[k]$ or 
\item[(2)] $d=r(d_1)$ for some rule  $r=(A_0(p)\to B(f))$ and $d_1 \in D_{\G}(B,\xi,f(c_0))$.
\end{enumerate}
In case (1) we let $g(d)=r'(d'_1,\ldots,d'_k)$, where 
\begin{compactitem}
\item $r'=(Z_0(p) \rightarrow \sigma((A_1,A_0)(f_1),\ldots,(A_k,A_0)(f_k)))$ and \item we obtain $d'_i$ from $d_i$ by replacing every nonterminal $A$ by $(A,A_0)$.
\end{compactitem}
 In case (2) let $g(d)=r'(d'_1)$, where 
\begin{compactitem}
\item $r'=(Z_0(p)\to(B,A_0)(f))$ and 
\item we obtain $d'_1$ from $d_1$ as in case (1).
\end{compactitem}
By the construction of $\G'$ it should be clear that $g$ is a bijection and that  $\wt(d) = \wt'(g(d))$ for each $d \in D_{\G}(\xi)$.
Then it is easy to see that $\seml \G \semr = \seml \G'\semr$.
\end{proof}

\paragraph{Special cases.}\label{page:special-cases} Here we define and analyze four special cases of $(S,\Sigma,K)$-regular weighted tree languages: (i)~the unweighted case, (ii)~the storage-free case, (iii)~the unweighted and storage-free case, and (iv)~the string case. 

\

(i) $K= \mathbb{B}$: 
 A {\em regular tree grammar over $\Sigma$ with storage $S$} (for short: $(S,\Sigma)$-rtg) is an $(S,\Sigma,\mathbb{B})$-rtg. In the specification of an  $(S,\Sigma)$-rtg we  assume w.l.o.g. that each $k$-ary rule is mapped to $\all_{k,1}$, and hence we drop the weight function~$\wt$.

Let $\G$ be an $(S,\Sigma)$-rtg. Since $\wt(d)=1$ for each $d \in D_\G(\xi)$, we have that $\supp(\lBert \G \rBert) = \{\xi \in T_\Sigma \mid D_\G(\xi) \not= \emptyset\}$. We call this set the {\em tree language generated by $\G$} and denote it by $\mathcal L(\G)$. Moreover, we say that $\G$ is \textit{unambiguous} if for each $\xi \in T_\Sigma$ we have $|D_\G(\xi)| \le 1$.

A tree language  $L\subseteq T_\Sigma$ is called {\em $(S,\Sigma)$-regular} if there is an $(S,\Sigma)$-rtg $\G$ such that $\mathcal L(\G) = L$. We denote the class of all $(S,\Sigma)$-regular tree languages by $\Reg(S,\Sigma)$.
 
We note that each $(S,\Sigma)$-rtg $\G$ is an $\mathrm{RT}(S)$-transducer $\mathcal M_\G$ as defined in \cite[Def.~3.3]{engvog86} and in \cite[Def.~3.3]{engvog88} where $\mathrm{RT}$ means the class of regular tree grammars \cite{bra69,gecste84,gecste97}; the range of the translation induced by $\mathcal M_\G$ is $\mathcal L(\G)$. Vice versa, each $\mathrm{RT}(S)$-transducer $\mathcal M$ (with the definition of storage type of the present paper)  is an $(S,\Sigma)$-rtg if the Boolean expressions occurring in the rules of $\mathcal M$ are  predicates (and not arbitrary Boolean expressions over $P$).
 
Moreover, by \cite[Thm.~6.15]{engvog88}, we have that $\Reg(\PD-op^n,\Sigma)$ is the class of level-$n$ OI-tree languages for each $n \ge 0$ (denoted by $n$-$T$ in \cite{engvog88}; also cf. \cite{damgue81}). This hierarchy has been intensively investigated in \cite{dam82,damgue81}. The classes for $n=0$ and $n=1$ are the class of regular tree languages and of OI context-free tree languages, respectively (cf. \cite[Prop.~4.4]{engvog88}).

We recall from \cite[Thm.~7.8]{dam82} that the emptiness problem for level-$n$ OI-tree languages is decidable (for each $n \ge 0$).
  
\

(ii) $S=\TRIV$: A {\em $K$-weighted regular tree grammar over $\Sigma$} (for short: $(\Sigma,K)$-rtg) is a $(\TRIV,\Sigma,K)$-rtg. In the rules of a $(\Sigma,K)$-rtg we drop the always true predicate  from the left-hand side and the identity instruction from the right-hand side. 
A weighted tree language $s: T_\Sigma \rightarrow K$ is called {\em $(\Sigma,K)$-regular}  if there is a $(\Sigma,K)$-rtg $\G$ such that $s = \seml \G\semr$. We denote the class of all $(\Sigma,K)$-regular tree languages by $\Reg(\Sigma,K)$.

In \cite[Sec.~2.6]{fulstuvog12}, weighted tree automata were defined over M-monoids which need not be absorptive and complete. However, it is obvious that each chain-free $(\Sigma,K)$-rtg corresponds to a weighted tree automata over $\Sigma$ and $K$ (in the sense of \cite[Sec.~2.6]{fulstuvog12}), and vice versa. Thus we obtain the following characterization.

\begin{observation}\rm \label{ob:chain-free-Reg=Rec} $\Reg_{\text{nc}}(\Sigma,K)$ is the class of recognizable tree series over $\Sigma$ and $K$ defined in \cite{fulstuvog12}. 
\end{observation}

Now let, additionally, $K$ be an arbitrary complete semiring and $M_K$ be the \mmon{} associated with $K$. Then each chain-free $(\Sigma,M_K)$-rtg is essentially a semiring-weighted tree automaton in the sense of \cite{fulvog09new}, and vice versa.

  \

(iii) $K=\mathbb{B}$ and $S=\TRIV$: A {\em regular tree grammar over $\Sigma$} (for short: $\Sigma$-rtg) is a $(\TRIV,\Sigma,\mathbb{B})$-rtg.
A tree language  $L\subseteq T_\Sigma$ is called {\em $\Sigma$-regular} if there is a $\Sigma$-rtg $\G$ such that $\mathcal L(\G) = L$. We denote the class of all $\Sigma$-regular tree languages by $\Reg(\Sigma)$. 

Clearly, the class $\Reg(\Sigma)$  coincides with the class of recognizable (or: regular) tree languages over~$\Sigma$~\cite{gecste84}. Since finite-state tree automata can be determinized \cite[Thm.~2.2.6]{gecste84}, we can always assume that a $\Sigma$-rtg $\G$ is unambiguous.

\

(iv) String case: For semirings as weight structures, it is demonstrated in \cite[p. 324]{fulvog09new} that, roughly speaking,  (a)  weighted string automata (without storage) are the same as  (b) weighted tree automata over a monadic ranked alphabet. We can generalize this relation to (a')
weighted string automata (including $\varepsilon$-transitions) with arbitrary storage and (b') weighted regular tree grammars over a monadic ranked alphabet (with chain rules) with arbitrary storage. Moreover, in the tree case we can use the M-monoid associated to the semiring as weight structure.

\subsection{Elimination of finite storage types in the Boolean case}\label{elimstorage-new}\label{elimstorage}

It is easy to see that each $(S,\Sigma)$-rtg can be simulated by a $\Sigma$-rtg assuming that $S$ is a finite storage type.

\begin{lemma} \label{lm:finite-storage-new} Let $S $ be  finite. For each $(S,\Sigma)$-rtg $\G$ there is a $\Sigma$-rtg $\G'$ such that $\cal L(\G) = \cal L(\G')$. Moreover, if $\G$ is chain-free (resp., unambiguous), then so is $\G'$.
\end{lemma}
\begin{proof} Let $\G = (N,Z,R)$ be an $(S,\Sigma)$-rtg.  For the construction of a $\Sigma$-rtg $\G'$, we simply encode the finitely many  configurations into the new set of nonterminals. Formally, we construct $\G' = (N\times C,Z',R')$ such that $Z' = Z \times \{c_0\}$ and   
\begin{compactitem}
\item if $r = (A(p) \rightarrow \sigma(A_1(f_1),\ldots,A_k(f_k))$ is a rule in $R$, 
then for each $c \in C$ such that $p(c)= 1$ and $f_1(c), \ldots, f_k(c)$ are defined, the rule 
\[r' = ((A,c) \rightarrow \sigma((A_1,f_1(c) ),\ldots, (A_k,f_k(c)))\] is in $R'$,
\item if $r = (A(p) \rightarrow B(f))$ is in $R$, 
then for each $c \in C$ such that $p(c)= 1$ and $f(c)$ is defined, the rule 
\[r' = ((A,c) \rightarrow (B,f(c))\] is in $R'$.
\end{compactitem}
It is clear that $\G'$ is chain-free (resp., unambiguous) if $\G$ is so. 

For each $\xi \in T_\Sigma$, we define the mapping $\theta: D_\G(\xi) \rightarrow D_{\G'}(\xi)$ as follows. Let $d \in D_\G(\xi)$ and $(c_w \mid w \in \pos(\beta(d)))$ be the family of configurations determined by the $(\Delta_\G,c_0)$-behaviour $\beta(d)$ and $c_0$. We recall that $\beta$ is a tree relabeling and hence $\pos(\beta(d)) = \pos(d)$.  Then we define  $\pos(\theta(d)) = \pos(d)$. Moreover, for each $w \in \pos(d)$, 
\begin{compactitem}
\item if $d(w) = (A(p) \rightarrow \sigma(A_1(f_1),\ldots,A_k(f_k))$, 
then we define 
\[(\theta(d)(w)) = ((A,c_w) \rightarrow \sigma((A_1,c_{w1}),\ldots, (A_k,c_{wk}))\enspace,\]
\item if $d(w) = (A(p) \rightarrow B(f))$, then we define  
\[(\theta(d)(w)) = ((A,c_w) \rightarrow (B,c_{w1}))\enspace.\]
\end{compactitem}

 Due to the construction, it is clear that $\theta(d) \in D_{\G'}(\xi)$. 
  Moreover, $\theta$ is surjective. However, it is not necessarily injective, as can be seen if, e.g.,  two rules $A(p)\to B(f_1)$ and $A(p)\to B(f_2)$ are in $R$ and there exists a $c\in C$ with $p(c)=1$ and $f_1(c)=f_2(c)$.

Then it follows that $D_\G(\xi) \not= \emptyset$ iff $D_{\G'}(\xi) \not= \emptyset$ (where the surjectivity of $\theta$ is needed in the ``if''-implication). Thus $\cal L(\G) = \cal L(\G')$. 
\end{proof}

%%%%%%%%%%%%%%%%%%%%%%%%%%%%%%%%%%%

\section{Support}\label{support}

In this section we prove that,  for each complete, zero-sum free, and commutative strong bimonoid $K$ (cf. Section~\ref{sec:M-monoids}), the supports of $(S,\Sigma,M_K)$-regular weighted tree languages are $(S,\Sigma)$-regular tree languages. We follow the same approach as in \cite{K}, where the support theorem was proved for  weighted string automata without $\varepsilon$-transitions. We generalize this approach in a straightforward way to the case of weighted regular tree grammars with storage (including chain rules, which correspond to $\varepsilon$-transitions in string automata).

First we recall some definitions.
Let $(K, \cdot, 1)$ be a monoid. An element $0\in K$ with $0\not=1$ is called a \emph{zero} if $a\cdot 0 = 0\cdot a=0$ for every $a\in K$. For every $a_1,\ldots, a_n\in K$, we let $\langle a_1,\ldots,a_n\rangle$ denote the smallest submonoid of $K$ containing $a_1,\ldots,a_n$. For every $a \in K$ and $A \subseteq K$, we let $a \cdot A = \{a \cdot a' \mid a' \in A\}$. 

As defined by Kirsten \cite{K}, the \emph{zero generation problem (ZGP)} for a monoid $(K, \cdot, 1)$ with zero 0 is, given two integers $m,n\in\Nat$ and elements $a_1,\ldots,a_m,a_1',\dots,a_n'\in K$, the question whether $0\in a_1\cdot\ldots\cdot a_m\cdot\langle a_1',\ldots,a_n'\rangle$. 
For instance, if $K$ is idempotent and commutative, then it has a decidable ZGP, because in this case the set  $a_1\cdot\ldots\cdot a_m\cdot\langle a_1',\ldots,a_n'\rangle$ is finite.

For every tuples $\bar z=(z_1,\dotsc,z_n)\in\mathbb{N}^{n}$ and $\bar y = (y_1,\dotsc,y_n) \in\mathbb{N}^{n}$, we define $\bar z \le \bar y$ if $z_i \le y_i$ for all $i=1,\dotsc,n$.
Obviously, $\le$ is a partial order on $\mathbb{N}^{n}$. For a subset $M \subseteq \mathbb{N}^n$, an element $\bar z\in M$ is \emph{minimal in $M$} if $\bar y\le \bar z$ implies $\bar y=\bar z$ for every  $\bar y\in M$. We denote by $\min(M)$ the set of all minimal elements in $M$.
By Dickson's lemma \cite{dic13}, $\min(M)$ is finite (cf. \cite[Lm. 2.1]{K} and \cite {Kreuzer}). For every $\bar z \in \mathbb{N}^n$ and  $k \in \mathbb{N}$, we define the \emph{cut of $\bar z$ and $k$}, denoted by $\lfloor \bar z\rfloor_k$, to be the vector $\lfloor \bar z\rfloor_k \in \mathbb{N}^n$ with 
\[( \lfloor \bar z \rfloor_k)_i = \min \{z_i, k \}\]
as $i$-th component for each $i=1,\dotsc,n$.

Now let $(K,{\cdot},1)$ be a commutative monoid with a zero $0$. For every $a \in K$ and $z \in \mathbb{N}$, we let $a^z$ be the product of $z$ many $a$s. In particular, $a^0=1$.

Fix a tuple $\bar{a} = (a_1,\dotsc,a_n) \in K^n$. We define the mapping 
\[
\lBert \,.\, \rBert_{\bar{a}} \colon \mathbb{N}^n \to K
\ \text{  by letting } \
\lBert \bar z \rBert_{\bar{a}} = a_1^{z_1} \cdot \ldots \cdot a_n^{z_n}
\]
 for each $\bar z=(z_1,\dotsc,z_n) \in \mathbb{N}^n$. We put $\lBert 0 \rBert_{\bar{a}}^{-1} = \{ \bar z \in \mathbb{N}^n \mid \lBert \bar z \rBert_{\bar{a}} = 0 \}$. Note that $(0,\ldots,0) \not\in \lBert 0 \rBert_{\bar{a}}^{-1}$, because $a_1^0 \cdot \ldots \cdot a_n^0= 1$.  Clearly, if $\bar z, \bar y \in \mathbb{N}^n$ with $\lBert \bar z \rBert_{\bar{a}} = 0$ and $\bar z \le \bar y$, then also $\lBert \bar y \rBert_{\bar{a}} = 0$. Since $\min(\lBert 0 \rBert_{\bar{a}}^{-1})$ is finite, there is a smallest number $m \in \mathbb{N}$ satisfying $\min(\lBert 0 \rBert_{\bar{a}}^{-1}) \subseteq \{0,\dotsc,m\}^n$, and we define $\operatorname{dg}(\bar{a})=m$.

We state the following obvious connection between the defined concepts. 

\begin{observation}\rm \label{ob:three} Let $\bar a = (a_1,\ldots,a_n)$ be an element of $K^n$. Then the following three statements are equivalent:
\begin{compactenum}
\item $\lBert 0 \rBert_{\bar{a}}^{-1} = \emptyset$.
\item $\operatorname{dg}(\bar{a})=0$.
\item The submonoid $(\langle a_1,\ldots,a_n\rangle, \cdot,1)$ is zero-divisor free, i.e., for each $a,b \in \langle a_1,\ldots,a_n\rangle$ we have that $a \cdot b = 0$ implies that $a=0$ or $b=0$.
\end{compactenum}
\end{observation} 

 Moreover, we recall from \cite{K} the following statements.
 
\begin{lemma}\label{Kirsten2}(\cite[Lm. 4.1]{K}) For each  $\bar z \in \mathbb{N}^n$, we have $\lBert \bar z \rBert_{\bar{a}} = 0$ iff $\lBert \lfloor \bar z \rfloor_{\operatorname{dg}(\bar{a})} \rBert_{\bar{a}} = 0$ .
\end{lemma}

\begin{lemma}\label{Kirsten}(\cite[Lm. 4.2]{K})
Let $(K,{\cdot},1)$ be a commutative monoid with a zero, $n\in \Nat$, and $\bar{a} \in K^n$. If the ZGP for $K$ is decidable, then $\operatorname{dg}(\bar{a})$ is effectively computable.
\end{lemma}

Now we can prove the main theorem of this section.  We follow the proof and the construction of the corresponding results \cite[Thm.~3.1]{K} for weighted automata over semirings and \cite[Thm.~4.6]{goe16} for weighted tree automata over tv-monoids (also cf. \cite{DH} for a similar result for weighted unranked tree automata over bimonoids). Also we provide the correctness proof of the construction.

\begin{theorem}\label{th:support}
  Let $K$ be a complete, zero-sum free, and commutative strong bimonoid and $M_K$ be the \mmon{} associated with $K$.
  \begin{compactenum}
  \item (a) For every $(S,\Sigma,M_K)$-rtg $\G$, there is an $(S,\Sigma)$-rtg $\G'$ such that 
$L(\G')=\supp(\lBert \G\rBert)$. (b) Moreover, if $(K,{\cdot},1)$
    has a decidable ZGP, then $\G'$ can be constructed effectively.
  \item Assume that $\lvert \Sigma\ui 1 \rvert \ge 2$. If there is an effective
    construction of a $\Sigma$-rtg which generates
    $\operatorname{supp}(\lBert \mathcal{G} \rBert)$ from any
    given $(\Sigma,M_K)$-rtg $\mathcal{G}$, then $(K,{\cdot},1)$
    has a decidable ZGP.
  \end{compactenum}
\end{theorem}

\begin{proof}
First we prove 1(a). Let $\G = (N,Z,R,\wt)$ be an $(S,\Sigma,M_K)$-rtg. By Lemma \ref{lm:one-initial} we may assume that $Z\in N$. 
Since the weight of each rule is an operation $\mul_{k,a}$ for some $k\in \Nat$ and $a\in K$,
the weight of each derivation tree is the (bimonoid) multiplication of the $a$'s appearing in the rules in that tree. Since $K$ is commutative, this amounts to counting how many times each such $a$ occurs. 

Formally, we define the mapping $\wt_K\colon R \to K$ such that, for each $r\in R$, we let $\wt_K(r)=b$ if $\wt(r)=\mul_{k,b}$ for some $k\in\Nat$. We let $W = \wt_K(R)$ be the set comprising all elements of $K$ each of which corresponds to the weight of some rule of $\G$. Let $\bar{a} = (a_1,\dotsc,a_n) \in K^n$ be an enumeration of $W$. Then the weight of each derivation tree can be written in the form $\lBert \bar y \rBert_{\bar{a}}$ for some $\bar y\in \Nat^n$. Moreover, let $T=\{0,\dotsc,\operatorname{dg}(\bar{a})\}$. We define the mapping
\[
\begin{array}{ll}
{\oplus}\colon T^n \times W \to T^n \ & \text{  by letting }\\
\bar z \oplus a_i =\lfloor \bar y \rfloor_{\operatorname{dg}(\bar{a})} \ & \text{ for } \ \bar y =(y_1,\ldots,y_n)\in \mathbb{N}^n \ \text{ with } \ y_i = z_i + 1 \ \text{ and } \\
 & \ y_j = z_j
\ \text{  for each } \ j \ne i \enspace,
\end{array}
\]
and the mapping
\begin{align*}
& \bar{\oplus}\colon T^n \times T^n \to T^n \ \text{ by letting } \\
& \bar{z}\, \bar{\oplus} \,\bar{z}' =\lfloor \bar{y} \rfloor_{\operatorname{dg}(\bar{a})} \ \text{ for } 
\bar{y}=(y_1,\ldots,y_n) \in \mathbb{N}^n  \text{ with } y_i = z_i + z_i' \text{ for all } i\in[n]. 
\end{align*} 
With these mappings, we will be able to count, up to the threshold of $\operatorname{dg}(\bar{a})$, the number of occurrences of the $a_i$'s in the derivation trees of $\G$. 

Now we define an (unweighted) $(S,\Sigma)$-rtg $\G'$ which simulates the derivations of $\G$ and counts the elements of $K$ corresponding to the weights which occur in derivation trees as follows. Let $\G' = (N', Z', R')$ such that
  \begin{compactitem}
  \item $N' = N \times T^n$,
  \item $Z' = \{(Z,\bar z)\mid \bar z \in T^n, \lBert \bar z \rBert_{\bar{a}} \ne 0\}$,
  \item $R'$ is defined as follows:

\begin{compactitem}
\item Let  $r=(A(p)\to\alpha)$ be  a rule in $R$. Then $((A,\bar z)(p)\to \alpha)$ is in $R'$ where  $\bar z=(0,\ldots,0)\oplus \wt_K(r)$.

\item Let $r=(A(p) \rightarrow \sigma(A_1(f_1),\ldots,A_k(f_k)))$ be a rule in $R$. Moreover, let $\bar z_1, \ldots,\bar z_k \in T^n$. 
Then $((A,\bar z)(p) \rightarrow \sigma((A_1,\bar z_1)(f_1),\ldots,(A_k,\bar z_k)(f_k)))$ is in $R'$ where $\bar z= (\bar z_1\bar{\oplus}\ldots\bar{\oplus} \bar z_k)\oplus \wt_K(r)$.

\item Let $r=(A(p) \rightarrow B(f))$ be a rule in $R$. Moreover, let $\bar z_1 \in T^n$. 
 Then $((A,\bar z)(p) \rightarrow (B,\bar z_1)(f))\in R'$ where  $\bar z= \bar z_1\oplus \wt_K(r)$.
\end{compactitem}
  \end{compactitem}
Next we prove the equality  $L( \mathcal{G'} ) =\operatorname{supp}(\lBert \mathcal{G} \rBert)$. 

First, we prove that $ \supp(\lBert \mathcal{G} \rBert)  \subseteq  L( \mathcal{G'} )$. For this, we show the following Statement:

(*) For each $l\in\Nat_+$, $\xi\in T_\Sigma$, $A\in N$, $c\in C$, and $d\in D_\G(A,\xi,c)\colon $ if $\mathrm{size}(d)=l$, then there are $\bar{y}\in \mathbb{N}^n$ and $d'\in D_{\G'}((A,\ycut),\xi,c)$ such that $\mathrm{size}(d')=l$ and $\lBert\bar{y}\rBert_{\bar{a}}=\wt(d)$. 

We prove Statement (*) by strong induction on $l$.

Let $l=1$. Then $d\in D_\G(A,\xi,c)$ has to be of the form $r=(A(p)\to\alpha)$ for some $r\in R$ such that $p(c)=\true$. Then $\alpha\in\Sigma\ui 0$ and $\xi=\alpha$. Clearly, $\wt(d)=\wt_K(r)$. Now let $\bar{y}=(y_1,\ldots,y_n)$ be the tuple in $\Nat^n$ such that, for each $i\in[n]$, $y_i=1$ if $a_i=\wt_K(r)$, and $y_i=0$ otherwise. Obviously, $\lBert\bar{y}\rBert_{\bar{a}}=\wt_K(r)=\wt(d)$. Moreover, by construction, the rule $r'=((A,\ycut)(p)\to\alpha)$ is in $R'$. Since $p(c)=\true$, $r'$ is in $D_{\G'}((A,\ycut),\xi,c)$.

Now, let $l> 1$ and assume that Statement (*) holds for all $l'\in\Nat$ with $l'< l$. We consider two cases. 

\underline{Case 1:} There exist some $r=(A(p)\to B(f))$ in $R$ and $d_1\in D_\G(B,\xi,f(c))$ such that $d=r(d_1)$. Then $p(c)=\true$ and $f(c)$ is defined. Moreover, $l=\mathrm{size}(d_1)+1$ and $\wt(d)=\wt(d_1)\cdot \wt_K(r)$. By IH there exist $\bar{y}_1=(y_{1,1},\ldots,y_{1,n})$ in $\Nat^n$ and $d_1'\in D_{\G'}((B,\yycut),\xi,f(c))$ such that $\mathrm{size}(d_1')=\mathrm{size}(d_1)$ and $\lBert\bar{y}_1\rBert_{\bar{a}}=\wt(d_1)$. Then let $\bar{y}=(y_1,\ldots,y_n)$ in $\Nat^n$ such that, for each $i\in[n]$, $y_i=y_{1,i}+1$ if $a_i=\wt_K(r)$, and $y_i=y_{1,i}$ otherwise. Obviously, $\lBert\bar{y}\rBert_{\bar{a}}=\wt(d)$. Moreover, by construction, there is a rule $r'=((A,\ycut)(p)\to(B,\yycut)(f))$ in $R'$. Since $p(c)=\true$ and $f(c)$ is defined, $d'=r'(d_1')$ is an element in $D_{\G'}((A,\ycut),\xi,c)$ with $\mathrm{size}(d')=l$.

\underline{Case 2:} There exist some $k\geq 1$, $r=(A(p)\to\sigma(B_1(f_1),\ldots,B_k(f_k)))$ in $R$, and $d_i\in D_\G(B_i,\xi_i,f_i(c))$ for each $i\in[k]$ such that $d=r(d_1,\ldots,d_k)$. Hence, $p(c)=\true$ and $f_i(c)$ is defined for each $i\in[k]$. Then $\xi=\sigma(\xi_1,\ldots,\xi_k)$ and $l=\mathrm{size}(d_1) +\ldots + \mathrm{size}(d_k) +1$. Moreover, $\wt(d)=\wt(d_1)\cdot\ldots\cdot\wt(d_k)\cdot\wt_K(r)$. For each $i\in[k]$, by IH, there exist $\bar{y}_i=(y_{i,1},\ldots,y_{i,n})$ in $\Nat^n$ and $d_i'\in D_{\G'}((B_i,\yicut),\xi_i,f_i(c))$ such that $\mathrm{size}(d_i')=\mathrm{size}(d_i)$ and $\lBert\bar{y}_i\rBert_{\bar{a}}=\wt(d_i)$. Then let $\bar{y}=(y_1,\ldots,y_n)$ in $\Nat^n$ such that, for each $j\in[n]$, $y_j=y'_{j}+1$ if $a_j=\wt_K(r)$, and $y_j=y'_{j}$ otherwise, where $\bar{y}'=\bar{y}_1 + \ldots +\bar{y}_k$ with $+$ denoting the pointwise addition of tuples. It is not hard to see that $\lBert\bar{y}\rBert_{\bar{a}}=\wt(d)$. Furthermore, by construction there exists a rule $r'=((A,\ycut)(p)\to\sigma((B_1,\yycut)(f_1),\ldots,(B_l,\ykcut)(f_k)))$ in $R'$. Since $p(c)=\true$ and $f_i(c)$ is defined for each $i\in[k]$, $d'=r'(d_1',\ldots,d_k')$ is an element in $D_{\G'}((A,\ycut),\xi,c)$ with $\mathrm{size}(d')=l$.

This finishes the proof of Statement (*).

Now let $\xi\in \supp(\lBert\G\rBert)$. Then there is a derivation tree $d\in D_\G(Z,\xi,c_0)$ with $\wt(d)\neq 0$. By Statement (*) there are $\bar{y}\in \Nat^n$ and $d'\in D_{\G'}((Z,\ycut),\xi,c_0)$ such that $\lBert\bar{y}\rBert_{\bar{a}}=\wt(d)$.  Since $\lBert\bar{y}\rBert_{\bar{a}}\neq 0$, also $\lBert\ycut\rBert_{\bar{a}}\neq 0$ (by Lemma \ref{Kirsten2}). Thus, $(Z,\ycut)\in Z'$ and  $\xi$ is in $L(\G')$.

Secondly, we prove that $ L( \mathcal{G'} )  \subseteq \supp(\lBert \mathcal{G} \rBert)$.  For this, we show the following Statement: 

(**) For each $l\in\Nat_+$, $\xi\in T_\Sigma$, $A\in N$, $\bar{z}\in T^n$, $c\in C$, and $d'\in D_{\G'}((A,\bar{z}),\xi,c)$: if $\mathrm{size}(d')=l$, then there are $\bar{y}\in \Nat^n$ and $d\in D_{\G}(A,\xi,c)$ such that $\mathrm{size}(d)=l$, $\lBert\bar{y}\rBert_{\bar{a}}=\wt(d)$ and $\bar{z}=\lfloor\bar{y}\rfloor_{\mathrm{dg}(\bar{a})}$. 

Statement (**) can be proved by strong induction on $l$. Since the proof is very similar to that of Statement (*), we drop it here.

Now let $\xi\in L(\G')$. Then there is a derivation tree $d'\in D_{\G'}((Z,\bar{z}),\xi,c_0)$ for some $\bar{z}\in T^n$ with $\lBert\bar{z}\rBert_{\bar{a}}\neq 0$.  By Statement (**) there are $\bar{y}\in \Nat^n$ and $d\in D_{\G}(Z,\xi,c_0)$ such that $\lBert\bar{y}\rBert_{\bar{a}}=\wt(d)$ and $\bar{z}=\lfloor\bar{y}\rfloor_{\mathrm{dg}(\bar{a})}$.  Since $\lBert\bar{z}\rBert_{\bar{a}}\neq 0$ also $\lBert\bar{y}\rBert_{\bar{a}}\neq 0$ (by Lemma \ref{Kirsten2}). Thus, since $K$ is zero-sum free, $\xi$ is in $\supp(\lBert\G\rBert)$.

The proof of 1(b) follows from the fact that
if the ZGP for $(K,\cdot,1)$ is decidable, then by Lemma \ref{Kirsten} we can compute the number $\operatorname{dg}(\bar{a})$ and hence construct $\G'$ effectively.

For the proof of  2, we note the following. Due to the conditions  $\lvert \Sigma \ui 1\rvert \ge 2$, the weighted finite automaton constructed in the corresponding part of the proof of \cite[Thm.~3.1.]{K} can be simulated by an $(\Sigma,M_K)$-rtg, hence that proof can be adapted to this setting.
\end{proof}

We note that the construction in the proof of Theorem~\ref{th:support}(1) becomes very simple if $K$ is zero-divisor free. Then, by Observation~\ref{ob:three}, $\operatorname{dg}(\bar{a})=0$ for every $\bar a$, and hence $N'$ is essentially $N$ (and the same holds for $Z'$ and $Z$). Thus, the rules of $\G'$ are obtained from those of $\G$ simply by dropping the weights.

Also we note that the first part of  Theorem~\ref{th:support}(1) (choosing $S=\TRIV$) provides a sufficient condition for the recognizability of support tree languages, which is different from the following well-known fact: If $K$ is a zero-sum free and zero-divisor free (not necessarily commutative) semiring, then the support of $\lBert \G \rBert$ for any $(\Sigma,K)$-rtg $\G$  is a recognizable tree language (cf., e.g., \cite[Thm.~3.12]{fulvog09new}).

In the next example we illustrate the construction of Theorem \ref{th:support}.
 
\begin{example}\rm We consider the strong bimonoid $(K,\max,\cdot,0,1)$ where $K = \{i \in \mathbb{N} \mid 0 \le i \le 9\}$, $\max$ is extended to maximum over countable index sets in the obvious way, 
and $\cdot$ is the multiplication of natural numbers modulo 9; thus, e.g., $3 \cdot 4 = 12 \ (\mathrm{mod}\ 9) = 3$. Indeed, $K$ is complete, zero-sum free, and commutative. 
We note that $(K,\cdot,1)$ has a decidable ZGP.

We consider the ranked alphabet $\Sigma = \{\gamma^{(1)}, \alpha^{(0)}, \beta^{(0)}\}$ and the $(\Sigma,M_K)$-rtg $\G = (\{A\},A,R,\wt)$ 
where the rules in $R$ and the corresponding weights are:
\begin{align*}
r_1\colon & A \rightarrow \gamma(A), \  \wt(r_1) = \mul_{1,2}\\
r_2\colon & A \rightarrow \alpha, \  \wt(r_2) = \mul_{1,3}\\
r_3\colon & A \rightarrow \beta, \  \wt(r_3) = \mul_{1,6}\enspace.
\end{align*}

By direct inspection of $\G$ we can easily see that $\supp(\lBert \G\rBert)= T_\Sigma$.  

Using the notations in the proof of Theorem \ref{th:support}, we have $W= \{2,3,6\}$.
 We let $\bar{a}=(2,3,6)$. Then the set $\lBert 0\rBert^{-1}_{\bar{a}}$ contains, e.g., the following three elements: $(0,2,0)$, $(0,0,2)$, and $(0,1,1)$. Indeed, 
\[\min(\lBert 0\rBert^{-1}_{\bar{a}}) = \{(0,2,0), (0,0,2), (0,1,1)\}\enspace.
\] 
The degree of $\bar{a}$ is $\mathrm{dg}(\bar{a}) = 2$, because  $\min(\lBert 0\rBert^{-1}_{\bar{a}}) \subseteq \{0,1,2\}^3$  and $\min(\lBert 0\rBert^{-1}_{\bar{a}}) \not\subseteq \{0,1\}^3$.
Hence $T= \{0,1,2\}$. 

Applying the construction of Theorem \ref{th:support}, we obtain the following $\Sigma$-rtg $\G' = (N',Z',R')$ with 
\begin{compactitem}
\item $N' = \{A\} \times T^3$; we  abbreviate $(A,(n_1,n_2,n_3))$ by $(n_1,n_2,n_3)$;
\item $Z' = \{(0,0,0)\} \cup \{(n,1,0) \mid 0 \le n \le 2\} \cup \{(n,0,1) \mid 0 \le n \le 2\}$;
\item $R'$ contains the following eight useful rules:
\begin{align*}
(0,1,0) &\rightarrow \alpha  &          (0,0,1) &\rightarrow \beta\\
(1,1,0) &\rightarrow \gamma((0,1,0)) &  (1,0,1) &\rightarrow \gamma((0,0,1))\\
(2,1,0) &\rightarrow \gamma((1,1,0)) &  (2,0,1) &\rightarrow \gamma((1,0,1))\\
(2,1,0) &\rightarrow \gamma((2,1,0)) &  (2,0,1) &\rightarrow \gamma((2,0,1))\enspace.
\end{align*}
\end{compactitem}
Hence also $L(\G') = T_\Sigma$. \hfill $\Box$
\end{example}

From Theorem \ref{th:support}(1) and the fact, that the emptiness problem of iterated pushdown tree automata is decidable \cite[Thm.~7.8]{dam82}, we obtain the following result:

\begin{corollary}\label{cor:emptyness}
	Let $K$ be a complete, zero-sum free, and commutative strong bimonoid with a decidable ZGP and $M_K$ be the complete \mmon{} associated with $K$. Moreover, let $s\colon T_\Sigma \rightarrow M_K$ be $(\mathrm{P}^n,\Sigma,M_K)$-recognizable for some $n\in \mathbb{N}$. Then it is decidable whether $\supp(s)=\emptyset$.
\end{corollary}

As particular case of Corollary \ref{cor:emptyness}, let us consider a complete lattice $(V,\vee,\wedge,0,1)$ with $0$ and $1$ as smallest and largest elements, respectively. It is clear that $V$ is a complete, zero-sum free, commutative, and $\wedge$-idempotent strong bimonoid. Thus, the ZGP of the monoid $(V,\wedge,1)$ is decidable. Then, Corollary \ref{cor:emptyness} shows that, for every complete lattice $(V,\vee,\wedge,0,1)$ and $(\mathrm{P}^n,\Sigma,M_V)$-recognizable weighted tree language $s$, it is decidable whether $\supp(s)=\emptyset$.

Moreover, combining Theorem \ref{th:support}(1) and Lemma \ref{lm:finite-storage-new} we obtain the following result.

\begin{corollary}\label{cor:Sigma-regular} Let $S$ be finite and $K$  a complete, zero-sum free, and commutative strong bimonoid such that the ZGP of $(K,\cdot,1)$ is decidable. For every $(S,\Sigma,M_K)$-rtg $\G$, 
a $\Sigma$-rtg can effectively be constructed which generates $\supp(\beh{\G})$.
\end{corollary}

We finish this section by showing that, in general, the empty-support problem is not decidable. Let us recall that  $\mathbb{B}$ is the Boolean M-monoid.

\begin{lemma} There is a storage type $S$ and a ranked alphabet $\Sigma$ such that the empty-support problem for $(S,\Sigma,\mathbb{B})$-rtg  is undecidable.
\end{lemma}
\begin{proof} We reduce this decidability problem to the Post-correspondence problem.  Let $\mathrm{PCP} = (u_1,v_1) \ldots (u_n,v_n)$ be a Post-correspondence problem \cite{pos46} where $u_i,v_i$ are non-empty strings over some alphabet $\Delta$ for each $i\in[n]$. We construct the storage type $S_\mathrm{PCP} = (\Delta^* \times \Delta^*, P,F,(\varepsilon,\varepsilon))$ with $P = \{\mathrm{equal}, \true_{\Delta^* \times \Delta^*}\}$ and $F = [n]$ where for each $(u,v) \in \Delta^* \times \Delta^*$ we define
\begin{compactitem}
\item $\mathrm{equal}((u,v)) = 1$ iff $u=v$
\end{compactitem}
and for each $i \in [n]$:
\begin{compactitem}
\item $i((u,v)) = (u u_i,v v_i)$.
\end{compactitem}
We let $\Sigma$ be the ranked alphabet with $\Sigma = \Sigma^{(0)} = \{\#\}$.
We construct the $(S_\mathrm{PCP},\Sigma,\mathbb{B})$-rtg $\G = (\{Z\},Z,R,\wt)$ with the following rules:
$Z(\true_{\Delta^* \times \Delta^*}) \rightarrow Z(i)$ for each $i \in [n]$ and $Z(\mathrm{equal}) \rightarrow \#$. Moreover, the weight of each rule is $1$. 

Then it should be clear that $\sem{\G} = 1.\#$ if $\mathrm{PCP}$ has a solution, and 
$\widetilde{0}$ otherwise (where $\widetilde{0}: T_\Sigma \rightarrow M_K$ is the weighted tree language which maps each tree to $0$).   Hence $\supp(\sem{\G}) \not= \emptyset$ iff $\mathrm{PCP}$ has a solution.
\end{proof}

%%%%%%%%%%%%%%%%%%%%%%%%%%%%%%%%%%%%%%%%%%%%%%

\section{Decomposition results}
\label{sec:decomp}

In this section we will decompose the weighted tree language generated by an $(S,\Sigma,K)$-rtg in two different ways. As a combination of them, we can characterize elements of $\Reg(S,\Sigma,K)$ by elementary concepts: a tree transformation, an alphabetic mapping, and a regular tree language. 

\begin{figure}[t]\centering\footnotesize
	\begin{tikzpicture}[sibling distance=4em, level distance=3em,
  every node/.style = {align=center}]]
  \pgfdeclarelayer{bg}    % declare background layer
  \pgfsetlayers{bg,main}  % set the order of the layers (main is the standard layer)

  \node(x) {$\mhl{\sigma}$}
    child { node {$\mhl{\delta}$}
      child { node {$\mhl{\alpha}$} } }
    child { node {$\mhl{\alpha}$} };

 \begin{scope}[sibling distance=12em, xshift=30em,yshift=6.5em]
 \node(t) {$\langle(\ttop=\gamma_0,\push(\gamma)),*\rangle$}
   child { node {$\langle(\ttop=\gamma,\push(\gamma)\push(\gamma)),\mhl{\sigma}\rangle$}
    child { node {$\langle(\ttop=\gamma,\pop),*\rangle$}
      child { node {$\langle(\ttop=\gamma,\pop),\mhl{\delta}\rangle$}
        child { node {$\langle(\ttop=\gamma_0,\varepsilon),\mhl{\alpha}\rangle$} } } }
    child { node {$\langle(\ttop=\gamma,\push(\gamma)),*\rangle$}
      child { node {$\langle(\ttop=\gamma,\pop),*\rangle$}
        child { node {$\langle(\ttop=\gamma,\pop),*\rangle$}
        	child { node {$\langle(\ttop=\gamma,\varepsilon),\mhl{\alpha}\rangle$} } } } } };
 \end{scope}
\begin{pgfonlayer}{bg}
 \draw[bend angle=20, bend left, black!30,dashed] (x) to (t-1);
 \draw[bend angle=20, bend left, black!30,dashed] (x-1) to (t-1-1-1);
 \draw[bend angle=10, bend right, black!30,dashed] (x-1-1) to (t-1-1-1-1);
 \draw[bend angle=10, bend right, black!30,dashed] (x-2) to (t-1-2-1-1-1);
 \end{pgfonlayer}
\end{tikzpicture}
\caption{A tree $\xi\in T_\Sigma$ on the left side, an element $\zeta$ of $\B_\Delta(\xi)$ on the right side for the ranked alphabet  $\Delta$ corresponding to $\Sigma$, $P' =\{\ttop=\gamma_0,\ttop=\gamma\}$, and $F'=\{\push(\gamma),\pop\}$, and the bijection $\theta_{\xi,\zeta}$.}
\label{Fig1}
\end{figure}
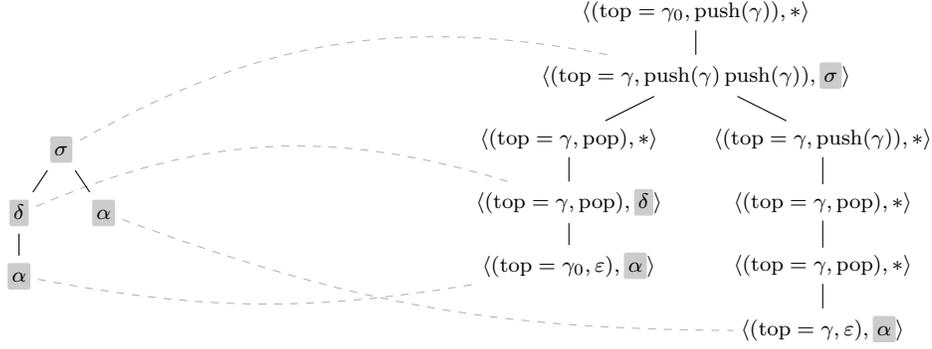

 First, we separate the storage and second we separate weights.
For this we need the concept of behaviour on a tree $\xi \in T_\Sigma$.  Since we need this concept also in Section~\ref{sec:BET}, we place its definition  here (and not inside Section~\ref{sec:sep-storage}). 
Intuitively, a behaviour on $\xi$ is a tree that is obtained from $\xi$ by adding to the label of each position $w$ a pair $(p,f_1\ldots f_k)$ of predicate $p$ and instructions $f_1,\ldots,f_k$ if $w$ has $k$ successors, and inserting an arbitrarily long, but finite sequence of unary symbols of the form $\langle( p,f),*\rangle$ above each position of $\xi$. Figure \ref{Fig1} gives a first rough impression (where the storage type is $\PD-op^1$). A behaviour on $\xi$ can be seen as a trace of a derivation tree of an $(S,\Sigma,K)$-rtg for $\xi$ in which the occurrences of nonterminals  are dropped; the unary symbols $\langle( p,f),*\rangle$ represent applications of chain rules.

Formally, let $\Sigma$ be a ranked alphabet and $P' \subseteq P$ and $F' \subseteq F$ be finite sets. Moreover, let $\Delta$ be the ranked alphabet corresponding to $\Sigma$, $P'$, and $F'$ (cf. Section~\ref{subsec:storage}). Furthermore, let $*$ be a symbol of rank 1 such that $* \not\in \Sigma$. We define 
the \textit{$\Sigma$-extension of $\Delta$}, denoted by  $\DS$, to be the ranked alphabet 
where $\DS\ui 1 =\Delta\ui 1\times(\Sigma\ui 1\cup\{*\})$ and $\DS\ui k =\Delta\ui k\times\Sigma{\ui k}$ for $k\neq 1$.

Additionally, let $\mathcal{R}$ be the term rewriting system having the rules:
\begin{align*}
\sigma(x_1,\ldots,x_k) &\to \langle (p,f),*\rangle(\sigma(x_1,\ldots,x_k)),\\
&\textrm{for every }  k\geq 0, \sigma\in\Sigma^{(k)}, \text{ and } (p,f)\in\Delta^{(1)}, \text{ and}\\
\sigma(x_1,\ldots,x_k)&\to \langle (p,f_1\ldots f_k), \sigma \rangle(x_1,\ldots ,x_k),\\
&\textrm{for every }  k\geq 0, \sigma\in\Sigma^{(k)}, \text{ and }(p,f_1\ldots f_k)\in\Delta^{(k)}.
\end{align*}
Then we define the mapping $\B_\Delta: T_\Sigma \rightarrow \mathcal{P}(T_{\DS})$ for each $\xi\in T_\Sigma$ by 
\[
\B_\Delta(\xi)=\{\zeta\in T_{\DS} \mid \xi\Rightarrow_\mathcal{R}^{*} \zeta \text{ and }\pr_1(\zeta)\in \B(\Delta)\}
\]
and we call the set $\B_\Delta(\xi)$ the set of \textit{$\Delta$-behaviours on $\xi$}.

Let $\Theta= \DS\setminus (\Delta\ui 1\times\{*\})$. It is clear that, for each $\xi \in T_\Sigma$ and $\zeta \in \B_\Delta(\xi)$, there is a unique bijection $\theta: \pos(\xi) \rightarrow \pos_{\Theta}(\zeta)$ which preserves the lexicographic order, i.e., if $w_1 \le_{\mathrm{lex}} w_2$, then $\theta(w_1) \le_{\mathrm{lex}} \theta(w_2)$ for every $w_1,w_2 \in \pos(\xi)$. We denote this bijection by $\theta_{\xi,\zeta}$. In Figure \ref{Fig1} we illustrate a $\Delta$-behaviour $\zeta$ on some tree $\xi\in T_\Sigma$ (for the storage type $\PD-op^1$) and the bijection $\theta_{\xi,\zeta}$.

\subsection{Separating storage}
\label{sec:sep-storage}

Inspired by the decomposition of CFT($S$)-transducers into an approximation and a macro tree transducer (see \cite[Thm.~3.26]{engvog86}), we can decompose an $(S,\Sigma,K)$-rtg into the mapping $\B_\Delta$ (for appropriate $\Delta$) and a $(\langle \Delta,\Sigma \rangle,K)$-rtg.

\begin{definition}\rm\label{df:related}
Let $\G=(N,Z,R,\wt)$ be an $(S,\Sigma,K)$-rtg. Moreover, let  $P' \subseteq P$ and $F' \subseteq F$ be finite subsets, $\Delta$  the ranked alphabet corresponding to $\Sigma$, $P'$ and $F'$, and   $\G' = (N',Z',R',\wt')$ a chain-free $(\DS,K)$-rtg. We say that $\G$ and $\G'$ are \emph{related} if
\begin{compactitem}
\item $\Delta=\Delta_\G$,
\item $N=N'$ and $Z = Z'$,
\item there is a bijection between $R$ and $R'$ such that the following holds:
\begin{compactitem}
\item \(r =  (A(p) \rightarrow \sigma(B_1(f_1),\ldots,B_k(f_k)))\) in $R$ corresponds to\\ 
\(r' = (A \rightarrow \langle(p,f_1\ldots f_k),\sigma\rangle(B_1,\ldots,B_k))\) in $R'$ and
\item \(r = (A(p) \rightarrow B(f))\) in $R$  corresponds to  \\
\(r' = (A \rightarrow \langle(p,f),*\rangle(B)\) in $R'$;
\item $\wt'(r') = \wt(r)$ for each $r$ and $r'$ in this bijection.
\end{compactitem} 
\end{compactitem}
\end{definition}

\begin{lemma}\label{lm:related} Let $\G$ be an $(S,\Sigma,K)$-rtg. Moreover, let $P' \subseteq P$ and $F' \subseteq F$ be finite subsets and $\Delta$ the ranked alphabet corresponding to $\Sigma$, $P'$, and $F'$. Also let $\G'$  be a chain-free $(\DS,K)$-rtg.
If $\G$ and $\G'$ are related, then   $\seml \G\semr = \B_\Delta ; \seml \G'\semr$.
\end{lemma}
\begin{proof} Let $\xi \in T_\Sigma$. 
It is obvious that there is a one-to-one correspondence between the sets $D_{\G}(\xi)$ and $\{(\zeta,d') \mid \zeta \in \B_\Delta(\xi), d' \in D_{\G'}(\zeta)\}$. Moreover, if $d$ and  $(\zeta,d')$ correspond to each other, then $\wt(d) = \wt'(d')$.  

Then we can calculate as follows for each $\xi \in T_\Sigma$:
\begin{align*}
\seml \G\semr(\xi) 
& = \sum_{d \in D_\G(\xi)} \wt(d)
 = \sum_{\zeta \in \B_\Delta(\xi)}\; \sum_{d' \in D_{\G'}(\zeta)} \wt'(d')\\
& = \sum_{\zeta \in \B_\Delta(\xi)} \lBert \G' \rBert(\zeta) 
= (\B_\Delta ; \seml \G'\semr)(\xi)\enspace.
\end{align*}

\vspace{-5mm}

\end{proof}

\begin{theorem}\label{th:sep-storage} For every $s\colon T_\Sigma\to K$ the following two statements are equivalent:
	\begin{compactitem}
		\item[(i)] $s$ is $(S,\Sigma,K)$-regular.
		\item[(ii)] There are finite sets $P' \subseteq P$ and $F' \subseteq F$ and there is a chain-free $(\langle \Delta,\Sigma\rangle,K)$-rtg $\G$ such that $\Delta$ is  the ranked alphabet corresponding to $\Sigma$, $P'$, and $F'$ and $s=\B_\Delta ; \seml \G \semr$.
	\end{compactitem}
\end{theorem}

\begin{proof} ``(i) $\Rightarrow$ (ii)'':  Let $\G$ be a  $(S,\Sigma,K)$-rtg. Then we can easily construct  a chain-free $(\langle \Delta_\G,\Sigma\rangle,K)$-rtg $\G'$ such that $\G$ and $\G'$ are related. Lemma \ref{lm:related} implies $\seml \G\semr = \B_{\Delta_\G}; \seml \G'\semr$.

``(ii) $\Rightarrow$ (i)'': Let $P' \subseteq P$ and $F' \subseteq F$ be finite sets and  $\Delta$ be the ranked alphabet corresponding to $\Sigma$, $P'$, and $F'$. Moreover, let $\G$ be a chain-free $(\langle \Delta,\Sigma\rangle,K)$-rtg. Then we can easily construct  an $(S,\Sigma,K)$-rtg $\G'$ such that $\G'$ and $\G$ are related. Lemma \ref{lm:related} implies that $\B_\Delta; \seml\G\semr = \seml \G'\semr$.
\end{proof}

%%%%%%%%%%%%%%%%%%%%%%%%%%%%%%%%%%%%%%%%%%%%

\subsection{Separating weights}\label{weightsep}

Let $\Theta$ be a ranked alphabet and let $h=(h_k\mid 0\leq k\leq \maxrk(\Theta))$ be a family of mappings such that 
\begin{align*}
h_1&\colon\Theta\ui 1\to  \Omega^{(1)} \cup (\Omega^{(1)} \times \Sigma \ui 1) \ \text{ and }\\
h_k&\colon \Theta\ui k\to \Omega\ui k \times \Sigma \ui k \ \text{ for $k\neq 1$}\enspace.
\end{align*}
Then the \emph{alphabetic mapping induced by $h$} is  $h'\colon T_\Theta\to K[T_\Sigma]$ defined as follows: for every $k\in\Nat$, $\theta\in\Theta\ui k$, and $\zeta_1,\ldots,\zeta_k\in T_\Theta$ we let
\[h'(\theta(\zeta_1,\ldots,\zeta_k))=
\begin{cases}
        \omega(a_1).\xi_1 &\text{ if } k=1 \text{ and } h_1(\theta)=\omega \\
	\omega(a_1,\ldots,a_k).\sigma(\xi_1,\ldots,\xi_k) &\text{ if } h_k(\theta)=(\omega,\sigma),
\end{cases}\]
where $h'(\zeta_i)=a_i.\xi_i$ for each $i\in[k]$. In the sequel we identify $h$ and $h'$. Now let $L \subseteq T_\Theta$. We define the weighted tree language $h(L): T_\Sigma \rightarrow K$  by 
\[
h(L) = \sum_{\zeta \in L} h(\zeta)\enspace. 
\]

The following theorem shows how to decompose an $(S,\Sigma,K)$-rtg into an alphabetic mapping and an unambiguous and  chain-free $(S,\Theta)$-rtg. This theorem is inspired by \cite[Lm.~3 and Lm.~4]{drovog13} and \cite[Th.~6]{hervog15} and uses the same proof technique.

\begin{theorem}\label{sep-weights}
	For every $s\colon T_\Sigma\to K$ the following two statements are equivalent:
	\begin{compactitem}
		\item[(i)] $s=\lBert \G\rBert$ for some $(S,\Sigma,K)$-rtg $\G$.
		\item[(ii)] There are a ranked alphabet $\Theta$, an unambiguous and chain-free $(S,\Theta)$-rtg $\H$, and an alphabetic mapping $h\colon T_\Theta\to K[T_\Sigma]$ such that $s=h(\cal L(\H))$.
\end{compactitem}
Moreover, if in (i) $\G$ is chain-free, then in (ii) $h_1(\Theta^{(1)}) \subseteq \Omega^{(1)} \times \Sigma^{(1)}$, and vice versa.
\end{theorem}

\begin{proof} 
 (i)$\Rightarrow$(ii): Let  $\G = (N,Z,R,\wt)$. 
As before we view $R$ as ranked alphabet by associating rank $k$  with a rule $r \in R$ if its right-hand side contains $k$ nonterminal occurrences. We choose $\Theta= R$. Moreover, we let $\H=(N,Z,R')$ be the $(S,R)$-rtg and $h\colon T_{R}\to K[T_\Sigma]$ be the alphabetic mapping such that 
 \begin{compactitem}
 	\item if $r=(A(p)\to\sigma(A_1(f_1),\ldots,A_k(f_k)))$ is in $R$, then let\\ $r'=(A(p)\to r(A_1(f_1),\ldots,A_k(f_k)))$ be in $R'$ and $h_k(r)=(\wt(r),\sigma)$, and
 	\item if $r=(A(p)\to B(f))$ is in $R$, then let $r'=(A(p)\to r(B(f))$ be in $R'$ and $h_1(r)=\wt(r)$.
 \end{compactitem}
If $\G$ is chain-free, then $h_1(R^{(1)}) \subseteq \Omega^{(1)} \times \Sigma^{(1)}$. Obviously, $\H$ is  chain-free. It is also easy to see that $\H$ is unambiguous because the tree relabeling $\mu\colon T_R\to T_{R'}$ defined by the correspondence $r\mapsto r'$ is a bijection and, for every $d\in \cal L(\H)$, the only derivation tree of $\H$ for $d$ is $\mu(d)$. Moreover,
\begin{equation}\tag{$\dagger$}
\mathcal{L}(\H)=\bigcup_{\xi\in T_\Sigma} D_\G(\xi). 
\end{equation}
\noindent In fact, each $d\in \cal L(\H)$ is a derivation tree of $\G$ for  $\pi(d)$, where $\pi\colon T_R\to T_\Sigma$ is the mapping defined on page \pageref{page:pi}. Moreover, if $d\in D_\G(\xi)$ for some $\xi\in T_\Sigma$, then $\mu(d)\in T_{R'}$ is the (only) derivation tree of $\H$ for $d$, hence $d\in \cal L(\H)$. Finally, we note that for every  $d\in \cal L(\H)$ and $\xi\in T_\Sigma$ we have
\begin{equation}\tag{$*$} (h(d))(\xi)=
	\begin{cases}\wt(d) &\text{ if } d\in D_\G(\xi)\\
	0 &\text{ otherwise.}
	\end{cases}
\end{equation}
Then we have
 \begin{align*}
 \lBert\G\rBert(\xi) &= \sum_{d\in D_\G(\xi)}\wt(d) = \sum_{d\in L(\H)\colon d\in D_\G(\xi)}(h(d))(\xi)
 = \sum_{d\in \cal L(\H)} (h(d))(\xi)
 =(h(\cal L(\H)))(\xi)
 \end{align*}
 where the second equality is justified by $(\dagger)$ and $(*)$.

 (ii)$\Rightarrow$(i): Let $\H=(N',Z',R')$ be an unambiguous chain-free $(S,\Theta)$-rtg and let $h\colon T_{\Theta}\to K[T_\Sigma]$ be an alphabetic mapping. We construct an $(S,\Sigma,K)$-rtg $\G$ such that $\lBert\G\rBert=h(\cal L(\H))$. The idea for the construction is to code the preimage of $h$ in the nonterminals of $\G$ as in \cite[Lm. 4]{drovog13}.
 We let $\G=(N,Z,R,\wt)$ with $N=N'\times \Theta$, $Z=Z'\times\Theta$ and $R$ and $\wt$ are defined as follows. 
 \begin{compactitem}
 	\item If $A(p)\to\theta(A_1(f_1),\ldots,A_k(f_k))$ is in $R'$ with $h(\theta)=(\omega,\sigma)$, then for every $\theta_1,\ldots,\theta_k\in\Theta$ the rule $r=((A,\theta)(p)\to\sigma((A_1,\theta_1)(f_1),\ldots,(A_k,\theta_k)(f_k)))$ is in $R$ and $\wt(r)=\omega$.
 	\item If $A(p)\to\theta(A_1(f_1))$ is in $R'$ with $h(\theta)=\omega$, then for every $\theta_1\in\Theta$ the rule $r=((A,\theta)(p)\to (A_1,\theta_1)(f_1))$ is in $R$  and $\wt(r)=\omega$.
 \end{compactitem}
If  $h_1(\Theta^{(1)}) \subseteq \Omega^{(1)} \times \Sigma^{(1)}$, then $\G$ is chain-free.

 For every $\zeta\in \cal L(\H)$ let us denote by $d_\H(\zeta)$ the unique derivation tree of $\H$ for $\zeta$ (note that $\H$ is unambiguous). Moreover, for every $\xi\in T_\Sigma$, let
us denote by $D_\H(h^{-1}(\xi))$ the set
\[\{ d_\H(\zeta) \mid \zeta\in \mathcal{L}(\H), h(\zeta)(\xi)\neq 0\}.\]

Let  $d=d_\H(\zeta)$, for some $\zeta \in \cal L(\H)$, and  let $d'\in D_\G(\xi)$. We say that \emph{$d$ corresponds to $d'$} if $\pos(d)=\pos(d')$ and for each $w\in\pos(d)$:
 \begin{compactitem}
 	\item if $d(w)=(A(p)\to\theta(A_1(f_1),\ldots,A_k(f_k)))$ with $h(\theta)=(\omega,\sigma)$, then $d'(w)=((A,\theta)(p)\to\sigma((A_1,\zeta(w1))(f_1),\ldots,(A_k,\zeta(wk))(f_k)))$, and
 	\item if $d(w)=(A(p)\to\theta(B(f)))$ with $h(\theta)=\omega$, then $d'(w)=((A,\theta)(p)\to(B,\zeta(w1)))$.
 \end{compactitem}
 It is not hard to see that the above correspondence is a bijection between $D_\H(h^{-1}(\xi))$ and $\{d'\in D_\G(\xi)\mid \wt(d')\ne 0\}$ for every $\xi\in T_\Sigma$. Moreover, if
$d_\H(\zeta)$ corresponds to $d'$, then  $\wt(d')=(h(\zeta))(\xi)$.
 Then, for every $\xi\in T_\Sigma$, we have
 \begin{align*}
 (h(\cal L(\H)))(\xi) &= \sum_{\zeta\in \cal L(\H)} (h(\zeta))(\xi) =
\sum_{\substack{\zeta\in \cal L(\H):\\h(\zeta)(\xi)\neq 0}} (h(\zeta))(\xi)
=^{(\dagger)}\sum_{\substack{\zeta\in \cal L(\H):\\ d_\H(\zeta)\in D_\H(h^{-1}(\xi)) }} (h(\zeta))(\xi)\\
 & =^{(*)}\sum_{\substack{d'\in D_\G(\xi):\\\wt(d')\ne 0}}\wt(d')
=\sum_{d'\in D_\G(\xi)}\wt(d')=\lBert\G\rBert(\xi),
 \end{align*}
 where $(\dagger)$ holds because for $\zeta\in \cal L(\H)$ we have $(h(\zeta))(\xi)=0$ if $d_\H(\zeta)\not\in D_\H(h^{-1}(\xi))$;
and $(*)$ holds due to the bijection described above. 
\end{proof}

Now we can prove that, even in the weighted case, we can eliminate finite storage types. 

\begin{corollary}\label{lm:dropping-finite-storage} Let $S$ be finite.
\begin{compactenum}
\item $\Reg(S,\Sigma,K) \subseteq \Reg(\Sigma,K)$ and $\Regnc(S,\Sigma,K) \subseteq \Regnc(\Sigma,K)$.
\item If $S_{\true,\id}=S$, then $\Reg(\Sigma,K) \subseteq \Reg(S,\Sigma,K)$ and $\Regnc(\Sigma,K) \subseteq \Regnc(S,\Sigma,K)$.
\end{compactenum}
\end{corollary}

\begin{proof} First we prove 1. Let $\G$ be an $(S,\Sigma,K)$-rtg. By Theorem \ref{sep-weights}(i)$\Rightarrow$(ii), there are a ranked alphabet $\Theta$, an unambiguous and  chain-free $(S,\Theta)$-rtg $\H$, and an alphabetic mapping $h\colon T_\Theta\to K[T_\Sigma]$ such that $s=h(\cal L(\H))$. By Lemma \ref{lm:finite-storage-new} there is a $\Theta$-rtg $\H'$ such that $\cal L(\H) = \cal L(H')$. Moreover, $\H'$ is unambiguous and chain-free. Hence, by Theorem \ref{sep-weights}(ii)$\Rightarrow$(i) (with $S=\TRIV$), we obtain that $s \in  \Reg(\Sigma,K)$.

Now let, additionally, $\G$ be chain-free. Then $h_1(\Theta^{(1)}) \subseteq \Omega^{(1)} \times \Sigma^{(1)}$ and thus, by the same argumentation as above, we obtain that $s \in \Regnc(\Sigma,K)$. 

Next we prove  2. Let $\G = (N,Z,R,\wt)$ be a $(\Sigma,K)$-rtg. Then we construct the $(S,\Sigma,K)$-rtg $\G' =(N,Z,R',\wt')$ such that 
\begin{compactitem}
\item if $r = (A \rightarrow \sigma(A_1,\ldots,A_k))$ is a rule in $R$,\\ then $r' = (A(\true_C) \rightarrow \sigma(A_1(\id_C),\ldots,A_k(\id_C)))$ is in $R'$,
\item if $r = (A \rightarrow B)$ is a rule in $R$, \\
then $r' = (A(\true_C) \rightarrow B(\id_C))$ is in $R'$, and
\item  in both cases we let $\wt'(r') = \wt(r)$.
\end{compactitem}
If $\G$ is chain-free, then so is $\G'$. For each $\xi \in T_\Sigma$, there is a bijection $\theta: D_\G(\xi) \rightarrow D_{\G'}(\xi)$  such that $\wt'(\theta(d)) = \wt(d)$ for each $d \in D_\G(\xi)$. This implies that $\seml \G' \semr = \seml \G \semr$. 
\end{proof}

From Corollary \ref{lm:dropping-finite-storage} we immediately obtain the following result.

\begin{corollary}\label{th:finite-storage-new} If $S$ is finite and $S_{\true,\id} = S$,
then $\Reg(S,\Sigma,K) = \Reg(\Sigma,K)$ and $\Regnc(S,\Sigma,K) = \Regnc(\Sigma,K)$.
\end{corollary}

%%%%%%%%%%%%%%%%%%%%%%%%%%%%%%%%%%%%%%%%%%%%%%%%%%%%

\subsection{Combination of separation results}\label{combsep}

In this section we combine the separation of storage with the separation of weights. In this way, we can characterize each element in $\Reg(S,\Sigma,K)$ by elementary concepts: a tree transformation $\B_\Delta$, an alphabetic mapping $h$, and an element in $\Reg(\Theta)$.

\begin{theorem}\label{thm:combsep} For every $s\colon T_\Sigma\to K$ the following two statements are equivalent:
	\begin{compactitem}
		\item[(i)] $s$ is $(S,\Sigma,K)$-regular.
		\item[(ii)] $s=\B_\Delta ; h(\cal L(\H))$ for some finite sets $P'\subseteq P$, $F'\subseteq F$, ranked alphabet $\Delta$ corresponding to $\Sigma$, $P'$, and $F'$, ranked alphabet $\Theta$, unambiguous and chain-free $\Theta$-rtg $\H$, and alphabetic mapping $h\colon T_\Theta\to K[T_{\langle \Delta,\Sigma\rangle}]$.
	\end{compactitem}
\end{theorem}

\begin{proof}
	(i) $\Rightarrow$ (ii): By Theorem \ref{th:sep-storage} there are finite sets $P' \subseteq P$ and $F' \subseteq F$ and there is a chain-free $(\langle \Delta,\Sigma\rangle,K)$-rtg $\G$ such that $\Delta$ is  the ranked alphabet corresponding to $\Sigma$, $P'$, and $F'$ and $s=\B_\Delta ; \seml \G \semr$.
	\newline\indent According to Theorem \ref{sep-weights} there are a ranked alphabet $\Theta$, an unambiguous and chain-free $\Theta$-rtg $\H$, and an alphabetic mapping $h\colon T_\Theta\to K[T_{\langle \Delta,\Sigma\rangle}]$ such that $\lBert \G\rBert=h(\cal L(\H))$.

	(ii) $\Rightarrow$ (i): By Theorem \ref{sep-weights} we have that $h(\cal L(\H))$ is $(\langle \Delta,\Sigma\rangle,K)$-recognizable. Then, by  Theorem \ref{th:sep-storage}, $\B_\Delta ; h(\cal L(\H))$ is $(S,\Sigma,K)$-recognizable.
\end{proof}

\begin{example}\rm Here we show a slightly more complex example of a weighted tree language $s$, we show a weighted regular tree grammar $\G$ with storage, and we apply the decompositions inherent in the proof of Theorem \ref{thm:combsep} to $\G$ in order to show that $s = \sem{\G}$. 

First, we define the weighted tree language $s$. For this, we let $\Sigma=\{\sigma\ui 2,\delta\ui 2,\#\ui 0\}$. Roughly speaking, the mapping $s$ maps each tree $\xi \in T_\Sigma$ to its degree of unbalancedness \cite[Example 1]{stuvogfue09} if $\xi$ has the following path property, and to $0$ otherwise. A tree $\xi$ has the path property if, for each position $w$ of $\xi$, the number of occurrences of $\sigma$'s above $w$ or at $w$ is greater than or equal to the number of occurrences of  $\delta$'s above $w$ or at $w$.

The degree of unbalancedness is formalized as follows. For each tree $\xi\in T_\Sigma$ and each $w\in\pos(\xi)$ we define 
\[
\mathrm{ubal}(\xi,w)=
\begin{cases} |\height(\xi_{|w1}) - \height(\xi_{|w2})| &\text{ if } \xi(w) \in\{\sigma,\delta\},\\
0 &\text{ otherwise, }
\end{cases}
\]
and we define $\mathrm{ubal}(\xi)=\max(\mathrm{ubal}(\xi,w)\mid w\in\pos(\xi))$, which is the \emph{degree of unbalancedness of~$\xi$}. (At the end of this example we will use $\mathrm{ubal}$ for another ranked alphabet.)

The path property can be captured by the formal string language 
\[L=\{w\in \{\sigma,\delta\}^*\{\#\}\mid \forall u \in\{\sigma,\delta\}^*\{\#\}: \text{if } u \preceq w, \text{ then } |u|_\sigma\geq|u|_\delta\}\enspace.\]
Then we define the weighted tree language $s\colon T_\Sigma\to \Nat$ for each $\xi\in T_\Sigma$ by
\[
s(\xi)=
\begin{cases}
	\mathrm{ubal}(\xi) &\text{ if } \ourpath(\xi)\subseteq L,\\
	0 &\text{ otherwise. }\\
\end{cases}
\]

Second, we define a  $(\COUNT,\Sigma,K_{\max})$-rtg $\G$ (for some specific M-monoid $K_{\max}$) which generates $s$  (cf. Section \ref{subsec:storage} for the definition of $\COUNT$; we abbreviate the predicate $\true_\mathbb{N}$ by $\true$). For this, we define the complete M-monoid
\[
K_{\max}=(\Nat\cup\{\infty\},\max,0,\Omega)
\]
where $\max$ is extended  to maximum over countable index sets in the obvious way.
Moreover, we let  
\[
\Omega=\{0_k\mid k\in \Nat\}\cup\{1_0\} \cup \{\pi_1\ui 2,\pi_2\ui 2,\mathrm{diff}\ui 2,\mathrm{ht}\ui 2\}
\]
where $1_0$ is the $0$-ary function defined by $1_0()=1$, and for every $a,b\in\Nat \cup \{\infty\}$ we define the binary operations by case analysis as follows. 

Case 1: If $a=0$ or $b=0$, then $ \pi_1(a,b) = \pi_2(a,b)= \mathrm{diff}(a,b)= \mathrm{ht}(a,b)= 0$. (These definitions guarantee that each binary operation of $\Omega$ is absorptive.) 

Case 2: If $a \in \mathbb{N}\setminus\{0\}$ and $b \in \mathbb{N}\setminus \{0\}$, then 
$\pi_1(a,b)=a$,  $\pi_2(a,b) =b$, $\mathrm{diff}(a,b)=|a-b|$, and $\mathrm{ht}(a,b)=1+\max(a,b)$.

Case 3: Otherwise we define $\pi_1(a,b)$, $\pi_2(a,b)$, $\mathrm{diff}(a,b)$, and $\mathrm{ht}(a,b)$ arbitrarily.

We note that Cases 1 and 3 will not occur in computations in our example.

Now we construct the $(\COUNT,\Sigma,K_{\max})$-rtg $\G=(N,U,R,\wt)$ with the following intuition. As in \cite[Example 1]{stuvogfue09}, we use two nonterminals, i.e., $N=\{H,U\}$, such that for each tree $\xi \in T_\Sigma$ and derivation tree $d$ of $\xi$, there is exactly one position $w \in \pos(d)$ such that at $w$ and each of its predecessors a rule with left-hand side nonterminal $U$ is applied and at each other position a rule with left-hand side nonterminal $H$ is applied. Then the weight of this derivation tree is $\mathrm{ubal}(\xi,w)$. Since the position $w$ is chosen nondeterministically, the value $\max(\mathrm{ubal}(\xi,w)\mid w\in\pos(\xi))$  is computed, which is the degree of unbalancedness of the whole tree $\xi$. The path property is checked by using the storage $\COUNT$: on each generation of a $\sigma$-labeled position, the counter is incremented, and on each generation of a $\delta$-labeled position it is decremented.  
\begin{align*}
	r_1&:H(\true)\to \# \ , \ 1_0 & r_2&:U(\true)\to \#\ , \ 0_0\\
r_3&:H(\true)\to \sigma(H(\inc),H(\inc)) \ , \ \mathrm{ht} &
	r_4&:U(\true)\to \sigma(H(\inc),H(\inc))\ , \ \mathrm{diff}\\
	r_5&:U(\true)\to \sigma(H(\inc),U(\inc)) \ , \ \pr_2 & 
	r_6&:U(\true)\to \sigma(U(\inc),H(\inc)) \ , \ \pr_1\\
r_7&:H(\true)\to \delta(H(\dec),H(\dec)) \ , \ \mathrm{ht} &
	r_8&:U(\true)\to \delta(H(\dec),H(\dec)) \ , \ \mathrm{diff}\\
	r_9&:U(\true)\to \delta(H(\dec),U(\dec)) \ , \ \pr_2 & 
	r_{10}&:U(\true)\to \delta(U(\dec),H(\dec)) \ , \ \pr_1\enspace.
	\end{align*}

\begin{figure}[t]
	\scriptsize\centering
\begin{tikzpicture}[sibling distance=16em, level distance=3.5em,
  every node/.style = {align=center}]]

\pgfdeclarelayer{bg}    % declare background layer
  \pgfsetlayers{bg,main}  % set the order of the layers (main is the standard layer)
 
 \node {$U(\true)\to \mhl{\sigma}(U(\inc),H(\inc))$}
    child { node {$U(\true)\to \mhl{\delta}(H(\dec),H(\dec))$}
      child { node {$H(\true)\to \mhl{\#}$} } 
      child{ node {$H(\true)\to \mhl{\sigma}(H(\inc),H(\inc))$} child{ node {$H(\true)\to \mhl{\#}$} }child{ node {$H(\true)\to \mhl{\#}$} }}}
     child { node {$H(\true)\to \mhl{\#}$} } ;

\end{tikzpicture}
\caption{\label{fig5} A derivation tree $d\in D_{\G}(\xi)$ for the tree $\xi = \sigma(\delta(\#,\sigma(\#,\#)),\#)$.}
\end{figure}
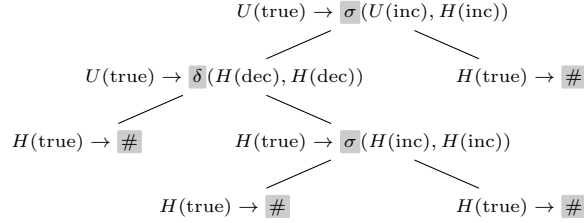

Indeed, the grammar $\G$ does not use the predicate $\zero$. In Figure \ref{fig5} we show a derivation tree $d$ of $\G$ for the tree $\xi = \sigma(\delta(\#,\sigma(\#,\#)),\#)$ with $\wt(d) =1$.

Third, we apply to the $(\COUNT,\Sigma,K_{\max})$-rtg $\G$ the decompositions that are inherent in the proof of Theorem \ref{thm:combsep} consecutively, i.e., the separation of the storage (cf. Theorem \ref{th:sep-storage}(i)$\Rightarrow$(ii))  and the separation of the weights (cf. Theorem \ref{sep-weights}(i)$\Rightarrow$(ii)).  Since we use the storage to check the path property and the weights to calculate the degree of unbalancedness, Theorem \ref{thm:combsep} suggests that we can prove $s = \sem{\G}$ in two independent steps. For this, let $\xi \in T_\Sigma$.

\underline{Step 1:} Applying the construction of the storage separation theorem (cf. Theorem \ref{th:sep-storage}(i)$\Rightarrow$(ii)), we obtain the following objects:

\begin{compactenum}
\item the ranked alphabet $\Delta_\G$ corresponding to $\G$ with 
$P'=\{\true\}$, $F'=\{\inc,\dec\}$, and 
\begin{align*}
\Delta_\G= \ \ &\{(\true,\varepsilon)^{(0)}\} \cup \{(\true,f)^{(1)} \mid f \in F'\} \cup \{(\true,f_1f_2)^{(2)} \mid f_1,f_2 \in F'\} \ \text{ and }
\end{align*}

\item the chain-free $(\langle \Delta_{\G},\Sigma\rangle,K_{\max})$-rtg $\G'=(N,U,R',\wt')$  where $R'$ consists of rules derived from $R$ by using enriched terminal symbols instead of storage predicates and instructions, e.g., the rule
\[r_3:H(\true)\to \sigma(H(\inc),H(\inc))\] of $\G$ is transformed into the rule 
\[r_3':H\to \langle(\true,\inc\inc),\sigma\rangle(H,H)\] 
of $\G'$ with $\wt'(r_3')=\wt(r_3)$. It is obvious how the other rules of $R'$ and $\wt'$  look like. 
\end{compactenum}
It is easy to see that $\supp(\sem{\G'}) \subseteq T_\Psi$ where $\Psi \subseteq \langle \Delta_\G,\Sigma\rangle$  is the ranked alphabet 
\[
\Psi = \{\langle(\true,\varepsilon),\#\rangle^{(0)} , \
\langle(\true,\inc\inc),\sigma\rangle^{(2)}, \ 
\langle(\true,\dec\dec),\delta\rangle^{(2)}
\}\enspace.
\]

By Theorem \ref{th:sep-storage}(i)$\Rightarrow$(ii) and the fact that  $\supp(\sem{\G'}) \subseteq T_\Psi$, we have
\[
\sem{\G}(\xi) = \max( \sem{\G'}(\zeta) \mid \zeta \in \mathcal{B}_{\Delta_\G}(\xi))
=  \max( \sem{\G'}(\zeta) \mid \zeta \in \mathcal{B}_{\Delta_\G}(\xi) \cap T_\Psi)\enspace.
\]
We distinguish two cases. First, let $\mathcal{B}_{\Delta_\G}(\xi) \cap T_\Psi = \emptyset$. Then $\sem{\G}(\xi) = \max( \sem{\G'}(\zeta) \mid \zeta \in \emptyset) = 0$. Moreover, there is no behaviour $b \in \mathcal{B}(\Delta_\mathcal{G})$ such that $\pos(b) = \pos(\xi)$ and $b$ increments at $\sigma$-labeled positions of $\xi$ and decrements at $\delta$-labeled positions of $\xi$. In other words, $\xi$ does not have the path property. Hence $s(\xi)=0$ and thus $\sem{\G}(\xi)=s(\xi)$.

Now let $\mathcal{B}_{\Delta_\G}(\xi)  \cap T_\Psi \not= \emptyset$. For each $\zeta \in \mathcal{B}_{\Delta_\G}(\xi) \cap T_\Psi$, we have that $\pos(\xi) = \pos(\zeta)$, because $\G$ does not have chain rules. Moreover,    
due to the  definition of the mapping $\mathcal{B}_{\Delta_\G}$, the set  $\mathcal{B}_{\Delta_\G}(\xi) \cap T_\Psi$ has one element, denoted by, say $\zeta_\xi$, that is defined for each $w \in \pos(\xi)$ by 
\[
\zeta_\xi(w) =
\left\{
\begin{array}{ll}
\langle(\true,\inc\inc),\sigma\rangle & \text{if } \xi(w)=\sigma\\
\langle(\true,\dec\dec),\delta\rangle & \text{if } \xi(w)=\delta\\
\langle(\true,\varepsilon),\#\rangle & \text{if } \xi(w)=\# \enspace.
\end{array}
\right.
\]
Figure \ref{fig6} shows $\zeta_\xi$ for the tree $\xi = \sigma(\delta(\#,\sigma(\#,\#)),\#)$.
Thus, 
\[
\sem{\G}(\xi) = \sem{\G'}(\zeta_\xi) \enspace.
\]

\begin{figure}[t]
	\scriptsize\centering
\begin{tikzpicture}[sibling distance=16em, level distance=3.5em,
  every node/.style = {align=center}]]

\pgfdeclarelayer{bg}    % declare background layer
  \pgfsetlayers{bg,main}  % set the order of the layers (main is the standard layer)
 
 \node {$\langle(\true,\inc\inc),\sigma\rangle$}
    child { node {$\langle(\true,\dec\dec),\delta\rangle$}
      child { node {$\langle(\true,\varepsilon),\#\rangle$} } 
      child{ node {$\langle(\true,\inc\inc),\sigma\rangle$} child{ node {$\langle(\true,\varepsilon),\#\rangle$} }child{ node {$\langle(\true,\varepsilon),\#\rangle$} }}}
     child { node {$\langle(\true,\varepsilon),\#\rangle$} } ;
\end{tikzpicture}
\caption{\label{fig6} The tree $\zeta_\xi \in \mathcal{B}_{\Delta_\G}(\xi) \cap T_\Psi$  for the tree $\xi = \sigma(\delta(\#,\sigma(\#,\#)),\#)$.}
\end{figure}
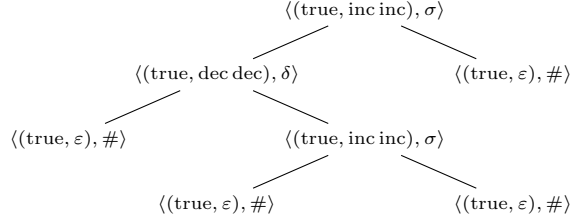

Since $\pr_1(\zeta_\xi)$ is a $\Delta_\G$-behaviour, we can consider the family $(c_w \mid w \in \pos(\pr_1(\zeta_\xi)))$ of configurations determined by $\pr_1(\zeta_\xi)$ and $0$ (the initial configuration of $\COUNT$). Then, by definition, $c_w \ge 0$ for each $w \in \pos(\pr_1(\zeta_\xi))$. Moreover, it is easy to see that $c_w$ is the difference between the number of occurrences of $\sigma$ above $w$ or at $w$ and the number of occurrences of $\delta$ above $w$ or at $w$. Hence, $\xi$ has the path property.

\underline{Step 2:} It remains to prove that, if $\xi$ has the path property, then $\sem{\G}(\xi) = \mathrm{ubal}(\xi)$.  So let $\xi \in T_\Sigma$ and $\xi$ have the path property.
 Applying the construction of the weight separation theorem to $\G'$ (cf. Theorem \ref{sep-weights}(i)$\Rightarrow$(ii)), we obtain the following objects:
\begin{compactenum}
\item the ranked alphabet $\Theta$, i.e., $\Theta=\{r_1'^{(0)},r_2'^{(0)}\} \cup\{r_i'^{(2)}\mid 3\leq i\leq 10\}$, 

\item the unambiguous and chain-free $\Theta$-rtg $\H=(N,U,R'')$ and the alphabetic mapping $h: T_\Theta \rightarrow K[T_\Sigma]$ where $R''$ contains the following rules; for each rule $r_i''$ we have indicated the pair $h_j(r_i'')$ (for appropriate rank $j\in\{0,2\}$) after the comma. 

\

\hspace*{-5mm}{%\small
$\begin{array}{lll}
	r_1'':H\to r_1' , \  (1,\langle (\true,\varepsilon),\#\rangle)  & 
        r_2'':U\to r_2' , \  (1,\langle (\true,\varepsilon),\#\rangle)\\
        r_3'':H\to r_3'(H,H)  , \ (\mathrm{ht},\langle(\true,\inc\inc),\sigma\rangle)  &
	r_4'':U\to r_4'(H,H)  , \  (\mathrm{diff},\langle(\true,\inc\inc),\sigma\rangle) \\
	r_5'':U\to r_5'(H,U)  , \ (\pr_2,\langle(\true,\inc\inc),\sigma\rangle)  & 
	r_6'':U\to r_6'(U,H)  , \ (\pr_1,\langle(\true,\inc\inc),\sigma\rangle) \\
        r_7'':H\to r_7'(H,H)  , \ (\mathrm{ht},\langle(\true,\dec\dec),\delta\rangle)  &
	r_8'':U\to r_8'(H,H)  , \ (\mathrm{diff},\langle(\true,\dec\dec),\delta\rangle) \\
	r_9'':U\to r_9'(H,U)  , \  (\pr_2,\langle(\true,\dec\dec),\delta\rangle)  & 
	r_{10}'':U\to r_{10}'(U,H)  , \  (\pr_1,\langle(\true,\dec\dec),\delta\rangle).
	\end{array}
$}
\end{compactenum}

\

 According to Theorem \ref{sep-weights}(i)$\Rightarrow$(ii),  we have that
\[
\seml \G'\semr(\zeta_\xi) = \max(h(d)(\zeta_\xi) \mid d \in \mathcal{L}(\mathcal{H}))\enspace.
\] 
 
Let $d \in  \mathcal{L}(\mathcal{H})$. As shown in the proof of Theorem \ref{sep-weights}(i)$\Rightarrow$(ii), the tree $d$ is a derivation tree of $\mathcal{G}'$ for $\pi(d)$ (where $\pi$ is the mapping defined on page \pageref{page:pi}). If $\pi(d) \not= \zeta_\xi$, then $h(d)(\zeta_\xi) = 0$ by definition of $h$. Thus we have 
\[
\seml \G'\semr(\zeta_\xi) = \max(h(d)(\zeta_\xi) \mid d \in \mathcal{L}(\mathcal{H}) \cap \pi^{-1}(\zeta_\xi))\enspace.
\] 
We note that $\pos(d) = \pos(\zeta_\xi) = \pos(\xi)$ for each $d \in \mathcal{L}(\mathcal{H}) \cap \pi^{-1}(\zeta_\xi)$.

Due to the behaviour of the nonterminals in $\mathcal{H}$, each tree in $\mathcal{L}(\mathcal{H}) \cap \pi^{-1}(\zeta_\xi)$ has a particular shape. More precisely,
\[
\mathcal{L}(\mathcal{H}) \cap \pi^{-1}(\zeta_\xi) = \{d_w \mid w \in \pos(\zeta_\xi)\}
\]
where, for each $w \in \pos(\zeta_\xi)$, we define the tree $d_w \in T_\Theta$  such that $\pos(d_w)=\pos(\zeta_\xi)$ and, for each $v \in \pos(d_w)$, we have
\[
d_w(v) = 
\left\{
\begin{array}{ll}
U \rightarrow (\zeta_\xi(v))\Big(U,H\Big)  & \text{ if } (\exists u \in \mathbb{N}^*): w = v1u\\
U \rightarrow (\zeta_\xi(v))\Big(H,U\Big)  & \text{ if } (\exists u \in \mathbb{N}^*): w = v2u\\
U \rightarrow (\zeta_\xi(v))\Big(H,H\Big)   & \text{ if } v=w \text{ and $v$ is not a leaf}\\
U \rightarrow \zeta_\xi(v)   & \text{ if } v=w \text{ and $v$ is a leaf}\\
H \rightarrow (\zeta_\xi(v))\Big(H,H\Big)  & \text{ if $v$ is not a leaf and ($v \not\preceq w$ or $\exists u \in \mathbb{N}^+: v=wu$) }\\
H \rightarrow \zeta_\xi(v)  & \text{ if $v$ is a leaf and ($v \not\preceq w$ or $\exists u \in \mathbb{N}^+: v=wu$) }\enspace.
\end{array}
\right.
\]
Figure \ref{fig7} shows the tree $d_w \in \mathcal{L}(\mathcal{H}) \cap \pi^{-1}(\zeta_\xi)$ for $w = 1$ and $\zeta_\xi$ as shown in Figure \ref{fig6}.

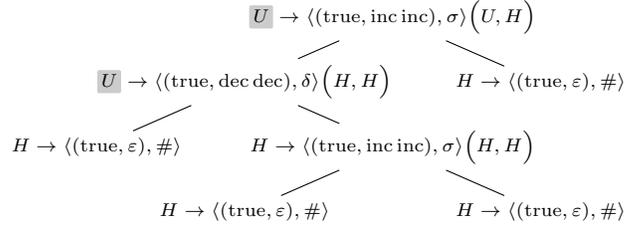
\begin{figure}[t]
	\scriptsize\centering
\begin{tikzpicture}[sibling distance=16em, level distance=3.5em,
  every node/.style = {align=center}]]

\pgfdeclarelayer{bg}    % declare background layer
  \pgfsetlayers{bg,main}  % set the order of the layers (main is the standard layer)
 
 \node {$\mhl{U}\to \langle(\true,\inc\inc),\sigma\rangle\Big(U,H\Big)$}
    child { node {$\mhl{U}\to \langle(\true,\dec\dec),\delta\rangle\Big(H,H\Big)$}
      child { node {$H\to \langle (\true,\varepsilon),\#\rangle$} } 
      child{ node {$H\to \langle(\true,\inc\inc),\sigma\rangle\Big(H,H\Big)$} child{ node {$H\to \langle (\true,\varepsilon),\#\rangle$} }child{ node {$H\to \langle (\true,\varepsilon),\#\rangle$} }}}
     child { node {$H\to \langle (\true,\varepsilon),\#\rangle$} } ;
\end{tikzpicture}
\caption{\label{fig7} The tree $d_w \in T_\Phi$ for $w = 1$ and $\zeta_\xi$ of Figure \ref{fig6}; $w$ is indicated by the grey shaded nonterminals.}
\end{figure}

Hence 
\[
\seml \G'\semr(\zeta_\xi) = \max(h(d_w)(\zeta_\xi) \mid w \in \pos(\zeta_\xi))\enspace.
\]

Due to the definition of $h$ it is also obvious that, for each $w \in \pos(\zeta_\xi)$,
\begin{align*}
h(d_w|_{w1})(\zeta_\xi|_{w1}) &= \mathrm{height}(\zeta_\xi|_{w1}), \\
h(d_w|_{w2})(\zeta_\xi|_{w2}) &= \mathrm{height}(\zeta_\xi|_{w2}), \\
h(d_w|_{w})(\zeta_\xi|_{w}) &= \mathrm{diff}(\mathrm{height}(\zeta_\xi|_{w1}),\mathrm{height}(\zeta_\xi|_{w2})) =  \mathrm{ubal}(\zeta_\xi,w), \\
h(d_w)(\zeta_\xi) &=  \mathrm{ubal}(\zeta_\xi,w)
\end{align*}
where we have used here $\mathrm{ubal}$ for the ranked alphabet $\Psi$.
\begin{align*}
 \seml \G'\semr(\zeta_\xi) 
= \max(\mathrm{ubal}(\zeta_\xi,w) \mid w \in \pos(\zeta_\xi)) = \mathrm{ubal}(\zeta_\xi)\enspace.
\end{align*}

Finally, we have $\sem{\G}(\xi) = \sem{\G'}(\zeta_\xi) =  \mathrm{ubal}(\zeta_\xi) = \mathrm{ubal}(\xi) = s(\xi)$, where the last but one equation holds because $\pos(\zeta_\xi) = \pos(\xi)$. 
\end{example}

%%%%%%%%%%%%%%%%%%%%%%%%%%%%%%%%%%%%%%%%%%%%%

\section{Elimination of chain rules}\label{elimchain}

It is well known that, for each $(\Sigma,M_K)$-rtg where $K$ is a complete semiring, an equivalent chain-free $(\Sigma,K)$-rtg can be constructed \cite{esikui03,fulmalvog09} by solving the corresponding classical algebraic path problem \cite{rot85}.
This is not true if we consider non-trivial storage types as the following theorem shows.

\begin{theorem}\label{lm:not-gen}  $\Reg(\PD-op^1,\Sigma) \setminus \bigcup_S \Reg_{\text{nc}}(S,\Sigma) \not= \emptyset$ where $S$ ranges over the set of all storage types.
\end{theorem}
\begin{proof} Let $\Sigma=\{\alpha\ui 0,\delta\ui 1,\sigma\ui 2\}$. The tree language
\[
L=\{ \sigma(\delta^n(\alpha),\delta^n(\alpha))\mid n\geq 0\}\enspace
\]
is in $\Reg(\PD-op^1,\Sigma)$, because it can be generated by the $(\PD-op^1,\Sigma)$-grammar which we obtain from $\G$ of Example \ref{ex:run} by dropping its weight structure and weight function (cf. \cite{gue83}). 

On the other hand, we can show by contradiction that
$L\not\in \Reg_{\text{nc}}(S,\Sigma)$ for any storage type $S$.
For this, we assume that there is a storage type $S$ and a chain-free $(S,\Sigma)$-rtg $\G = (N,Z,R)$ such that $\mathcal L(\G)=L$. By Lemma~\ref{lm:one-initial} we can assume that $Z$ is a single nonterminal. There are finitely many $\sigma$-rules in $R$, i.e., rules of the form
\begin{flalign*}
r=(Z(p) \rightarrow \sigma(A(f),B(g)))\enspace.\tag{1}
\end{flalign*}
Since $L$ is infinite, there is a $\sigma$-rule which is the root of some $d\in D_{\G}(\sigma(\delta^n(\alpha),\delta^n(\alpha)))$ for infinitely many $n$'s. Let (1) be such a rule.
Then  $p(c_0)=\true$ and $f(c_0)$ and 
$g(c_0)$ are defined
 (where $c_0$ is the initial configuration of $S$), and there are
\begin{compactitem}
\item integers $n,m\in\Nat$ with $n\ne m$,
\item $(A,f(c_0))$-derivation trees $d_n,d_m$ for $\delta^n(\alpha),\delta^m(\alpha)$, respectively,
\item $(B,g(c_0))$-derivation trees $d'_n,d'_m$ for $\delta^n(\alpha),\delta^m(\alpha)$, respectively,
\end{compactitem}
such that $r(d_n,d'_n)\in D_{\G}(\sigma(\delta^n(\alpha),\delta^n(\alpha)))$ and 
$r(d_m,d'_m)\in D_{\G}(\sigma(\delta^m(\alpha),\delta^m(\alpha)))$. But then 
$r(d_n,d'_m)\in D_{\G}(\sigma(\delta^n(\alpha),\delta^m(\alpha)))$ and hence $\sigma(\delta^n(\alpha),\delta^m(\alpha)) \in L$ with $m \not= n$. This is a contradiction to 
$L=\{ \sigma(\delta^n(\alpha),\delta^n(\alpha))\mid n\geq 0\}$. 
\end{proof}

Thus, in general, chain rules cannot be eliminated from $(S,\Sigma,K)$-rtg (even if  $K = \mathbb{B}$). But we can eliminate chain rules for particular $(S,\Sigma,K)$-rtg, which we will call 'simple'. There is another problem which concerns the weight algebra. For $(\Sigma,M_K)$-rtg where $K$ is a complete semiring, the eliminiation typically uses elements $a^* \in K$ (for some $a \in K$) to capture the weight of cycles of chain rules. Here $a^*$ is the sum of all powers $a^n$ of $a$ and the powers are defined by the multiplication of the semiring. In our setting we deal with M-monoids and, instead of the binary multiplication, we have operations with different arities. Thus, we will have to guarantee that the M-monoid is closed under iterated composition of operations. Let us now formalize these requirements on the $(S,\Sigma,K)$-rtg and on the M-monoid.

We call an $(S,\Sigma,K)$-rtg \emph{simple}  if for each chain rule $A(p) \rightarrow B(f)$ we have $p=\true_C$ and $f =\id_C$. Let $\Reg_{\mathrm{simple}}(S,\Sigma,K)$ denote the class of all weighted tree languages generated by simple $(S,\Sigma,K)$-rtgs.

Now we introduce the necessary closure properties of $K$.
An operation $\omega\in\Ops\ui k(K)$ for $k\geq 1$ is \emph{completely distributive} if 
\[\omega(\sum_{i_1\in I_1}a_{i_1},\ldots,\sum_{i_k\in I_k}a_{i_k})=\sum_{i_1\in I_1}\ldots\sum_{i_k\in I_k}\omega(a_{i_1},\ldots,a_{i_k})\] for all countable index sets $I_j$ and family $(a_{i_j}\in K\mid i_j\in I_j)$, $j\in [k]$. We say that \emph{$K$ is completely distributive} if each operation in $\Omega$ is completely distributive. (This concept was introduced as complete DM-monoid in \cite{kui98new}, cf. also \cite{esikui03}.)

Let $k\geq 0$, $I$ be a countable index set,  and $(\omega_i\in \Ops\ui k(K) \mid i\in I)$ a family of operations. We define the operation $\sum_{i\in I}\omega_i\in \Ops\ui k(K)$ by letting
\[\big(\sum_{i\in I}\omega_i\big)(a_1,\ldots,a_k)=\sum_{i\in I}\omega_i(a_1,\ldots,a_k)\]
for every $a_1,\ldots,a_k\in K$. We say that \emph{$K$ is completely 1-sum closed} if $\sum_{i\in I}\omega_i\in \Omega\ui 1$ for every 
countable index set $I$  and family $(\omega_i\in \Omega\ui 1 \mid i\in I)$.

Let $\omega\in \Ops\ui 1(K)$ and $\omega'\in \Ops\ui k(K)$ for some $k\geq 0$. The \emph{composition of $\omega$ and $\omega'$}
is the operation $\omega\circ \omega'\in \Ops\ui k$ defined by $(\omega\circ \omega')(a_1,\ldots,a_k)=\omega(\omega'(a_1,\ldots,a_k))$
for every $a_1,\ldots,a_k\in K$. We say that $K$ is \emph{$(1,k)$-composition closed} if $\omega\circ \omega'\in \Omega\ui k$ for every 
$\omega\in \Omega\ui 1$ and $\omega'\in \Omega\ui k$. Moreover, $K$ is \emph{$(1,*)$-composition closed}  if it is $(1,k)$-composition closed for every $k\geq 0$ (cf. \cite[Def. 4.3]{fulmalvog09}).

The following statement follows easily from the corresponding definitions.

\begin{observation}\rm\label{obs:operations}
For every countable index set $I$, family $(\omega_i\in \Ops\ui 1(K)\mid i\in I)$, $k\geq 0$, and 
$\omega\in \Ops\ui k(K)$, we have  $\sum_{i\in I}(\omega_i\circ \omega)=\big(\sum_{i\in I}\omega_i\big)\circ \omega$.
\end{observation}

Let $N$ be a finite set. Moreover, let $V$ and $W$ be $(N\times N)$-matrices over $\Ops\ui 1(K)$. We define the \emph{product $V\cdot W$ of $V$ and $W$} by $(V\cdot W)_{A,B}=\sum_{C\in N}V_{A,C}\circ W_{C,B}$ for every $A,B\in N$. Note that, although the set $N$ is not ordered, the expression $\sum_{C\in N}V_{A,C}\circ W_{C,B}$ is well defined, because the monoid $(\Ops\ui 1(K),+,0)$ is commutative. Moreover, for every $n\geq 0$ we define the $(N\times N)$-matrix $W^n$ over $\Ops\ui 1(K)$ by induction as follows: let  $W^0=E$ and $W^{n}=W\cdot W^{n-1}$ for every $n\geq 1$, where $E$ is the \emph{unit matrix over $\Ops\ui 1(K)$} defined by $E_{A,B}=\id_K$ if $A=B$ and $0_1$ otherwise for every $A,B\in N$. Finally, we define $W^*=\sum_{n\geq 0}W^n$, where $\big(\sum_{n\geq 0}W^n\big)_{A,B}=\sum_{n\geq 0}W^n_{A,B}$ for every $A,B\in N$.

Finally, we say that \emph{$K$ has identity} if $\id_K \in \Omega\ui 1$.

We call $K$ \emph{compressible} if it has identity, it is $(1,*)$-composition closed, completely 1-sum closed, and completely distributive.
Each M-monoid associated with a complete semiring is compressible.  The fact that such an M-monoid is completely 1-sum closed and completely distributive can be derived from the generalized distributivity law of the complete semiring. In particular, the Boolean M-monoid is compressible. 

First we show that if $K$ is compressible and $S$ contains the always true predicate and the identity instruction, then chain rules can be eliminated from simple $(S,\Sigma,K)$-rtgs.

Similar results for the elimination of chain rules (or $\varepsilon$-transitions) in the weighted case have been proved in \cite[Thm.~3.2]{esikui03} and  \cite[Lm.~3.2]{fulmalvog11}. In fact, in \cite[Thm.~3.2]{esikui03} it was shown that $\varepsilon$-transitions can be eliminated from weighted tree automata over commutative and continuous semirings. The same was shown for weighted tree automata over commutative and complete semirings in \cite[Lm.~3.2]{fulmalvog11}. The second result generalizes the first because any continuous semiring is complete \cite[Prop. 2.2]{esikui03}. The next lemma generalizes further \cite[Lm.~3.2]{fulmalvog11} because simple $(S,\Sigma,K)$-rtgs, where $K$ is a compressible M-monoid, generalize weighted tree automata over commutative and complete semirings.

\begin{theorem}\label{lm:reg=nc-reg}  If $S_{\true,\id}=S$ and $K$ is compressible, then $\Reg_{\mathrm{simple}}(S,\Sigma,K) = \Regnc(S,\Sigma,K)$.
\end{theorem}
\begin{proof} \sloppy Since each  chain-free $(S,\Sigma,K)$-rtg is simple, we only have to prove $\Reg_{\mathrm{simple}}(S,\Sigma,K) \subseteq \Regnc(S,\Sigma,K)$. Let $\G = (N,Z,R,\wt)$ be a simple $(S,\Sigma,K)$-rtg. Recall that $P_\G$ and $F_\G$ are the finite sets of predicates and instructions, respectively, which occur in $\G$.  Without loss of generality we can assume that for each $k \in \mathbb{N}$, $\sigma \in \Sigma^{(k)}$, $A, B_1,\ldots,B_k \in N$, $p \in P_\G$, and $f_1,\ldots,f_k \in F_\G$,  
there is a rule $r = (A(p) \rightarrow \sigma(B_1(f_1),\ldots,B_k(f_k)))$ in $R$. If there is no such rule, then we can add it to $R$ and let $\wt(r) = 0_k$.
In a similar way, we can assume that for each $A, A' \in N$ 
there is a rule $r = (A(\true_C) \rightarrow A'(\id_C))$ in $R$. 

  We let $W$ be the $(N \times N)$-matrix over  $\Omega^{(1)}$ such that
\[
W_{A,A'} = \wt(A(\true_C) \rightarrow A'(\id_C))
\]
 for each $A,A' \in N$. 

We construct the chain-free $(S,\Sigma,K)$-rtg $\G' = (N,Z,R',\wt')$ as follows.
 For each $k \in \mathbb{N}$, $\sigma \in \Sigma^{(k)}$, $A, B_1,\ldots,B_k \in N$, $p \in P_\G$, and $f_1,\ldots,f_k \in F_\G$ there is a rule $r' = (A(p) \rightarrow \sigma(B_1(f_1),\ldots,B_k(f_k)))$ in $R'$ and 
\[
\wt'(r') = \sum_{A'\in N} \Big( (W^*)_{A,A'} \circ \wt(A'(p) \rightarrow \sigma(B_1(f_1),\ldots,B_k(f_k)))\Big).
\]
Since $K$ has identity, it is (1,1)-composition closed, and completely 1-sum closed, each entry of the matrix $W^*$ is in $\Omega^{(1)}$.
Moreover, since $K$ is  $(1,k)$-composition closed, the right-hand side of the above equality is an operation in $\Omega^{(k)}$.
Hence, $\wt'(r')$ is well defined.

We define the family $\textrm{eff} = (\textrm{eff}_{A,\xi,c} \mid \xi \in T_\Sigma, A\in N, c \in C)$ of mappings 
\[\textrm{eff}_{A,\xi,c}: D_{\G}(A,\xi,c) \rightarrow D_{\G'}(A,\xi,c)\]
 as follows. Let $\xi=\sigma(\xi_1,\ldots,\xi_k)$ and $d\in D_{\G}(A,\xi,c)$. Then 
\begin{compactitem}
\item there are $n\ge 0$ and rules $r_1 = (A_1(\true_C)\to A_2(\id_C))$, \ldots, $r_n=(A_n(\true_C)\to A_{n+1}(\id_C))$, and $A_{n+1}(p) \rightarrow \sigma(B_1(f_1),\ldots,B_k(f_k))$
in $R$ such that  $p(c)=\true$ and 
\item for each $i\in[k]$ the function $f_i$ is defined on $c$ and  there is a derivation tree $d_i\in D_{\G}(B_i,\xi_i,f_i(c))$
\end{compactitem}
 such that
\[
d=r_1 \ldots r_n (A_{n+1}(p) \rightarrow \sigma(B_1(f_1),\ldots,B_k(f_k)))\Big(d_1,\ldots,d_k\Big)\enspace.
\]
 We define
\[\textrm{eff}_{A,\xi,c}(d)=(A_{1}(p) \rightarrow \sigma(B_1(f_1),\ldots,B_k(f_k)))(d'_1,\ldots,d'_k),\] where $d'_i=\textrm{eff}_{B_i,\xi_i,f_i(c)}(d_i)$ for each $i\in[k]$. Note that $\lhs_N(d'_i(\varepsilon))=B_i$.

Then we can prove:
\begin{align*}
\seml \G \semr(\xi) 
&= \sum_{d \in D_{\G}(\xi)} \wt(d) 
= \sum_{A\in Z} \sum_{d \in D_{\G}(A,\xi,c_0)} \wt(d) \\[3mm]
&= \sum_{A \in Z} \ \sum_{d' \in D_{\G'}(A,\xi,c_0)} \  \sum_{\substack{d \in D_{\G}(A,\xi,c_0):\\\textrm{eff}_{A,\xi,c_0}(d)=d'}} \wt(d)
\\[3mm]
&=^{(*)} \sum_{A \in Z} \ \sum_{d' \in D_{\G'}(A,\xi,c_0)} \wt'(d')
= \sum_{d' \in D_{\G'}(\xi)} \wt'(d') = \seml \G' \semr(\xi)\enspace. 
\end{align*}

At $(*)$ we have used the following statement:
for each $\xi \in T_\Sigma$, $A\in N$, $c \in C$, and  $d' \in D_{\G'}(A,\xi,c)$: 
\[\sum_{\substack{d \in D_{\G}(A,\xi,c):\\\textrm{eff}_{A,\xi,c}(d)=d'}} \wt(d) = \wt'(d')\enspace.\]
We prove this statement by induction on $\xi$. We only show the induction step because it contains the base of the induction.

Let $\xi = \sigma(\xi_1,\ldots,\xi_k)$ for some $k\in \Nat$. Let $A \in N$, $c \in C$, and $d' \in D_{\G'}(A,\xi,c)$. Then
\begin{compactitem}
\item there is a rule $r' = (A(p) \rightarrow \sigma(B_1(f_1),\ldots,B_k(f_k)))$ in $R'$ such that $p(c)=\true$ and $f_i(c)$ is defined  and
 \item  for each $i \in [k]$ there is a $d_i' \in D_{\G'}(B_i,\xi_i,f_i(c))$ 
\end{compactitem}
such that $d' = r'(d_1',\ldots,d_k')$. Then we can calculate as follows (by abbreviating $\wt(B(\true_C) \rightarrow B'(\id_C))$ by $\wt(B,B')$ and $\wt(B(p) \rightarrow \sigma(B_1(f_1),\ldots,B_k(f_k)))$ by $\wt(B(p),\sigma(B_1(f_1)\ldots,B_k(f_k)))$; moreover, we use $\bigcirc_{j=1}^n \omega_j$ to denote the composition $\omega_1 \circ \ldots \circ \omega_n$ of $n$ unary operations):
\begin{align*}
  & \sum_{\substack{d \in D_{\G}(A,\xi,c):\\\textrm{eff}_{A,\xi,c}(d)=r'(d_1',\ldots,d_k')}} \wt(d)\\[3mm]
&= \sum_{n \in \mathbb{N}} \sum_{\substack{A_1,\ldots,A_{n+1} \in N: \\ A_1=A }}  \
\sum_{\substack{d_1 \in D_{\G}(B_1,\xi_1,f_1(c)),\ldots,d_k \in D_{\G}(B_k,\xi_k,f_k(c)):\\ \forall i \in [k]: \textrm{eff}_{B_i,\xi_i,f_i(c)}(d_i)=d_i' }} \\%[-2mm]
&\hspace*{10mm}\bigcirc_{j=1}^{n} \wt(A_j,A_{j+1}) \circ \wt(A_{n+1}(p), \sigma(B_1(f_1),\ldots,B_k(f_k))) \Big(\wt(d_1),\ldots,\wt(d_k)\Big)\\[5mm]
&= \sum_{n \in \mathbb{N}} \sum_{\substack{A_1,\ldots,A_{n+1} \in N: \\ A_1=A }} \
 \bigcirc_{j=1}^{n} \wt(A_j,A_{j+1}) \circ \wt(A_{n+1}(p), \sigma(B_1(f_1),\ldots,B_k(f_k)))\\[-5mm]
&\hspace*{40mm} \Big(\sum_{\substack{d_1 \in D_{\G}(B_1,\xi_1,f_1(c)):\\ \textrm{eff}_{B_1,\xi_1,f_1(c)}(d_1)=d_1' }}\wt(d_1),\ldots,\sum_{\substack{d_k \in D_{\G}(B_k,\xi_k,f_k(c)):\\ \textrm{eff}_{B_k,\xi_k,f_k(c)}(d_k)=d_k' }}\wt(d_k)\Big)\\
& \hspace*{7mm}\textrm{(because $K$ is completely distributive)}\\[4mm]
&= \sum_{n \in \mathbb{N}} \sum_{\substack{A_1,\ldots,A_{n+1} \in N: \\ A_1=A }} \
 \bigcirc_{j=1}^{n} \wt(A_j,A_{j+1}) \circ \wt(A_{n+1}(p), \sigma(B_1(f_1),\ldots,B_k(f_k)))\\[-5mm]
&\hspace*{80mm} \Big(\wt'(d_1'),\ldots,\wt'(d_k')\Big)\\
& \hspace*{7mm}\textrm{(due to I.H.)}\\[5mm]
&= \sum_{A'\in N}  \sum_{n \in \mathbb{N}} \sum_{\substack{A_1,\ldots,A_{n+1} \in N: \\ A_1=A, A_{n+1}=A' }} \
 \bigcirc_{j=1}^{n} \wt(A_j,A_{j+1}) \circ \wt(A'(p), \sigma(B_1(f_1),\ldots,B_k(f_k)))\\[-5mm]
&\hspace*{80mm}  \Big(\wt'(d_1'),\ldots,\wt'(d_k')\Big)\\
& \hspace*{7mm}\text{(by renaming of $A_{n+1}$ by $A'$)}\\[5mm]
&= \sum_{A'\in N}  \bigg( \Big(\sum_{n \in \mathbb{N}} \sum_{\substack{A_1,\ldots,A_{n+1} \in N: \\ A_1=A, A_{n+1}=A' }} \
 \bigcirc_{j=1}^{n} \wt(A_j,A_{j+1})\Big) \circ \wt(A'(p), \sigma(B_1(f_1),\ldots,B_k(f_k)))\bigg)\\[-5mm]
&\hspace*{80mm}  \Big(\wt'(d_1'),\ldots,\wt'(d_k')\Big)\\
&\hspace*{7mm}\text{(by Observation \ref{obs:operations})}\\[5mm]
&= \sum_{A'\in N}  \bigg(\Big(\sum_{n \in \mathbb{N}} (W^n)_{A,A'}\Big) \circ \wt(A'(p), \sigma(B_1(f_1),\ldots,B_k(f_k)))\bigg)  \Big(\wt'(d_1'),\ldots,\wt'(d_k')\Big)\\
&\hspace*{7mm}\text{(by definition of $W^n$)}\\[5mm]
&= \sum_{A'\in N}  \bigg((W^*)_{A,A'} \circ \wt(A'(p), \sigma(B_1(f_1),\ldots,B_k(f_k)))\bigg)  \Big(\wt'(d_1'),\ldots,\wt'(d_k')\Big)\\[5mm]
&= \bigg(\sum_{A'\in N}  (W^*)_{A,A'} \circ \wt(A'(p), \sigma(B_1(f_1),\ldots,B_k(f_k)))\bigg)  \Big(\wt'(d_1'),\ldots,\wt'(d_k')\Big)\\
&\hspace*{7mm}\text{(by Observation \ref{obs:operations})}\\[5mm]
&=  \wt'(r') \circ \Big(\wt'(d_1'),\ldots,\wt'(d_k')\Big)\\
&\hspace*{7mm}\text{(by construction of $\G'$)}\\[3mm]
&= \wt'(r'(d_1',\ldots,d_k')).
\end{align*}
\end{proof}

\pagebreak

We can instantiate the previous theorem to (1) the trivial storage type and (2) the Boolean M-monoid and obtain the following corollary.
\begin{corollary} \label{cor:chain-freeness-TRIV} $\ $
\begin{compactenum}
\item If $K$ is compressible, then $\Reg(\Sigma,K) = \Regnc(\Sigma,K)$.
\item If $S_{\true,\id}=S$, then $\Reg_{\mathrm{simple}}(S,\Sigma) = \Regnc(S,\Sigma)$. In particular, for each $n \in \mathbb{N}$ we have $\Reg_{\mathrm{simple}}(\PD-op^n,\Sigma) = \Regnc(\PD-op^n,\Sigma)$.
\end{compactenum}
\end{corollary}
\begin{proof}  First we prove Statement 1. Since $\TRIV_{\true,\id}=\TRIV$ and each $(\Sigma,K)$-rtg is simple, the statement follows from Theorem \ref{lm:reg=nc-reg}.

Then we prove Statement 2 as follows. Since the Boolean M-monoid $\mathbb{B}$ is compressible, the first statement follows from Theorem \ref{lm:reg=nc-reg}. Then the second  follows from the first one, because $(\PD-op^n)_{\true,\id} = \PD-op^n$. 
\end{proof}

For an arbitrary compressible M-monoid, we can even go beyond the trivial storage type and prove the following chain rule elimination result for particular finite storage types.

\begin{corollary}\label{cor:chain-freeness-finite}   If $K$ is compressible, $S$ is finite, and  $S_{\true,\id}=S$, then $\Reg(S,\Sigma,K) = \Reg_{\textrm{simple}}(S,\Sigma,K) = \Regnc(\Sigma,K)$.
\end{corollary}
\begin{proof} We have $\Reg(S,\Sigma,K) = \Reg(\Sigma,K) =  \Regnc(\Sigma,K)$ by 
Corollary \ref{th:finite-storage-new}
and Corollary \ref{cor:chain-freeness-TRIV}(1), respectively. The inclusion $\Regnc(\Sigma,K)  \subseteq  \Regnc(S,\Sigma,K)$ follows from 
Corollary \ref{lm:dropping-finite-storage}(2), 
and the inclusions $\Regnc(S,\Sigma,K) \subseteq \Reg_{\textrm{simple}}(S,\Sigma,K)$ and $\Reg_{\textrm{simple}}(S,\Sigma,K) \subseteq \Reg(S,\Sigma,K)$ are obvious. 
\end{proof}

%%%%%%%%%%%%%%%%%%%%%%%%%%%%%%%%%%%%%%%%%%%%

\section{B\"uchi-Elgot-Trakhtenbrot theorem}
\label{sec:BET}

By the classical results of B\"uchi \cite{buc60,buc62}, Elgot \cite{elg61}, and Trakhtenbrot \cite{tra61}, recognizable languages are the same as languages definable in monadic second order logic (MSO-logic). We call this characterization B\"uchi-Elgot-Trakhtenbrot theorem and in this section we present a corresponding one for  weighted tree languages generated by $(S,\Sigma,K)$-rtg (including chain rules). For this, we will introduce a weighted MSO-logic with storage behaviour, where the weights are taken from a complete M-monoid. Each formula, called expression,  of this logic is interpreted over finite, labeled, and ordered trees. Our new weighted MSO-logic generalizes (i)~the weighted MSO-logic with storage behaviour of \cite{vogdroher16} by considering trees as models, and (ii)~M-expressions of \cite{fulstuvog12,fulvog15} by adding storage type. Our B\"uchi-Elgot-Trakhtenbrot theorem states that  weighted tree languages generated by $(S,\Sigma,K)$-rtg  are the same as weighted tree languages definable by expressions. Thus, our  result generalizes the corresponding one of \cite{vogdroher16} in the sense that we allow chain rules on the rtg side (which corresponds to $\varepsilon$-transitions for the automata considered in \cite{vogdroher16}).

We note that in \cite[Ch.~7]{drovog11} an alternative weighted MSO-logic was used for the characterization of the class $\Regnc(\Sigma,M_K)$ where $K$ is an arbitrary semiring. In its turn, that logic is based on the weighted MSO-logic in \cite{drogas05,drogas07,drogas09} for weighted string automata. For a recent survey we refer to \cite{gasmon15}.

Since the formulas of our new weighted MSO-logic generalize M-expressions of~\cite{fulstuvog12}, we recall the syntax of M-expressions. For the definitions of all semantic notions, like variable assignment, correspondence between (i) pairs of a tree and a variable assignment and (ii)~trees over some extended ranked alphabet, and semantics of M-expressions we refer to~\cite{fulstuvog12}; here we only recall some of them.

First, we recall the (unweighted) MSO-logic for trees \cite{gecste97}.  For this, let $\Theta$ be an arbitrary ranked alphabet. We define the \emph{set of formulas of  MSO-logic over $\Theta$}, denoted by  $\MSO(\Theta)$, as the language generated by the following EBNF with nonterminals~$\psi$ and $\varphi$ and with initial nonterminal $\varphi$:
	\begin{align*}
                \psi &::= \myrm{label}_\sigma(x) \mid \myrm{edge}_i(x,y) \mid (x\in X)\\
		\varphi\,\,&::= \psi \mid \neg \varphi \mid (\varphi\vee\varphi) \mid (\varphi\wedge\varphi) \mid \exists x.\varphi \mid \forall x.\varphi \mid \exists X.\varphi \mid \forall X.\varphi
	\end{align*}
where $\sigma \in \Theta$ and $i \in [\maxrk_\Theta]$.  For any finite  set $\mathcal V$ of first-order or second-order variables we define the ranked alphabet $\Theta_{\mathcal V}$ by $\Theta_{\mathcal V}^{(k)} = \Theta^{(k)} \times {\mathcal P}({\mathcal V})$. We denote the set of all trees in $T_{\Theta_{\mathcal V}}$ in which each first-order variable of $\mathcal V$ occurs exactly once by $T_{\Theta_{\mathcal V}}^v$. It is well known that such trees can be identified with pairs of a tree and an assignment to variables in $\mathcal V$.
 For an MSO-formula $\varphi$ over $\Theta$ with free variables contained in $\mathcal V$, we let 
\[
\mathcal{L}_{\mathcal V}(\varphi)=\{\xi\in T_{\Theta_{\mathcal V}}^v\mid \varphi\models \xi\}\enspace,
\]
 the set of models of $\varphi$. The models operator $\models$ is defined straightforwardly as for (classical) MSO-formulas for the string case (cf. \cite[Ch.~II.2]{str94}).
	
Second, we recall from \cite[Def.~3.1]{fulstuvog12} the definition of M-expressions. Note that $K$ is some arbitrary complete \mmon{}.
 The atomic M-expressions have the form $\mathrm H(\omega)$ where $\omega=(\omega_\sigma\mid\sigma\in\Theta_{\mathcal U})$ is a $\Theta_{\mathcal U}$-family of operations for some  finite set $\mathcal U$ of  variables; moreover, we require that $\omega_\sigma\in\Omega^{(k)}$ for every $k\in\Nat$ and $\sigma\in\Theta_{\mathcal U}^{(k)}$.
 The set of \emph{M-expressions over $(\Theta,K)$}, denoted by $\MExp(\Theta,K)$, is the set of all formulas generated by the following EBNF with nonterminal~$E$:
\begin{align*}
		\textstyle E\,\, &::= \mathrm H(\omega)\mid (E+E) \mid (\varphi\rhd E) \mid \sum\nolimits_x E\mid \sum\nolimits_X E \enspace,
	\end{align*}
where $\omega$ is a $\Theta_{\mathcal U}$-family of operations in $\Omega$ for some finite set $\mathcal U$ of  variables, and $\varphi\in\MSO(\Theta)$. A sentence is an M-expression without free variables.

The semantics of $\mathrm H(\omega)$ with  $\omega=(\omega_\sigma\mid\sigma\in\Theta_{\mathcal U})$  is based on the following concept from universal algebra. Since $(K,\omega)$ is a $\Theta_{\mathcal U}$-algebra, there is a unique $\Theta_{\mathcal U}$-homomorphism from the $\Theta_{\mathcal U}$-term algebra $T_{\Theta_{\mathcal U}}$ to $(K,\omega)$ \cite[Thm.~4]{wec92}; we denote this homomorphism by $\mathrm h_\omega$. Then the semantics of $\mathrm H(\omega)$ on a tree $\xi$ is obtained by applying $\mathrm h_\omega$ to $\xi$ (after adaptation of $\omega$ to the set of free variables occurring in $\xi$, cf. \cite[p.~249]{fulstuvog12}).  

Intuitively, the semantics of $(\varphi\rhd e)$ on a tree $\xi$ is the semantics of $e$ on $\xi$ if $\xi$ is a model of $\varphi$, otherwise it is $0$. Formulas of the form $(e_1+e_2)$, $\sum\nolimits_x e$, and $\sum\nolimits_X e$ are interpreted by employing the summation operation of $K$ in the usual way (viewing  $\sum\nolimits_x e$ and $\sum\nolimits_X e$ as weighted first-order existential quantification and weighted second-order existential quantification, respectively). 

The semantics of a sentence $e \in \MExp(\Theta,K)$ is a weighted tree language $\seml e \semr: T_{\Theta} \rightarrow K$ defined in \cite[Def.~3.3]{fulstuvog12}.
% In particular, the semantics of $\mathrm H(\omega)$ uses the homomorphism $\mathrm h_\omega$.
 We say that a weighted tree language $s: T_\Theta \rightarrow K$ is \emph{M-definable} if there is a sentence $e$ such that  $\seml e \semr = s$.
 We denote by $\M(\Theta,K)$ the class of all weighted tree languages which are M-definable by some sentence $e \in \MExp(\Theta,K)$.

We recall the main theorem of \cite{fulstuvog12} (using Observation~\ref{ob:chain-free-Reg=Rec}).

\begin{theorem}\cite[Thm.~4.1]{fulstuvog12}\label{th:main-fulstuvog12} 
$\Regnc(\Theta,K) = \M(\Theta,K)$ for each ranked alphabet $\Theta$.
\end{theorem}

Now we define the main logic of this paper: (weighted) expressions with behaviour. We recall that $S = (C,P,F,c_0)$ denotes an arbitrary storage type and $\Sigma$ is a ranked alphabet. In the sequel, we let $P' \subseteq P$ and $F' \subseteq F$ be finite non-empty sets and we let $\Delta$ be the ranked alphabet corresponding to $\Sigma$, $P'$, and $F'$ (cf. Section~\ref{subsec:storage}).

 In a similar spirit as in \cite{vogdroher16}, an expression is an existentially quantified M-expression where the quantification runs over the set $\B_\Delta(\xi)$ of $\Delta$-behaviours on the tree $\xi\in T_\Sigma$ over which the expression is interpreted. The involved M-expression is over $(\DS,K)$.

\begin{definition}\rm\label{df:(Delta,Sigma,K)-expressions}
We define the set of \emph{expressions over $\Sigma$ and $K$ with $\Delta$-behaviours} (for short: $(\Delta,\Sigma,K)$-expressions) to be the set of all formulas of the form
\begin{align*}
\sum\nolimits^\textrm{beh}\nolimits e 
\end{align*}
where $e \in \MExp(\DS,K)$ is a sentence.  
		The \emph{semantics of $\sum\nolimits^{\mathrm{beh}}\nolimits e$} is the weighted tree language $\lBert \sum\nolimits^{\textrm{beh}}\nolimits e\rBert: T_{\Sigma}\rightarrow K$ defined by
\begin{align*}
\lBert \sum\nolimits^{\textrm{beh}}\nolimits e\rBert = \B_\Delta; \lBert e\rBert\; . \tag*{$\Box$}
\end{align*}
\end{definition}

Let $s: T_\Sigma \rightarrow K$ be a weighted tree language. We say that $s$  is \emph{$(\Delta,\Sigma,K)$-definable} if there is a $(\Delta,\Sigma,K)$-expression $\sum\nolimits^{\mathrm{beh}}\nolimits e$ with $\lBert \sum\nolimits^{\mathrm{beh}}\nolimits e \rBert = s$.
Moreover, $s$ is \emph{$(S,\Sigma,K)$-definable} if there are non-empty finite subsets $P' \subseteq P$ and $F' \subseteq F$ such that $s$ is $(\Delta,\Sigma,K)$-definable where $\Delta$ is the ranked alphabet corresponding to $\Sigma$, $P'$, and $F'$. We denote the \emph{class of all $(S,\Sigma,K)$-definable weighted tree languages} by $\Def(S,\Sigma,K)$.

Now we want to compare $(\Delta,\Sigma,K)$-expressions of Definition \ref{df:(Delta,Sigma,K)-expressions} with $(\Delta,\Sigma,K)$-expressions of \cite[Def.~5 and 6]{vogdroher16}. (Since in \cite{vogdroher16} $\Omega$ refers to a finite subset of $P \times F$, whereas in this paper $\Omega$ is used as set of operations of an M-monoid, we have replaced in the notations of \cite{vogdroher16} $\Omega$ by $\Delta$.) The following table gives an overview on  this comparison.

\

{%\small
\begin{tabular}{lll}
& \cite{vogdroher16} & present paper \\[1mm]\hline
structures: & string $u \in \Sigma^*$ & tree $\xi \in T_\Sigma$ \\[2mm]
behaviours: & $b \in \Delta^*$ & $\zeta \in \B_\Delta(\xi) \subseteq T_{\langle \Delta,\Sigma\rangle}$ \\
           &  with $\pos(b) = \pos(u)$ & and  $\pos(\xi)$ can be embedded\\
           &&into  $\pos(\zeta)$ (cf. Sect. \ref{sec:decomp})\\[2mm] 
weight algebra: & valuation monoid $(K,+,0,\mathrm{Val})$ & complete M-monoid $(K,+,0,\Omega)$\\[2mm]
expressions: & $\sum^{\textrm{beh}}\nolimits_B e$ &  $\sum\nolimits^{\textrm{beh}}\nolimits e$\\
& where $e$ is a B-expression & where $e$ is an M-expression \\
& over $(\Delta,\Sigma,K)$ &  of \cite{fulstuvog12} over $(\langle \Delta,\Sigma\rangle,K)$\\[3mm]
semantics: & $\lBert \sum^{\textrm{beh}}\nolimits_B e\rBert(u) =$ &
 $\lBert \sum\nolimits^{\textrm{beh}}\nolimits e\rBert(\xi) =$ \\ 
& \hfill $\sum\limits_{b \in \B(\Delta,|u|)} \lBert e\rBert_{\{B\}}\bigl(u, [B\mapsto b ]\bigr)$ & \hspace*{6mm} $\sum\limits_{\zeta \in \B_\Delta(\xi)} \lBert e\rBert(\zeta)$
\end{tabular}
}

\

There are two significant differences between the two approaches:

1) The behaviour $b$ in \cite{vogdroher16} has the same shape as the structure $u$ over which the formula is interpreted (i.e., $\pos(b)=\pos(u)$). In the present paper, the behaviour $\zeta$ is obtained by replacing, at each position $w \in \pos(\xi)$, the tree fragment $\xi(w)(x_1,\ldots,x_k)$ by the tree fragment
\[\langle(p_1,g_1),*\rangle\Big(\ldots\langle(p_n,g_n),*\rangle \big(\langle(p,f_1\ldots f_k),\xi(w)\rangle(x_1,\ldots,x_k)\big)\ldots\Big)
\]
for some $n\in \mathbb{N}$, $\langle (p_i,g_i),*\rangle \in \Delta^{(1)} \times \{*\}$, and $\langle(p,f_1\ldots f_k),\xi(w)\rangle \in \Delta^{(k)} \times \Sigma^{(k)}$. Then we might say that the sequence
\[
\langle(p_1,g_1),*\rangle\ldots\langle(p_n,g_n),*\rangle \langle(p,f_1\ldots f_k),\xi(w)\rangle
\]
is the \emph{segment of $\zeta$ belonging to $w$}. 

In particular, $\xi$ can be embedded into $\zeta$, i.e., there is a unique bijection $\theta_{\xi,\zeta}\colon\pos(\xi) \rightarrow \pos_\Theta(\zeta)$ which preserves the lexicographic order, where 
$\Theta= \DS\setminus (\Delta\ui 1\times\{*\})$ (cf. Section \ref{sec:decomp} and Fig.~\ref{Fig1}). This extension of $\xi$ will allow us to cope with chain rules in our B\"uchi-Elgot-Trakhtenbrot theorem.

2) B-expressions of \cite{vogdroher16} are almost the same as M-expressions except that, in addition,  there is a  variable $B$, which is assigned to a behaviour, and in a B-expression of the form $\varphi \rhd e$, the guard formula $\varphi$ may contain atomic formulas of the form $(B(x)=(p,f))$; they  allow to check whether, for a variable assignment $\rho$, the behaviour $\rho(B)$ carries the symbol $(p,f)$ at position $\rho(x)$. In M-expressions such atomic formulas do not occur, and the M-expression $e$ in the formula  $\sum\nolimits^{\textrm{beh}}\nolimits e$ is a sentence, i.e., does not contain any free variable (in particular, it does not contain the variable $B$). The information about the behaviour is coded into the tree $\zeta\in T_{\langle \Delta,\Sigma\rangle}$ over which the M-expression $e$ is interpreted.
For instance, an atomic formula of the form $(B(x)=(p,f))$ occurring in a B-expression will be represented by the formula $\bigvee_{\sigma \in \Sigma^{(1)}} \myrm{label}_{\langle (p,f),\sigma \rangle}(x)$ in the M-expression (note that this replacement only makes sense if the variable $x$ is associated with a unary position). 
 This modular definition of syntax and semantics makes it possible to apply directly results on M-expressions known from the literature. In particular, when we will prove the B\"uchi-Elgot-Trakhtenbrot theorem, there is no need for a proof by induction on the structure of M-expressions to show that such formulas induce recognizability/regularity (as it was necessary for B-expression  in \cite{vogdroher16}); instead, we can directly apply the appropriate result from the literature.

\begin{example}\rm Here we show an example of our new logic.
	For this, recall from Example \ref{ex:run} the $(\PD-op^1,\Sigma,M_{\Nat_\infty})$-recognizable tree language $s$, which maps each tree of the form $\sigma(\delta^n(\alpha),\delta^n(\alpha))$ to $8^n$ ($n\ge 0$) and any other tree to 0. This tree language can be defined by a $(\Delta,\Sigma,M_{\Nat_\infty})$-expression $\sum\nolimits^\textrm{beh}\nolimits e $ for the ranked alphabet  $\Delta$ corresponding to $\Sigma$, $P' =\{\true,\ttop=\gamma_0,\ttop=\gamma\}$, and $F'=\{\push(\gamma),\pop\}$ and an appropriate sentence $e\in \MExp(\DS,M_{\Nat_\infty})$.

For the specification of $e$ we need some auxiliary formulas. We define the 
binary relation $\mathrm{edge}(x,y)$ by
\[\mathrm{edge}(x,y)=\edge_1(x,y)\lor\ldots\lor\edge_{\mathrm{maxrk(\DS)}}(x,y)\enspace.
\]
Moreover, we define the binary relation $\mathrm{edge}^+(x,y)$ to be the  transitive closure of $\mathrm{edge}(x,y)$; recall from \cite[p. 43]{CouEng12} that it can be defined by a formula in MSO($\DS$).
Finally, we denote the formula $\neg \varphi \vee \psi$ by $\varphi \rightarrow \psi$ for every $\varphi, \psi \in \mathrm{MSO}(\DS)$.

	Now we construct
	\[e=\varphi\rhd \mathrm H(\omega) \qquad \text{with} \qquad \varphi=\exists x.(\varphi_{\mathrm{label}}(x)\land\varphi_{\mathrm{above}}(x)\land\varphi_{\mathrm{below}}(x)).\]

	Intuitively, for every $\xi\in T_\Sigma$ and behaviour $\zeta\in\B_\Delta(\xi)$ on $\xi$, the formula $\varphi$ determines a position $v \in \pos(\zeta)$ such that the following holds (recall the definition of $\theta_{\xi,\zeta}$ from the beginning of Section \ref{sec:decomp}): 
\begin{compactitem}
\item $(\zeta,[x \mapsto v]) \models \varphi_{\mathrm{label}}(x)\land\varphi_{\mathrm{above}}(x)$ iff the segment of $\zeta$ belonging to $\theta_{\xi,\zeta}^{-1}(v)$
has the form
\[
\langle(\true,\push(\gamma)),*\rangle^n \ \langle(\true,\id_{\Gamma^+}\id_{\Gamma^+}),\sigma\rangle
\]
for some $n \in \mathbb{N}$
where $\sigma$ is the label of $\xi$ at $\theta_{\xi,\zeta}^{-1}(v)$

\item $(\zeta,[x\mapsto v]) \models \varphi_{\mathrm{below}}(x)$ iff
 for each position $v' \in \pos(\zeta)$ below $v$, 
the segment of $\zeta$ belonging to $\theta_{\xi,\zeta}^{-1}(v')$
has the form 
\[
\langle (\ttop=\gamma,\pop),\delta\rangle \  \textrm{ or } \  
\langle (\ttop=\gamma_0,\varepsilon),\alpha\rangle\enspace.
\]
\end{compactitem}

	Formally, let 
	\begin{compactitem}
		\item $\varphi_{\mathrm{label}}(x)=\mathrm{label}_{\langle(\true,\id_{\Gamma^+}\id_{\Gamma^+}),\sigma\rangle}(x)$,
		\item $\varphi_{\mathrm{above}}(x)=\forall y. \mathrm{edge}^+(y,x)\to 
\mathrm{label}_{\langle(\true,\push(\gamma)),*\rangle}(y),$
		\item $\varphi_{\mathrm{below}}(x)=\forall y. \mathrm{edge}^+(x,y)\to(\mathrm{label}_{\langle(\ttop=\gamma,\pop),\delta\rangle}(y)\lor\mathrm{label}_{\langle(\ttop=\gamma_0,\varepsilon),\alpha\rangle}(y))$.
	\end{compactitem}
	Moreover, we define the $\DS$-family of operations $\omega$ by letting
	\begin{align*}
		&\omega_{\langle(\true,\id_{\Gamma^+}\id_{\Gamma^+}),\sigma\rangle}=\mul_{2,1},\\
		&\omega_{\langle(\true,\push(\gamma)),*\rangle}=\omega_{\langle(\ttop=\gamma,\pop),\delta\rangle}=\mul_{1,2},\\
		&\omega_{\langle(\ttop=\gamma_0,\varepsilon),\alpha\rangle}=\mul_{0,1}.
	\end{align*}

This finishes the construction of $e = \varphi \rhd \mathrm H(\omega)$. Next we show that $\seml \sum\nolimits^\textrm{beh}\nolimits e \semr =s$. For this, let $\xi \in T_\Sigma$. Then:
\begin{align*}
\seml \sum\nolimits^\textrm{beh}\nolimits e \semr( \xi) 
&= (\B_\Delta;\seml e\semr)(\xi) = \sum_{\zeta \in \B_\Delta(\xi)} \seml e\semr(\zeta)\\
&= \sum_{\zeta \in \B_\Delta(\xi)} \seml \varphi\rhd \mathrm H(\omega)\semr(\zeta) 
= \sum_{\zeta \in \B_\Delta(\xi) \cap \mathcal{L}_\emptyset(\varphi)} \seml \mathrm H(\omega)\semr(\zeta) \enspace.
\end{align*}
Now we analyse the set of models of $\varphi$. If $\xi\in\supp(s)$, i.e., $\xi=\sigma(\delta^n(\alpha),\delta^n(\alpha))$ for some $n\geq 0$, then the set $\B_\Delta(\xi)\cap\mathcal{L}_\emptyset(\varphi)$ consists of exactly one tree, denoted as $\zeta_\xi$, that is determined as follows. 

	\begin{compactitem}
		\item If $n=0$, then $\zeta_\xi=\langle b_1,\sigma\rangle\big(\langle b_2,\alpha\rangle,\langle b_2,\alpha\rangle\big)$, and
		\item if $n\geq 1$, then $\zeta_\xi=\langle b_3,*\rangle^n\big(\langle b_4,\sigma\rangle(\zeta',\zeta')\big)$  
	\end{compactitem}
	where $\zeta'=\langle b_5,\delta\rangle^n\big(\langle b_2,\alpha\rangle\big)$ and $b_1=(\true,\id_{\Gamma^+}\id_{\Gamma^+})$, $b_2=(\ttop=\gamma_0,\varepsilon)$, $b_3=(\true,\push(\gamma))$,  $b_4=(\true,\id_{\Gamma^+}\id_{\Gamma^+})$, and $b_5=(\ttop=\gamma,\pop)$.  Thus 
\[\sum_{\zeta \in \B_\Delta(\xi) \cap \mathcal{L}_\emptyset(\varphi)} \seml \mathrm H(\omega)\semr(\zeta)=\lBert \mathrm H(\omega)\rBert(\zeta_\xi)\enspace.\] 
Moreover, we have 
\[\lBert \mathrm H(\omega)\rBert(\zeta_\xi) = \underbrace{\mul_{1,2}(\ldots \mul_{1,2}(}_{n}
\mul_{2,1}(u,u) \underbrace{)\ldots)}_{n}
\]
with $u = \underbrace{\mul_{1,2}(\ldots \mul_{1,2}(}_{n}\mul_{0,1}\underbrace{) \ldots )}_{n}$ for every $n \ge 0$. Since $u=2^n$, we obtain:
 \begin{align*}
\seml \sum\nolimits^\textrm{beh}\nolimits e \semr( \xi) 
= \seml \mathrm H(\omega)\semr(\zeta_\xi) = 8^n = s(\xi) \enspace.
\end{align*}

If  $\xi\notin\supp(s)$, then $\B_\Delta(\xi)\cap\mathcal{L}_\emptyset(\varphi)=\emptyset$ and thus we have
 \begin{align*}
\seml \sum\nolimits^\textrm{beh}\nolimits e \semr( \xi) 
= \sum_{\zeta \in \emptyset} \seml \mathrm H(\omega)\semr(\zeta) = 0 \enspace.
\end{align*}

Hence $\seml \sum\nolimits^\textrm{beh}\nolimits e \semr = s$.
\end{example}

%%%%%%%%%%%%%%%%%%%%%%%%%%%%%%%%%%%%%%%%%%%%

As first result we prove a B\"uchi-Elgot-Trakhtenbrot theorem for $(S,\Sigma,K)$-rtgs.

\begin{theorem}\label{th:main} $\Reg(S,\Sigma,K) = \Def(S,\Sigma,K)$.
\end{theorem}

\begin{proof} First we prove $\Reg(S,\Sigma,K) \subseteq \Def(S,\Sigma,K)$. Let $s$ be an $(S,\Sigma,K)$-regular weighted tree language.  By Theorem~\ref{th:sep-storage} there are finite sets $P' \subseteq P$ and $F' \subseteq F$ and there is a chain-free $(\langle \Delta,\Sigma\rangle,K)$-rtg $\G$ such that $\Delta$ is the ranked alphabet corresponding to $\Sigma$, $P'$, and $F'$ and $s = \B_\Delta ; \seml \G \semr$. By Theorem \ref{th:main-fulstuvog12}, there is a sentence $e \in \MExp(\langle \Delta,\Sigma\rangle,K)$ such that $\seml \G \semr = \seml e\semr$. Then, by Definition \ref{df:(Delta,Sigma,K)-expressions}, we have that $s$ is $(S,\Sigma,K)$-definable.

By reading the above proof backwards, we obtain the proof of $\Def(S,\Sigma,K) \subseteq \Reg(S,\Sigma,K)$.
\end{proof}

As second result we prove that expressions with behaviours generalize  M-expressions as defined in \cite{fulstuvog12,fulvog15}. 

\begin{theorem}\label{lm:TRIV-M=Def} Let $s: T_\Sigma \rightarrow K$ be a weighted tree language. Then the following two statements hold.
\begin{compactenum}
\item If $s =\lBert e \rBert$ for some sentence $e \in \MExp(\Sigma,K)$, then $s$ is $(\mathrm{TRIV},\Sigma,K)$-definable.

\item If $K$ is compressible and  $s$ is $(\mathrm{TRIV},\Sigma,K)$-definable, then $s =\lBert e \rBert$ for some sentence $e \in \MExp(\Sigma,K)$.
\end{compactenum}
\end{theorem}
\begin{proof} We abbreviate the only predicate $\true_{\{c\}}$ and the only instruction $\id_{\{c\}}$ of $\TRIV$ by $\true$ and $\id$, respectively. Let $\Delta$ be the ranked alphabet corresponding to $\Sigma$, $\{\true\}$, and $\{\id\}$ and $\DS$ as defined in Subsection \ref{subsec:storage}.

For each  $\xi \in T_\Sigma$, the set $\B_\Delta(\xi)$ contains exactly one element $\zeta$ with $\pos(\zeta)=\pos(\xi)$.
We denote this element by $\zeta_\xi$. Then for each $w \in \pos(\xi)$ we have $\zeta_\xi(w) = \langle (\true,\id\ldots\id),\xi(w)\rangle$ where the number of occurrences of $\id$ is the rank of $\xi(w)$.

Proof of  1:  Let $e \in \MExp(\Sigma,K)$ be a sentence. Let us construct the formula $\bar{e}\in \MExp(\DS,K)$ that can be obtained from $e$ by replacing each subformula of the form $\mathrm{label}_\sigma(x)$ by $\mathrm{label}_{\langle(\true,\id\ldots\id),\sigma\rangle}(x)$ where the number of occurrences of $\id$ is the rank of $\sigma$.  Then $\lBert \bar{e}\rBert\bigl(\zeta_\xi)
= \lBert e\rBert(\xi)$ for every  $\xi \in T_\Sigma$. Moreover, let $e'=\varphi\rhd \bar{e}$, where 
  $\varphi=\neg\exists x.  \mathrm{label}_{\langle(\true,\id),*\rangle}(x)$. Then
  \[\mathcal{L}_\emptyset(\varphi)=\{\zeta\in T_{\DS}\mid\neg\exists w\in\pos(\zeta):\zeta(w)=\langle(\true,\id),*\rangle\}.\]
 Thus, for each $\xi \in T_\Sigma$ we have
\begin{align*}
\lBert \sum\nolimits^{\textrm{beh}}\nolimits e' \rBert(\xi) 
& = \sum_{\zeta \in \B_\Delta(\xi)} \lBert e'\rBert\bigl(\zeta\bigr)
 = \sum_{\zeta \in \B_\Delta(\xi)} \lBert \varphi\rhd\bar{e}\rBert\bigl(\zeta\bigr)
= \lBert \bar{e}\rBert\bigl(\zeta_\xi)
= \lBert e\rBert(\xi)\enspace,
\end{align*}
where the last but one equality is due to the fact that $\B_\Delta(\xi)\cap \mathcal{L}_\emptyset(\varphi)=\{\zeta_\xi\}$.

Proof of  2:  Let $s$ be   $(\mathrm{TRIV},\Sigma,K)$-definable. By Theorem~\ref{th:main} we have that $s$ is $(\mathrm{TRIV},\Sigma,K)$-regular. Since $K$ is compressible, Lemma~\ref{lm:reg=nc-reg} implies that $s$ is even chain-free $(\mathrm{TRIV},\Sigma,K)$-regular.
By Theorem \ref{th:main-fulstuvog12} we obtain that there is a sentence $e \in \MExp(\Sigma,K)$ such that $\seml e \semr = s$.
\end{proof}

\acknowledgments We thank the anonymous reviewers for their helpful comments.

\nocite{*}
%\bibliographystyle{abbrvnat}
% use the following instead if you encounter problems 
\bibliographystyle{alpha}
\label{sec:biblio}

\end{document}